\documentclass
  [
    DIV=14,
    fontsize=12,
    abstract=true, 
    bibliography=totoc
  ]
  {scrartcl}

\setkomafont{date}{\small\rmfamily}
\setkomafont{author}{\large\sffamily}

\usepackage{authblk}

\newcommand
{\NameEmail}[2]
{%
  \stackunder%
  {#1}%
  {%
   \clap{%
   \normalfont%
   \footnotesize%
   \texttt{#2}%
  }}%
  \hspace{-3.5pt}%
}

\newcommand\keywords[1]{%
  \begingroup
  \renewcommand\thefootnote{}
  \footnote{{Keywords:} #1}%
  \addtocounter{footnote}{-1}
  \endgroup
}
\newcommand\MSCcodes[1]{%
  \begingroup
  \renewcommand\thefootnote{}
  \footnote{{MSC codes:} #1}%
  \addtocounter{footnote}{-1}
  \endgroup
}
\newcommand\funding[1]{%
  \begingroup
  \renewcommand\thefootnote{}
  \footnote{{Funding:} #1}%
  \addtocounter{footnote}{-1}
  \endgroup
}

\usepackage{scrlayer-scrpage}
\clearpairofpagestyles 
\ihead{} 
\ohead{\headmark}
\automark{section}

\NewCommandCopy{\oldparagraph}{\paragraph}
\RenewDocumentCommand{\paragraph}{s t{!} O{#4} m}{%
  \IfBooleanTF{#1}
    {\IfBooleanTF{#2}{\oldparagraph*{#4}}{\oldparagraph*{#4.}}}
    {\IfBooleanTF{#2}{\oldparagraph[#3]{#4}}{\oldparagraph[#3]{#4.}}}%
}

\usepackage{caption}
\usepackage
  [leftcaption]
  {sidecap}

\usepackage{floatpag}

\usepackage{scalerel}

\usepackage{enumitem}
\setlist[enumerate,1]{label=\textbf{\textup{(\roman*)}}}
\setlist[enumerate,2]{label=\textbf{\textup{(\alph*)}}}
\setlist{
  topsep=1pt,
  itemsep=2pt,
  leftmargin=1cm,
  before=\leavevmode
}

\usepackage[
  backend=biber,
  style=alphabetic,
  giveninits=true, 
  backref=true
]{biblatex}
\usepackage[T1]{fontenc}
\usepackage{lmodern}
\usepackage{anyfontsize}

\usepackage{multirow}
\usepackage{hhline}

\usepackage{tabularray}
\UseTblrLibrary{booktabs}

\usepackage{float}
\usepackage{graphicx}

\usepackage{adjustbox}
\usepackage{tcolorbox}

\newtcolorbox{standout}{
  colback=gray!15,
  boxrule=0pt,
  left=.3cm,
  right=.3cm,
  top=.18cm,
  bottom=.18cm,
  boxsep=0pt
}

\usepackage{amsmath}
\allowdisplaybreaks
\usepackage{mathtools}
\usepackage{amssymb}
\usepackage{amsthm}
\usepackage{xfrac}
\usepackage{stmaryrd} 
\usepackage{stackengine} 
\usepackage[safe]{tipa} 
\usepackage{centernot}
\usepackage[
  cal=euler,
  scr=dutchcal
]
 {mathalpha} 

 \newcommand{\bracket}[3]{%
  \stretchleftright
    {#1}
    {%
      \ensurestackMath{\addstackgap[1pt]{#2}}%
      \vrule width 0pt depth 2pt height 0pt
    }
    {#3}%
} 
\newcommand{\scaledbracket}[3]{%
  \ThisStyle{%
    \stretchleftright
      {#1}
      {
        \ensurestackMath{\addstackgap[1pt]{\SavedStyle #2}}%
        \vrule width 0pt depth 1.5pt height 0pt
      }
      {#3}%
  }%
}
\newcommand{\bracketmid}[4]{%
  \stretchleftright{#1}{%
    \ensurestackMath{%
      \addstackgap[1pt]{#2}%
      \,\stretchrel*{|}{\addstackgap[1pt]{#2#3}}\,%
      \addstackgap[1pt]{#3}%
    }%
  }{#4}%
}

\theoremstyle{plain}
\newtheorem{theorem}{Theorem}[section]
\newtheorem{lemma}[theorem]{Lemma}
\newtheorem{proposition}[theorem]{Proposition}

\theoremstyle{definition}
\newtheorem{definition}[theorem]{Definition}

\newtheorem{notation}[theorem]{Notation}

\theoremstyle{remark}
\newtheorem{remark}[theorem]{Remark}
\usepackage[
  colorlinks=true,
  linkcolor=darkgreen,
  citecolor=darkgreen,
  urlcolor=darkgreen
]{hyperref}
\usepackage{xurl} 

\DeclareSourcemap{
  \maps[datatype=bibtex]{
    \map{
      \step[fieldsource=url, final=true]
      \step[fieldset=verba, origfieldval, final=true]
      \step[fieldsource=verba, match=\regexp{\Ahttps?://}, replace={}]
    }
  }
}
\DeclareFieldFormat{url}{%
  \mkbibacro{URL}\addcolon\space
  \href{#1}{\nolinkurl{\thefield{verba}}}%
}

\usepackage{cleveref}
\crefname{equation}{}{}
\crefname{section}{\S}{\S\S}
\crefname{subsection}{\S}{\S\S}
\crefname{subsubsection}{\S}{\S\S}
\crefname{definition}{Def.}{Defs.}
\crefname{theorem}{Thm.}{Thms.}
\crefname{corollary}{Cor.}{Cors.}
\crefname{lemma}{Lem.}{Lems.}
\crefname{proposition}{Prop.}{Props.}
\crefname{remark}{Rem.}{Rems.}
\crefname{notation}{Ntn.}{Ntns.}
\crefname{fact}{Fact}{Facts}
\crefname{example}{Ex.}{Exs.}
\crefname{figure}{Fig.}{Figs.}
\crefname{table}{Tab.}{Tabs.}
\crefname{footnote}{ftn.}{ftns.}
\Crefname{footnote}{Ftn.}{Ftns.}

\usepackage{tikz}
\usetikzlibrary{
  cd,
  calc,
  nfold,
  arrows.meta,
  backgrounds,
  decorations,
  decorations.pathmorphing, 
}

\tikzcdset{
  arrow style=tikz, 
  diagrams={
    cramped,
    baseline=
    {([yshift=-2pt]current bounding box.center)},
    >={
      Computer Modern Rightarrow[
        length=4pt, width=4pt
      ]
    },
    hook/.style={
      {Hooks[right, length=2pt, width=8pt]}->
    },
    hook'/.style={
      {Hooks[left, length=2pt, width=8pt]}->
    },
    shorten=0pt,
  }
}

\usepackage{xcolor}
\definecolor{darkblue}{rgb}{0.05,0.25,0.65}
\definecolor{darkgreen}{RGB}{20,140,10}
\definecolor{lightgray}{rgb}{0.9,0.9,0.9}
\definecolor{darkorange}{RGB}{200,100,5}
\definecolor{darkyellow}{rgb}{.91,.91,0}
\definecolor{lightpurple}{RGB}{225, 220, 185}

\let\originalsslash\sslash
\renewcommand{\sslash}{\mathord{\originalsslash}}

\makeatletter
\newcommand{\cpt}{\mathpalette\cpt@inner\relax}
\newcommand{\cpt@inner}[2]{%
  \scalebox{0.5}[0.9]{$#1\cup$}
  #1\{\infty\}
}
\makeatother

\makeatletter
\newcommand{\plus}{\mathpalette\sqcpt@inner\relax}
\newcommand{\sqcpt@inner}[2]{%
  \scalebox{0.5}[0.9]{$#1\sqcup$}
  #1\{\infty\}
}
\makeatother

\tikzset{
  snake left/.style={
    rounded corners,
    to path={
      let \p1 = (\tikztostart.east),
          \p2 = (\tikztotarget.west),
          \p3 = ($(\p1)!0.5!(\p2)$),
          \n1 = {8pt} 
      in
      (\p1)
      -- (\x1 + \n1, \y1)
      -- (\x1 + \n1, \y3)
      -- (\x2 - \n1, \y3) \tikztonodes
      -- (\x2 - \n1, \y2)
      -- (\p2)
    }
  }
}

\tikzset{
  uphordown/.style={
    rounded corners,
    to path={
      let \p1 = (\tikztostart.north),
          \p2 = (\tikztotarget.north),
          \n1 = {max(\y1,\y2) + 8pt}
      in
      (\p1)
      -- (\x1, \n1)
      -- (\x2, \n1) \tikztonodes 
      -- (\p2)
    }
  }
}

\tikzset{
  downhorup/.style={
    rounded corners,
    to path={
      let \p1 = (\tikztostart.south),
          \p2 = (\tikztotarget.south),
          \n1 = {min(\y1,\y2) - 8pt}
      in
      (\p1)
      -- (\x1, \n1)
      -- (\x2, \n1) \tikztonodes 
      -- (\p2)
    }
  }
}

\tikzset{
  rightvertleft/.style={
    rounded corners,
    to path={
      let \p1 = (\tikztostart.east),
          \p2 = (\tikztotarget.east),
          \n1 = {max(\x1,\x2) + 8pt}
      in
      (\p1)
      -- (\n1, \y1)
      -- (\n1, \y2) \tikztonodes 
      -- (\p2)
    }
  }
}

\tikzset{
  leftvertright/.style={
    rounded corners,
    to path={
      let \p1 = (\tikztostart.west),
          \p2 = (\tikztotarget.west),
          \n1 = {min(\x1,\x2) - 8pt}
      in
      (\p1)
      -- (\n1, \y1)
      -- (\n1, \y2) \tikztonodes 
      -- (\p2)
    }
  }
}

\newcommand{\inlinetikzcd}[1]{\begin{tikzcd}[sep=small, ampersand replacement=\&]#1\end{tikzcd}}

\newcommand{\acts}{%
  \hspace{1.3pt}%
  \raisebox{1.2pt}{%
    \rotatebox[origin=c]{90}{$%
      \curvearrowright%
    $}%
  }%
  \hspace{.7pt}%
}

\renewcommand{\setminus}{-}

\newcommand{\defneq}{\mathrel{\equiv}}

\newcommand{\Top}{\mathrm{Top}}

\newcommand{\R}{\mathbb{R}}
\newcommand{\LL}{\mathbb{L}}

\newcommand{\Cyc}
  {\operatorname{Cyc}}
\newcommand{\SmlCyc}
  {\operatorname{SmlCyc}}
\newcommand{\MinCyc} 
  {\operatorname{MinCyc}}
\newcommand{\MassMinCyc}
  {\operatorname{mMinCyc}}
\newcommand{\MinTor} 
  {\operatorname{MinTor}}

\newcommand
  {\BorcherdsAlgebra}
  {\mathfrak{bor}}
\newcommand
  {\bIIA}
  {\BorcherdsAlgebra^{+}_{\mathrm{IIA}}}
\newcommand
  {\bIIAMassive}
  {\BorcherdsAlgebra^{+}_{\mathrm{mIIA}}}
\newcommand
  {\FreeDGLie}
  {\mathrm{Free}_{\mathrm{dgL}}}

\newcommand
  {\TheInclusion}{\iota_{\mathrm{min}}}

\makeatletter
\newcommand{\mathcolorbox}[2]{\mathpalette{\@mathcolorbox{#1}}{#2}}
\newcommand{\@mathcolorbox}[3]{\colorbox{#1}{$\m@th#2#3$}}
\makeatother
\newcommand{\graybox}[1]{%
  \begingroup
  \setlength{\fboxsep}{1.5pt}
  \mathcolorbox{black!10}{#1}%
  \endgroup
}
\newcommand{\massive}[1]
  {\graybox{#1}}

\newcommand
  {\Romans}{\varphi_{-1}}

\newcommand{\AttachingSpace}[1]
  {%
    D^3
      \coprod^{#1}_{S^2} 
    BS^1
  }

\begin{document}

\setlength{\abovedisplayskip}{3.5pt}
\setlength{\belowdisplayskip}{3.5pt}
\setlength{\abovedisplayshortskip}{-5pt}
\setlength{\belowdisplayshortskip}{3pt}


\title
{%
Flux-Quantized Massive Type IIA
\\ 
Borcherds Hierarchy and D6-Cobordism
}

\author[1]
{\NameEmail
  {Nikita Golub}
  {ng3733@nyu.edu}
}

\author[1,2]
{\NameEmail
  {Hisham Sati}
  {hsati@nyu.edu}
}

\author[1]
{\NameEmail
  {Urs Schreiber}
  {us13@nyu.edu}
}

\affil[1]{Mathematics Program and Center for Quantum and Topological Systems (CQTS), \newline New York University Abu Dhabi, UAE}
\affil[2]{The Courant Institute of Mathematical Sciences, \newline New York University, New York, USA}

\maketitle

\begin{abstract}
We find a global completion of (massive) IIA supergravity  that in the massless sector lifts to the cohomotopical completion of 11D supergravity, and which quantizes at once all fluxes/charges in the (massive) IIA Borcherds hierarchy, including, besides all RR-charges, also all NS-charges, the dual dilaton charge and the infinite tower of higher global Borcherds symmetries. We find that the nonabelian cohomology theory of this charge hierarchy---a massive deformation of \emph{minimal cyclic 4-Cohomotopy}---is a nonabelian deformation of twisted K-theory which is equivalently the (unstable) cobordism cohomology theory of D6-brane/KK-monopole worldvolumes.
\end{abstract}


\MSCcodes{
  Primary:
 81T30, 
 81T70, 
 19L50, 
 55N20, 
 55P35, 
 Secondary:
 55Q55, 
 55S25, 
 53C08, 
 55P62. 
}

\funding{
  by \textit{Tamkeen UAE} under the \textit{Abu Dhabi Research Institute Grant} \texttt{CG008}
}

\newpage

\setcounter{tocdepth}{2}
\tableofcontents

\section
{Motivation}
\label{Motivation}

To start with, recall that the charges $\chi$ of \emph{singular} or \emph{black} branes%
\footnote{Singular branes are those whose locus is a gravitational or higher-electromagnetical singularity and as such deleted from spacetime; as opposed to solitonic branes which are finite concentrations of flux vanishing at infinity, cf. \cite[\S 2.2]{SS25-Flux}.
} 
are encoded by \emph{classifying spaces} $\mathcal{A}$ (cf. \cref{HigherGaugeFields,HomotopyTheory}) for the corresponding nonabelian cohomology theories \cref{NonabelianCohomologyFunctor}, $\chi \in H^1(X^d;\Omega \mathcal{A})$ (cf. review in \parencites[\S 3.2]{SS25-Flux}[\S 4.2]{GSS26-SuGra}[\S 4]{SS26-HigherGauge}).

\paragraph
{The Example of Maxwell Theory}
This is well familiar in the archetypical example of 4D Maxwell theory subject to the usual \emph{Dirac charge quantization}\footnote{
  Beware that the term ``quantization'' in ``charge quantization'' refers to \emph{discretization}, namely of the set of charges as classical observables, not to quantum observables. Charge quantization is applied to classical field theories and serves to globally define their field content before enhancing  that to the corresponding quantum (field) theory.
} 
condition (\parencites{Dirac1931}{WuYang76}, cf. \parencites{Alvarez1985}[\S 7.1]{Brylinski1993}[\S 16.4e]{Frankel2011}):
Here the classifying space is that of the gauge group $\mathrm{U}(1)$, denoted 
\begin{equation}
  \mathcal{A} 
    \defneq 
  B \mathrm{U}(1) \simeq B^2 \mathbb{Z}
  \mathrlap{\,.}
\end{equation}
On a globally hyperbolic 4D spacetime of the form $X^{1,3} \simeq \mathbb{R}^{1,0} \times X^3$, the total magnetic charge is characterized by the homotopy class of a classifying map from $X^3$ to $B^2 \mathbb{Z}$, which corresponds (cf. \parencites[p. 263]{FomenkoFuchs2016}[Cor. 12.1.20]{AguilarGitlerPrieto2002}) to an element in the integral 2-cohomology of the spatial Cauchy surface $X^3$:
\begin{equation}
\label{Ordinary2Cohomology}
  \mathcolor{gray}{\substack{
    \text{homotopy}
    \\
    \text{classes of maps}
  }}
  \;\;
  \pi_0\, \mathrm{Map}\bracket({
    X^3
    ,\,
    B^2 \mathbb{Z}
  })
  \simeq
  H^2\bracket({
    X^3
    ;\,
    \mathbb{Z}
  })
  \;\;
  \mathcolor{gray}{\substack{
    \text{integral} 
    \\
    \text{cohomology}
  }}
  \mathrlap{\,.}
\end{equation}
For instance, when $X^3$ is the near horizon spacetime of a \emph{magnetic monopole}, a magnetic Reissner--Nordstr{\"o}m spacetime (cf. \parencites{Carter1968}{Carter2009}), then $X^3 \simeq \mathbb{R}^3 \setminus \{0\}$ is the complement of the singularity in Euclidean topological space. Since this is homotopy equivalent to the \emph{linking sphere} around the (deleted) monopole (cf. \parencites[\S~1.2.1]{Marolf2012}),
\begin{equation}
  \mathbb{R}^3 \setminus \{0\}
  \underset
    {\mathrm{homeo}}
    {\simeq}
  S^2 
  \times 
  \mathbb{R}^1_{> 0}
  \underset
    {\mathrm{hmtp}}
    {\simeq}
  S^2
  \mathrlap{\,,}
\end{equation}
the magnetic charges are integers associated with this sphere:
\begin{equation}
  \label{MagneticAndElectricCharge}
  H^2\bracket({
    \mathbb{R}^3
    \setminus
    \{0\}
    ;\,
    \mathbb{Z}
  })
  \simeq
  H^2\bracket({
    S^2
    ;\,
    \mathbb{Z}
  })
  \simeq
  \mathbb{Z}
  \,.
\end{equation}

If one demands that also the singular electric charges (electric monopoles) are ``quantized'' (discretized) in this way, then the relevant classifying space becomes 
\begin{equation}
  \mathcal{A} 
    \defneq 
  B^2 \mathbb{Z} \times B^2 \mathbb{Z} 
    \simeq 
  B^2\bracket({ \mathbb{Z}^2 })
  \mathrlap{\,,}
\end{equation}
and the corresponding cohomology theory is two copies of integral cohomology:
\begin{equation}
  \pi_0
  \,
  \mathrm{Map}\bracket({
    X^3, B^2 \mathbb{Z}^2
  })
  \simeq
  H^2\bracket({
    X^3
    ;\,
    \mathbb{Z}^2
  })
  \simeq
  \underbrace{
  H^2\bracket({
    X^3
    ;\, 
    \mathbb{Z}
  })
  }_{\mathrm{magnetic}}
  \times
  \underbrace{
  H^2\bracket({
    X^3
    ;\, 
    \mathbb{Z}
  })
  }_{\mathrm{electric}}
  \mathrlap{\,.}
\end{equation}
This duality-symmetric quantization of electromagnetic charges is the classical sector \parencites[(1.26)]{FreedMooreSegal2007} of \parencites[(3.27)]{FreedMooreSegal2007b}, 
also considered in \parencites[Rem. 2.3]{BeckerBeniniSchenkelSzabo2017}, and as a charge quantization of 4D supergravity in \parencites{LazaroiuShahbazi2018}{LazaroiuShahbazi2022a}{LazaroiuShahbazi2022b}.

\paragraph
{The Example of Type IIA Charge in K-Theory---Hypothesis K}
\label{HypothesisK}

The joint quantization of both magnetic and electric charges may be more familiar in the context of the proposal that type IIA D$p$-brane charges are quantized in K-theory (\parencites{MinasianMoore1997}{Witten1998}{FreedWitten1999}{MooreWitten2000}{FreedHopkins2000}{BouwknegtMathai2000}, cf. \parencites[\S~3.7]{Freed2001}{Freed2002}{Szabo2002}{Fredenhagen2008}{DistlerFreedMoore2011}{GS22-KTheory}[\S~4.1]{SS25-Flux}). Here the classifying space for K-theory is $\mathcal{A} \defneq B \mathrm{U} \massive{\times \mathbb{Z}} := \bracket({\bigcup_{n} B \mathrm{U}(n) }) \massive{\times \mathbb{Z}}$, which predicts that not only the magnetic D4-, D6-$\massive{\text{, and D8}}$-branes but also the electric $\massive{\text{D(-2)}}$-, D0- and D2-branes carry integer charges:
\begin{equation}
  \label{KTheoryBraneCharges}
  \pi_0\, 
  \mathrm{Map}\big({
    \underbrace{
    \mathbb{R}^9 
      \setminus 
    \mathbb{R}^p
    }_{\simeq S^{8-p}}
    ,\,
    B \mathrm{U}
    \massive{\times \mathbb{Z}}
  }\big)
  \simeq
  \begin{cases}
    \mathbb{Z} & \text{if } p \in 
    \{
      \massive{-2},
      0,2,4,6,
      \massive{8}
    \}
    \\
    0 & \text{otherwise}
  \end{cases}
  \mathrlap{\,.}
\end{equation}
On the other hand, on the quantization of NS-brane charges the K-theory proposal treats only the NS5-brane charge, by positing that in their presence the classifying space is that of twisted K-theory, $\mathcal{A} \defneq B \mathrm{U} \sslash B \mathrm{U}(1)$, which gives
\begin{equation}
\label
{BraneChargesInTwistedKTheory}
  \pi_0\, \mathrm{Map}\big({
    \underbrace{
    \mathbb{R}^9 
      \setminus 
    \mathbb{R}^p
    }_{\simeq S^{8-p}}
    ,\,
    \bracket({
      B \mathrm{U}
      \massive{\times \mathbb{Z}}
    })
    \sslash
    B \mathrm{U}(1)
  }\big)
  \simeq
  \left\{
  \begin{array}{cll}
    \mathbb{Z} & \text{if } p \in 
    \{
      \massive{-2},
      0,2,4,6,
      \massive{8}
    \}
    &
    \substack{
      \text{D-branes}
    }
    \\
    \mathbb{Z} & \text{if } p = 5
    &
    \substack{
      \text{NS5-brane}
    }
    \\
    0 & \text{otherwise}
    &
    \substack{
      \text{nothing else}
    }
  \end{array}
  \right.
  \mathrlap{\,.}
\end{equation}
Hence the NS1-brane charge---the string charge no less---is missed by brane charge quantization in K-theory. This already shows that the K-theory proposal is not the last word on the quantization of type II brane charges. Our concern here is to further explore its possible enhancements.

\paragraph
{Fluxes Sourced by Charges}

The key to the question of which charge quantization laws---embodied by classifying spaces $\mathcal{A}$---are in principle admissible, is their relation to the systems of flux densities that are \emph{sourced} by these charges.

This is again most familiar in the case of 4D Maxwell theory. Here the magnetic flux density is a closed differential 2-form $B_2 \in \Omega^2_{\mathrm{cl}}\bracket({X^3})$, and its \emph{total flux} is the non-torsion charge in question. On a Cauchy surface $X^3$ this form is independent of the electric flux density $E_2 \in \Omega^2_{\mathrm{cl}}\bracket({X^3})$, which similarly corresponds to the electric charge \cref{MagneticAndElectricCharge}. We may regard the pair $(B_2, E_2)$ as a single closed 2-form with coefficients in $\mathbb{R}^2$ and hence furthermore as a closed 1-form, with coefficients in the $L_\infty$-algebra $\mathbb{R}^2[1]$ obtained by shifting $\mathbb{R}^2$ up by one degree:
\begin{equation}
\label{FluxQuantizationSchematicsForMaxwell}
  \begin{tikzcd}[
    sep=0pt
  ]
  \Omega^1_{\mathrm{cl}}\bracket({
    X^3
    ;\,
    \mathbb{R}^2[1]
  })
  \ar[rr]
  &&
  \pi_0
  \mathrm{Map}\bracket({
    X^3
    ,\,
    B^2 \mathbb{R}^2
  })
  &&
  \pi_0
  \mathrm{Map}\bracket({
    X^3
    ,\,
    B^2 \mathbb{Z}^2
  })
  \mathrlap{\,.}
  \ar[ll]
  \\
  \underset
  {
    \text{flux densities}
  }
  {
  \bracket({
    B_2, E_2
  })
  }
  &\longmapsto&
  \underset
  {
    \text{total fluxes}
  }
  {
  \bracket({
    \int_{(\text{-})} B_2,
    \int_{(\text{-})} E_2
  })
  }
  =
  \underset
  {
    \text{real charges}
  }
  {
  ({
    \chi_{\mathrm{mg}} \otimes \mathbb{R},
    \chi_{\mathrm{el}} \otimes \mathbb{R}
  })
  }
  &\longmapsfrom&
  \underset
  {
    \text{charges}
  }
  {
  ({
    \chi_{\mathrm{mg}},
    \chi_{\mathrm{el}}
  })
  }
  \end{tikzcd}
\end{equation}
This is the gist of Dirac flux/charge quantization: The total fluxes reflect the real charges. What makes this tick is that:
\begin{equation}
  \left(
  \begin{aligned}
    \mathrm{d}\, B_2 & = 0
    \\
    \mathrm{d}\, E_2 & = 0
  \end{aligned}
  \right)
  \Rightarrow
  \left[
  \begin{aligned}
    B_2
    \\
    E_2
  \end{aligned}
  \right]
  \in
  H^1_{\mathrm{dR}}\bracket({
    X^3
    ;\,
    \mathbb{R}^2[1]
  })
  \simeq
  H^2\bracket({
    X^3
    ;\,
    \mathbb{Z}^2
  })  
  \otimes 
  \mathbb{R}
  \mathrlap{\,,}
\end{equation}
hence that:
\begin{standout}
  The Bianchi/Gauss laws on fluxes 
  characterize cocycles 
  in the real form of the 
  cohomology theory of charges.
\end{standout}

\paragraph
{Whitehead $L_\infty$-Algebras}%
\hspace{-15pt}\footnote{
 We make free use of concepts and notation from higher Lie theory as summarized in \cref{app:HigherLieAlgebras}.
}
This relation between Bianchi/Gauss laws on flux densities and admissible cohomology theories of charges holds very generally:

For any connected classifying space $\mathcal{A}$%
\footnote{
  Throughout we demand that classifying spaces $\mathcal{A}$ are connected or have a set of connected components identified with $\mathbb{Z}$, and that connected components are \emph{nilpotent} and \emph{of finite rational type}, cf. \parencites[Def. 5.1]{FSS23-Char}.
}
  there is (\cref{MinimalSullivanModel,WHBracketLinfinityAlgebras}) an $L_\infty$-algebra structure
  \begin{equation}
  \label{TheWhiteheadBracketAlgebra}
    \bracket({\mathfrak{l}\mathcal{A}, \bracket({ [\text{-}], [\text{-},\text{-}], [\text{-},\text{-},\text{-}], \cdots  })})
    \in
    L_\infty
    \mathrm{Alg}
  \end{equation}
  on the real homotopy groups of its loop space
  \begin{equation}
  \label
  {UnderlyingChOfWhiteheadLInfinity}
    \bracket({
      \mathfrak{l}\mathcal{A}
    })_\bullet
    =
    \pi_\bullet
    \bracket({
      \Omega \mathcal{A}
    })
    \otimes_{_{\mathbb{Z}}}
    \mathbb{R}
  \end{equation}
  whose brackets are the \emph{higher Whitehead brackets} of $\mathcal{A}$ (cf. \cite[Props.~5.11--5.13 \& Rem.~5.4]{FSS23-Char}). 
  
 In particular, the cochain cohomology of the Chevalley-Eilenberg algebra $\mathrm{CE}(\text{-})$ \cref{TheCEAlgebra} of $\mathfrak{l}\mathcal{A}$ reproduces the real cohomology of $\mathcal{A}$:
  \begin{equation}
  \label
  {CohomologyOfCEOfWhiteheadLinfinity}
    H^\bullet\bracket({
      \mathrm{CE}\bracket({
        \mathfrak{l}\mathcal{A}
      })
    })
    \simeq
    H^\bullet\bracket({
      \mathcal{A}
      ;\,
      \mathbb{R}
    })
    \mathrlap{\,.}
  \end{equation}

It is \emph{this} $L_\infty$-algebra \cref{TheWhiteheadBracketAlgebra} which characterizes flux species and their Bianchi/Gauss laws which are sourced by the charges that are classified by $\mathcal{A}$ (\parencites{SS24-Phase}, exposition in \parencites[\S 3]{SS25-Flux}[\S\S 3--4]{SS26-HigherGauge}), in generalization of \cref{FluxQuantizationSchematicsForMaxwell}:
\begin{equation}
\label
{GeneralFluxQuantizationSchematics}
  \begin{tikzcd}[
    sep=0pt
  ]
  \Omega^1_{\mathrm{cl}}\bracket({
    X^d
    ;\,
    \mathfrak{l}\mathcal{A}
  })
  \ar[rr]
  &&
  \pi_0
  \mathrm{Map}\bracket({
    X^d
    ,\,
    \mathcal{A}_\mathbb{R}
  })
  &&
  \pi_0
  \mathrm{Map}\bracket({
    X^d
    ,\,
    \mathcal{A}
  })
  \mathrlap{\,.}
  \ar[ll]
  \\
  F :=
  \bracket({
    F^{(i)}_{\mathrm{deg}_i}
  })_{i \in I}
  &\longmapsto&
    {[F]}
   =
  \chi \otimes \mathbb{R}
  &\longmapsfrom&
  \chi
  \\[-4pt]
  \substack{
    \text{\color{gray}flux densities}
  }
  &&
  \substack{
    \text{\color{gray}total fluxes}
  }
  \;\;\;
  \substack{
    \text{\color{gray}real charges}
  }
  &&
  \substack{
    \text{\color{gray}charges}
  }
  \end{tikzcd}
\end{equation}

\paragraph
{Rederiving the K-Theory Conjecture}

For example, if one considers the RR-fluxes $F_{2p}$ and (only) the NS $H_3$-flux in (non-massive) type IIA, then admissible charge classifying spaces $\mathcal{A}$ must reproduce their Gauss/Bianchi identities, in that
\begin{equation}
  \Omega^1_{\mathrm{cl}}\bracket({
    X^9
    ;\,
    \mathfrak{l}\mathcal{A}
  })
  \simeq
  \left\{
  \begin{aligned}
    \mathcolor{darkgreen}{F_2}
    & 
    \in 
    \Omega^2_{\mathrm{dR}}\bracket({X^9})
    \\
    \mathcolor{darkgreen}{H_3}
    & 
    \in 
    \Omega^3_{\mathrm{dR}}\bracket({X^9})
    \\
    \mathcolor{darkgreen}{F_4}
    & 
    \in 
    \Omega^4_{\mathrm{dR}}\bracket({X^9})
    \\
    \mathcolor{darkgreen}{F_6}
    & 
    \in 
    \Omega^6_{\mathrm{dR}}\bracket({X^9})
    \\
    \mathcolor{darkblue}{F_8}
    & 
    \in 
    \Omega^8_{\mathrm{dR}}\bracket({X^9})
  \end{aligned}
  \;\middle\vert\;
  \begin{aligned}
    \mathcolor{darkgreen}{\mathrm{d}\, F_2} 
     & 
     \mathcolor{darkgreen}{= 0}
    \mathclap{\phantom{\Omega^8_{\mathrm{dR}}\bracket({X^9})}}
    \\
    \mathcolor{darkgreen}{\mathrm{d}\, H_3} & 
      \mathcolor{darkgreen}{= 0}
    \mathclap{\phantom{\Omega^8_{\mathrm{dR}}\bracket({X^9})}}
    \\
    \mathcolor{darkgreen}{\mathrm{d}\, F_4} 
      & 
      \mathcolor{darkgreen}{= H_3 \, F_2}
    \mathclap{\phantom{\Omega^8_{\mathrm{dR}}\bracket({X^9})}}
    \\
    \mathcolor{darkgreen}{\mathrm{d}\, F_6} 
      & 
      \mathcolor{darkgreen}{= H_3 \, F_4}
    \mathclap{\phantom{\Omega^8_{\mathrm{dR}}\bracket({X^9})}}
    \\
    \mathcolor{darkblue}{\mathrm{d}\, F_8} 
      & 
      \mathcolor{darkblue}{= H_3 \, F_6}
    \mathclap{\phantom{\Omega^8_{\mathrm{dR}}\bracket({X^9})}}
  \end{aligned}
  \right\}
  \mathrlap{\,.}
\end{equation}
There are always infinitely many admissible solutions $\mathcal{A}$ to such a condition, but a suggestive solution is the classifying space for rank-0 twisted K-theory, whose Whitehead-bracket $L_\infty$-algebra is of just this form, in that its minimal Sullivan model is%
\footnote{
  Here $\mathrm{Free}_{\mathrm{dgcA}}(S)$ denotes the free dgc-algebra on a set $S$ of graded generators $s_{\mathrm{deg}(s)} \in S$, and the quotient in \cref{SullivanModelOfTwistedKTheory} is by the indicated differential ideal. 
}
\begin{equation}
  \label{SullivanModelOfTwistedKTheory}
  \mathrm{CE}\bracket({
    \mathfrak{l}
    \bracket({
      B \mathrm{U}
      \sslash
      B \mathrm{U}(1)
    })
  })
  \simeq
  \mathrm{Free}_{\mathrm{dgcA}}
  \left(
  \begin{aligned}
    \mathcolor{darkgreen}{f_2}
    \\
    \mathcolor{darkgreen}{h_3}
    \\
    \mathcolor{darkgreen}{f_4}
    \\
    \mathcolor{darkgreen}{f_6}
    \\
    \mathcolor{darkblue}{f_8}
    \\
    \vdots
  \end{aligned}
  \right)
  \Big/
  \left(
  \begin{aligned}
    \mathcolor{darkgreen}
      {\mathrm{d}\, f_2} 
      & 
      \mathcolor{darkgreen}{= 0}
    \\
    \mathcolor{darkgreen}{\mathrm{d}\, h_3} 
      & 
      \mathcolor{darkgreen}{= 0}
    \\
    \mathcolor{darkgreen}{\mathrm{d}\, f_4} 
      & 
      \mathcolor{darkgreen}{= h_3 \, f_2}
    \\
    \mathcolor{darkgreen}
      {\mathrm{d}\, f_6} 
      & 
      \mathcolor{darkgreen}{= h_3 \, f_4}
    \\
    \mathcolor{darkblue}
      {\mathrm{d}\, f_8} 
      & 
      \mathcolor{darkblue}{= h_3 \, f_6}
    \\
    \mathcolor{darkblue}{\vdots}
  \end{aligned}
  \right)
  \mathrlap{\,.}
\end{equation}
This shows that and how twisted K-theory is an admissible flux/charge quantization law for the $F_{2p}$ and (only) the $H_3$-fluxes. The popular proposal that this is the ``right'' flux/charge quantization law for type IIA we may call \emph{Hypothesis K}.

\paragraph
{The Missing Fluxes}
But this proposal (\emph{Hypothesis K}, that type II charges are in twisted K-theory) is at least incomplete: While it does reflect both magnetic and electric RR-fluxes $F_{2p}$ it does \emph{not} reflect the electric $H_7$-flux and its Bianchi identity (cf. \cite[\S 22.1.3]{Ortin2015} after \parencites{GianiPernici1984}{HuqNamazie1985}{DallAgataEtAl1998}[\S 3]{CremmerJuliaLuPope1998}[(5.9)]{LavrinenkoEtAl1999}[(2.13) \& (2.14)]{GreitzHowe2011}):
\begin{equation}
  \mathcolor{darkorange}{%
  \mathrm{d}\, H_7
  =
  \sfrac{1}{2} \, F_4 \, F_4
  -
  F_2 \, F_6}
  \mathrlap{\,.}
\end{equation}

As such, \emph{Hypothesis K} is manifestly incompatible with lifts to 11D, since in 11D $F_6$ and $H_7$ unify into the electric C-field flux $G_7$, which makes it impossible there to quantize one but not the other.

In fact, there are yet further flux species in type IIA that one may want to subject to flux quantization. The next one is the dual dilaton flux $H_9$, whose Bianchi identity is the equation of motion of the dilaton (\parencites[(2.14)]{GreitzHowe2011}):
\begin{equation}
\label{TheDilatonFlux}
  \mathcolor{magenta}{%
  \mathrm{d}\, H_9
  =
  H_3\, H_7 
  + \sfrac{1}{2} \, F_4 \, F_{6}
  - \sfrac{3}{2} \, F_2 \, F_8}
  \mathrlap{\,.}
\end{equation}
Beyond that there is an infinite tower of further fluxes that are visible on superspace and/or make themselves felt as higher global symmetries.

\paragraph
{Massive IIA}

Last but not least, in parallel to the infinite extension of the IIA Borcherds hierarchy of flux species, there is the corresponding deformation to the Romans massive IIA theory (\parencites{Romans1986}{BergshoeffGreenPapadopoulosTownsend1995}{Tsimpis2005}). This step finally introduces the $F_0$-flux Hodge-dual on spacetime to the $F_{10}$, itself satisfying
\begin{equation}
\label{TheF0Bianchi}
  \massive{\mathrm{d}\, F_0
  \,=\, 0}
  \mathrlap{\,,}
\end{equation}
and further deforming the Bianchi identities throughout the Borcherds algebra (\parencites[\S 5.1]{LavrinenkoEtAl1999}, cf. \cref{OverviewDiagramMassive}), starting with the familiar
\begin{equation}
  \mathcolor{darkgreen}{\mathrm{d}\, F_2}
  \,=\,
  \massive{F_0 \, H_3}
  \mathrlap{\,.}
\end{equation}

Flux-quantization (discretization) of the locally constant function $F_0$ \cref{TheF0Bianchi} should make it integer-valued, the integer difference between connected components of spacetime (locally of the form $\mathbb{R}^{1,9} \setminus \mathbb{R}^{1,8}$) being the \emph{D8-brane charge} sitting in the gap between them \parencites{BergshoeffGreenPapadopoulosTownsend1995}{BergshoeffdeRooGreenPapadopoulosTownsend1996}.

\paragraph
{Aim}

Our aim is to establish consistent such flux/charge quantization (as explained in \parencites{SS25-Flux}{SS26-HigherGauge}) of the complete infinite Borcherds hierarchy of type IIA higher gauge fields (\cref{OverviewDiagram}) in a manner compatible with lifting to 11D supergravity, and its further extension to the massive case (\cref{OverviewDiagramMassive}). 

This complete problem has not been addressed before, as witnessed by how much the most popular proposal (namely charge quantization in twisted K-theory) misses (as highlighted in \cref{OverviewDiagram}) of the complete IIA field content, not to speak of the full IIA Borcherds hierarchy.

\begin{figure}[htbp]
\thisfloatpagestyle{empty}
\scalebox{.95}{
\begin{tikzpicture}

\node[
  rounded corners,
  draw=gray,
  scale=.8
] at (2,8.7) {%
  \def\arraystretch{.9}%
  \begin{tabular}{@{\hspace{0pt}}c@{\hspace{0pt}}}%
    \textbf{4-Cohomotopy}
  \end{tabular}
};
\node at (-1.1,8) {
  $
    \left\{
      \begin{aligned}
        G_4
        \\
        G_7
      \end{aligned}
      \,\middle\vert\,
      \begin{aligned}
        \mathrm{d}\, G_4 & = 0
        \\
        \mathrm{d}\, G_7 & = 
        \sfrac{1}{2}\, G_4 \, G_4
      \end{aligned}
    \right\}
    =
    \Omega^1_{\mathrm{cl}}\bracket({
     \text{-};
     \mathfrak{l}
     \bracket({
       S^4
      })
    })
  $
};

\node[
  rounded corners,
  draw=gray,
  scale=.8
] at (3.5,4.5) {%
  \def\arraystretch{.9}%
  \begin{tabular}{@{\hspace{0pt}}c@{\hspace{0pt}}}%
    \textbf{cyclic}
    \textbf{4-Cohomotopy}
  \end{tabular}
};
\node at (0,3.8) {
  $
    \left\{
      \begin{aligned}
        \mathcolor{darkgreen}{F_2}
        \\
        \mathcolor{darkgreen}{H_3}
        \\
        \mathcolor{darkgreen}{F_4}
        \\
        \mathcolor{darkgreen}{F_6}
        \\
        \mathcolor{darkorange}{H_7}
      \end{aligned}
      \,\middle\vert\,
      \begin{aligned}
        \mathcolor{darkgreen}{\mathrm{d}\, F_2} 
          & 
          \mathcolor{darkgreen}{= 0}
        \\
        \mathcolor{darkgreen}{\mathrm{d}\, H_3} 
          & 
        \mathcolor{darkgreen}{= 0}
        \\
        \mathcolor{darkgreen}{\mathrm{d}\, F_4} 
          & 
          \mathcolor{darkgreen}{= H_3 \, F_2}
        \\
        \mathcolor{darkgreen}{\mathrm{d}\, F_6} 
          & 
          \mathcolor{darkgreen}{= H_3 \, F_4}
        \\
        \mathcolor{darkorange}{\mathrm{d}\, H_7} 
          & 
          \mathcolor{darkorange}{
            =
            \sfrac{1}{2}\, 
            F_4\, F_4
            -
            F_2 \, F_6
          }
      \end{aligned}
    \right\}
    =
    \Omega^1_{\mathrm{cl}}\bracket({
     \text{-};
     \mathfrak{l}
     \bracket({
       \Cyc S^4
      })
    })
  $
};

\node[
  rounded corners,
  draw=gray,
  scale=.8
] at (7.9,-1.65) {%
  \def\arraystretch{.9}%
  \begin{tabular}{@{\hspace{0pt}}c@{\hspace{0pt}}}%
    \textbf{small cyclic}
    \\
    \textbf{4-Cohomotopy}
  \end{tabular}
};
\node at (.7,-1.7) {$
    \left\{
      \begin{aligned}
        \mathcolor{darkgreen}{F_2}
        \\
        \mathcolor{darkgreen}{H_3}
        \\
        \mathcolor{darkgreen}{F_4}
        \\
        \mathcolor{darkgreen}{F_6}
        \\
        \mathcolor{darkorange}{H_7}
        \\
        \mathcolor{darkblue}{F_8}
        \\
        \phantom{H_9}
        \\
        \mathcolor{darkblue}{F_{10}}
        \\
        \phantom{H_{11}^A}
        \\[-4pt]
        \mathcolor{purple}{H_{11}^B}
        \\
        \mathcolor{darkblue}{F_{12}}
        \\
        \smash{\vdots}
      \end{aligned}
      \,\middle\vert\,
      \adjustbox{raise=2pt}{$
      \begin{aligned}
        \mathcolor{darkgreen}{\mathrm{d}\, F_2} 
          & 
          \mathcolor{darkgreen}{= 0}
        \\
        \mathcolor{darkgreen}{\mathrm{d}\, H_3} & 
          \mathcolor{darkgreen}{= 0}
        \\
        \mathcolor{darkgreen}{\mathrm{d}\, F_4} & 
          \mathcolor{darkgreen}{= H_3 \, F_2}
        \\
        \mathcolor{darkgreen}{\mathrm{d}\, F_6} 
          & 
          \mathcolor{darkgreen}{= H_3 \, F_4}
        \\
        \mathcolor{darkorange}{\mathrm{d}\, H_7} 
          &
          \mathcolor{darkorange}{
          =
            \sfrac{1}{2}\, 
            F_4\, F_4
            -
            F_2 \, F_6
        }
        \\
        \mathcolor{darkblue}{\mathrm{d}\, F_8} & 
          \mathcolor{darkblue}{= H_3 \, F_6}
        \\
        \phantom{H_9}
        \\
        \mathcolor{darkblue}{\mathrm{d}\, F_{10}} & 
          \mathcolor{darkblue}{=  H_3 \, F_8}
        \\
        \phantom{H^A_{11}}
        \\
        \mathcolor{purple}{\mathrm{d}\, H^B_{11}} 
         & 
          \mathcolor{purple}{=  
          F_4 \, F_8
          - F_2 \, F_{10}
          - \sfrac{1}{2}\, F_6^2
        }
        \\
        \mathcolor{darkblue}{\mathrm{d}\, F_{12}} & 
          \mathcolor{darkblue}{=  H_3 \, F_{10}}
        \\
        \smash{\vdots}
      \end{aligned}
      $}
    \right\}
    =
    \Omega^1_{\mathrm{cl}}\bracket({
     \text{-};
     \mathfrak{l}
     \bracket({
       \SmlCyc S^4
      })
    })
$};

\node[
  rounded corners,
  draw=gray,
  scale=.8
] at (9.05,-9.2) {%
  \def\arraystretch{.9}%
  \begin{tabular}{@{\hspace{0pt}}c@{\hspace{0pt}}}%
    \textbf{minimal}
    \\
    \textbf{cyclic}
    \\
    \textbf{4-Cohomotopy}
  \end{tabular}
};
\node at (1.3,-9.3) {$
    \left\{
      \begin{aligned}
        \mathcolor{darkgreen}{F_2}
        \\
        \mathcolor{darkgreen}{H_3}
        \\
        \mathcolor{darkgreen}{F_4}
        \\
        \mathcolor{darkgreen}{F_6}
        \\
        \mathcolor{darkorange}{H_7}
        \\
        \mathcolor{darkblue}{F_8}
        \\
        \mathcolor{magenta}{H_9}
        \\
        \mathcolor{darkblue}{F_{10}}
        \\
        \mathcolor{magenta}{H_{11}^A}
        \\[-4pt]
        \mathcolor{purple}{H_{11}^B}
        \\
        \mathcolor{darkblue}{F_{12}}
        \\
        \smash{\vdots}
      \end{aligned}
      \,\middle\vert\,
      \adjustbox{raise=2pt}{$
      \begin{aligned}
        \mathcolor{darkgreen}{\mathrm{d}\, F_2} 
          & 
          \mathcolor{darkgreen}{= 0}
        \\
        \mathcolor{darkgreen}{\mathrm{d}\, H_3} & 
          \mathcolor{darkgreen}{= 0}
        \\
        \mathcolor{darkgreen}{\mathrm{d}\, F_4} & 
          \mathcolor{darkgreen}{= H_3 \, F_2}
        \\
        \mathcolor{darkgreen}{\mathrm{d}\, F_6} 
          & 
          \mathcolor{darkgreen}{= H_3 \, F_4}
        \\
        \mathcolor{darkorange}{\mathrm{d}\, H_7} 
          &
          \mathcolor{darkorange}{
          =
            \sfrac{1}{2}\, 
            F_4\, F_4
            -
            F_2 \, F_6
        }
        \\
        \mathcolor{darkblue}{\mathrm{d}\, F_8} & 
          \mathcolor{darkblue}{= H_3 \, F_6}
        \\
        \mathcolor{magenta}{\mathrm{d}\, H_9} & 
          \mathcolor{magenta}{=  H_3 \, H_7
          + \sfrac{1}{2}\, F_4 \, F_{6}
          - \sfrac{3}{2}\, F_2 \, F_8}
        \\
        \mathcolor{darkblue}{\mathrm{d}\, F_{10}} & 
          \mathcolor{darkblue}{=  H_3 \, F_8}
        \\
        \mathcolor{magenta}{\mathrm{d}\, H^A_{11}}
          & 
        \mathcolor{magenta}{ = H_3 \, H_9 
        - \sfrac{3}{2}\, F_2 \, F_{10}
        + \sfrac{1}{4} \, F_6^2}
        \\
        \mathcolor{purple}{\mathrm{d}\, H^B_{11}} 
         & 
          \mathcolor{purple}{=  
          F_4 \, F_8
          - F_2 \, F_{10}
          - \sfrac{1}{2}\, F_6^2
        }
        \\
        \mathcolor{darkblue}{\mathrm{d}\, F_{12}} & 
          \mathcolor{darkblue}{=  H_3 \, F_{10}}
        \\
        \smash{\vdots}
      \end{aligned}
      $}
    \right\}
    =
    \Omega^1_{\mathrm{cl}}\big(
     \text{-};
     \underbrace{
     \mathfrak{l}
     \bracket({
       \MinCyc S^4
      })
      }_{
        \substack{
          \cong \, \bIIA
          \\
          \text{\color{gray}%
            \rule{0pt}{10pt}(\cref{prop:RationalModelOfMinCyc})
          }
        }
      }
     \big)
$};

\node[
  rounded corners,
  draw=gray,
  scale=.8
] at (5.8,6.85) {%
  \def\arraystretch{.9}%
  \begin{tabular}{@{\hspace{0pt}}c@{\hspace{0pt}}}%
    \textbf{twisted}
    \\
    \textbf{K-theory}
  \end{tabular}
};
\node at (8,5.7) {$
    \Omega^1_{\mathrm{cl}}\bracket({
     \text{-};
     \mathfrak{l}
     \bracket({
       \begin{array}{c}
         B \mathrm{U} 
         \massive{\times \mathbb{Z}}
         \\
         \hline\hline
         B \mathrm{U}(1)
       \end{array}
      })
    })
    =
    \left\{
      \begin{aligned}
        \massive{F_0}
        \\
        \mathcolor{darkgreen}{F_2}
        \\
        \mathcolor{darkgreen}{H_3}
        \\
        \mathcolor{darkgreen}{F_4}
        \\
        \mathcolor{darkgreen}{F_6}
        \\
        \phantom{H_7}
        \\
        \mathcolor{darkblue}{F_8}
        \\
        \phantom{H_9}
        \\
        \mathcolor{darkblue}{F_{10}}
        \\
        \phantom{H_{11}^A}
        \\
        \phantom{H_{11}^B}
        \\
        \mathcolor{darkblue}{F_{12}}
        \\
        \vdots
      \end{aligned}
      \,\middle\vert\,
      \begin{aligned}
        \massive{\mathrm{d}\, F_0}
        & 
        \massive{= 0\mathclap{\phantom{F_0}}}
        \\
        \mathcolor{darkgreen}{\mathrm{d}\, F_2} & 
          \mathcolor{darkgreen}{= 0}
          \massive{+ F_0 \, H_3}
        \\
        \mathcolor{darkgreen}{\mathrm{d}\, H_3} 
          & 
          \mathcolor{darkgreen}{= 0}
        \\
        \mathcolor{darkgreen}{\mathrm{d}\, F_4} 
          & 
          \mathcolor{darkgreen}{= H_3 \, F_2}
        \\
        \mathcolor{darkgreen}{\mathrm{d}\, F_6} 
          & 
          \mathcolor{darkgreen}{= H_3 \, F_4}
        \\
        \phantom{H_7}
        \\
        \mathcolor{darkblue}{\mathrm{d}\, F_8} 
          & 
          \mathcolor{darkblue}{= H_3 \, F_6}
        \\
        \phantom{H_{9}}
        \\
        \mathcolor{darkblue}{\mathrm{d}\, F_{10}} & 
          \mathcolor{darkblue}{= H_3 \, F_8}
        \\
        \phantom{H_{11}^A}
        \\
        \phantom{H_{11}^B}
        \\
        \mathcolor{darkblue}{\mathrm{d}\, F_{12}} & 
          \mathcolor{darkblue}{= H_3 \, F_{10}}
        \\
        \vdots
      \end{aligned}
    \right\}
$};

\node[
  rotate=100
] at (5.8,-5.5) {
  \inlinetikzcd{
    {}
    \ar[r]
    \&[6.4cm]
    {}
  }
};

\node[
  rotate=100
] at (4.4,1.1) {
  \inlinetikzcd{
    {}
    \ar[r]
    \&[4cm]
    {}
  }
};
\node[
  rotate=85
] at (5.7,1.9) {
  \inlinetikzcd{
    {}
    \ar[r]
    \&[5.7cm]
    {}
  }
};

\end{tikzpicture}
}
\caption{\label{OverviewDiagram}%
Sets of closed differential forms $\Omega^1_{\mathrm{cl}}(\text{-}; \mathfrak{a})$ with coefficients in the real Whitehead $L_\infty$-algebras $\mathfrak{a} \defneq \mathfrak{l}\mathcal{A}$ of classifying spaces $\mathcal{A}$ for the indicated nonabelian cohomology theories.
The maps---induced from corresponding maps of classifying spaces (cohomology operations)---send generators to themselves when present, or to zero otherwise. From top to bottom this gives an increasingly good reflection of the flux species and their Bianchi identities in type IIA, culminating in the Borcherds hierarchy $\bIIA$, cf. \cref{MCircleReductionOfCharges,ObtainingBorcherdsIIA}. Furthermore, regarding the minimal cyclification as just the 0-component of a natural sequence of deformations enhances the flux content further to the  \emph{massive} IIA Borcherds hierarchy, cf. \cref{OverviewDiagramMassive,SystemOfClassifyingSpaces}.} 
\end{figure}

\begin{figure}[htbp]
\centering
\begin{tikzpicture}
\node at (0,0) {
$
\left\{
  \begin{aligned}
    \massive{F_0}
    \\
    \mathcolor{darkgreen}{F_2}
    \\
    \mathcolor{darkgreen}{H_3}
    \\
    \mathcolor{darkgreen}{F_4}
    \\
    \mathcolor{darkgreen}{F_6}
    \\
    \mathcolor{darkorange}{H_7}
    \\
    \mathcolor{darkblue}{F_8}
    \\
    \mathcolor{magenta}{H_9}
    \\
    \mathcolor{darkblue}{F_{10}}
    \\
    \mathcolor{magenta}{H_{11}^A}
    \\[-4pt]
    \mathcolor{purple}{H_{11}^B}
    \\
    \mathcolor{darkblue}{F_{12}}
    \\
    \smash{\vdots}
  \end{aligned}
  \,\middle\vert\,
  \adjustbox{raise=2pt}{$
  \begin{aligned}
    \massive{\mathrm{d}\, F_0}
      & \massive{= 0}
    \\
    \mathcolor{darkgreen}{\mathrm{d}\, F_2}
      & \mathcolor{darkgreen}{= 0}
        \;  \massive{+ F_0 \, H_3}
    \\
    \mathcolor{darkgreen}{\mathrm{d}\, H_3}
      & \mathcolor{darkgreen}{= 0}
    \\
    \mathcolor{darkgreen}{\mathrm{d}\, F_4}
      & \mathcolor{darkgreen}{= H_3 \, F_2}
    \\
    \mathcolor{darkgreen}{\mathrm{d}\, F_6}
      & \mathcolor{darkgreen}{= H_3 \, F_4}
    \\
    \mathcolor{darkorange}{\mathrm{d}\, H_7}
      & \mathcolor{darkorange}{=
          \sfrac{1}{2}\, F_4\, F_4
          - F_2 \, F_6}
        \; \massive{+ F_0 \, F_8}
    \\
    \mathcolor{darkblue}{\mathrm{d}\, F_8}
      & \mathcolor{darkblue}{= H_3 \, F_6}
    \\
    \mathcolor{magenta}{\mathrm{d}\, H_9}
      & \mathcolor{magenta}{= H_3 \, H_7
          + \sfrac{1}{2}\, F_4 \, F_6
          - \sfrac{3}{2}\, F_2 \, F_8}
        \; \massive{+ \sfrac{5}{2}\, F_0 \, F_{10}}
    \\
    \mathcolor{darkblue}{\mathrm{d}\, F_{10}}
      & \mathcolor{darkblue}{= H_3 \, F_8}
    \\
    \mathcolor{magenta}{\mathrm{d}\, H^A_{11}}
      & \mathcolor{magenta}{= H_3 \, H_9
          - \sfrac{3}{2}\, F_2 \, F_{10}
          + \sfrac{1}{4}\, F_6^2}
        \;  \massive{+ 4\, F_0 \, F_{12}}
    \\
    \mathcolor{purple}{\mathrm{d}\, H^B_{11}}
      & \mathcolor{purple}{= F_4 \, F_8
          - F_2 \, F_{10}
          - \sfrac{1}{2}\, F_6^2}
        \; \massive{+ F_0 \, F_{12}}
    \\
    \mathcolor{darkblue}{\mathrm{d}\, F_{12}}
      & \mathcolor{darkblue}{= H_3 \, F_{10}}
    \\
    \smash{\vdots}
  \end{aligned}
  $}
\right\}
  \;\cong\;
  \Omega^1_{\mathrm{cl}}\big({
    \text{-}
    ;
    \underbrace{
    \mathfrak{l}\bracket({
      \massive{\mathrm{m}}\!\MinCyc 
      S^4
    })}_{
      \substack{
        \cong \, \bIIAMassive
        \mathrlap{
        \text{\color{gray}%
        (\cref{prop:RationalModelOfMassiveMinCyc})}
        }
      }
    }
  }\big)
$
};

\begin{scope}[shift={(0,.6)}]
\node[
  rounded corners,
  draw=gray,
  scale=.8
] at (6.05,3.05) {%
  \def\arraystretch{.9}%
  \begin{tabular}{@{\hspace{0pt}}c@{\hspace{0pt}}}%
    \textbf{minimal cyclic}
    \\
    \textbf{4-Cohomotopy}
  \end{tabular}
};
\node at (5.8,1.75) {
  $
    \begin{tikzcd}[
      row sep=17pt
    ]
      \Omega^1_{\mathrm{cl}}
      \bracket({
        \text{-};
        \mathfrak{l}
        \bracket({
          \MinCyc S^4
        })
      })
      \ar[
        d,
        shift left=10pt,
        "{
          \substack{
            \text{massive}
            \\
            \phantom{\text{\color{gray}\cref{MassMinCyc}}}
          }
        }"{swap},
        "{
          \substack{
            \text{deformation}
            \\
            \text{\color{gray}\cref{MassMinCyc}}
          }
        }"{pos=.45}
      ]
      \\
      {}
    \end{tikzcd}
  $
};
\end{scope}

\node[
  rounded corners,
  draw=gray,
  scale=.8
] at (6.25,1.0) {%
  \def\arraystretch{.9}%
  \begin{tabular}{@{\hspace{0pt}}c@{\hspace{0pt}}}%
    \textbf{massive}
    \\
    \textbf{minimal cyclic}
    \\
    \textbf{4-Cohomotopy}
  \end{tabular}
};

\begin{scope}[shift={(.2,.25)}]
\node[
  rounded corners,
  draw=gray,
  scale=.8
] at (5.93,-3.8) {%
  \def\arraystretch{.9}%
  \begin{tabular}{@{\hspace{0pt}}c@{\hspace{0pt}}}%
    \textbf{twisted}
    \\
    \textbf{K-theory}
  \end{tabular}
};
\node at (5.8,-2.2) {
  \begin{tikzcd}
    {}
    \ar[
      d,
      shift left=8pt,
      "{
        \substack{
          \text{abelian}
          \\
          \phantom{\mathllap{
            \text{\color{gray}%
            (\cref{TheMassiveComparisonMap})
          }}}
        }
      }"{swap},
      "{
        \substack{
          \text{approximation}
          \\
          \text{\color{gray}%
            (\cref{TheMassiveComparisonMap})
          }
        }
      }"
    ]
    \\
    \Omega^1_{\mathrm{cl}}\bracket({
     \text{-};
     \mathfrak{l}
     \bracket({
       \begin{array}{c}
         B \mathrm{U} 
         \massive{\times \mathbb{Z}}
         \\
         \hline\hline
         B \mathrm{U}(1)
       \end{array}
      })
    })
  \end{tikzcd}
};
\end{scope}

\end{tikzpicture}
\caption{\label{OverviewDiagramMassive}
  The minimal cyclification of the 4-sphere is the 0-component of an integer family $\MassMinCyc S^4$ of deformations which reflect twists that prevent the M-circle fibers from collapsing as a principal bundle, cf. \cref{ThePartialMCircleBundle}. Closed differential forms with coefficients in the Whitehead dg-Lie-algebra of $\MassMinCyc S^4$ enhance the IIA Borcherds hierarchy from \cref{OverviewDiagram} to its Romans massive version, cf. \cref{MassiveBorcherds}.
}
\end{figure}
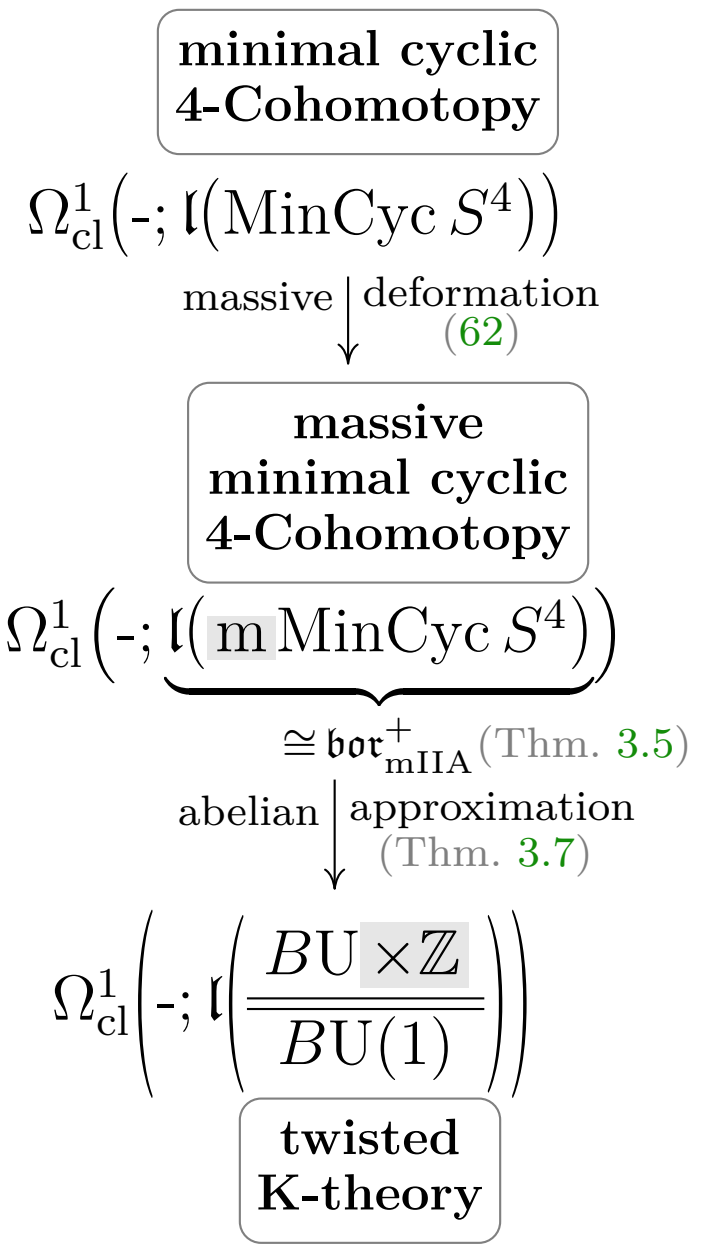

\section
{Methods}

We combine higher Lie algebra (cf. \cref{app:HigherLieAlgebras}) with (rational) homotopy and (unstable) generalized cohomology (cf. \cref{HomotopyTheory}):
\begin{description}
\item[In \cref{TheOvercompleteIIAFluxAlgebra}] we formulate the IIA Borcherds hierarchy (\cref{OverviewDiagram}) and its massive deformation (\cref{OverviewDiagramMassive}) as dg-Lie algebras and hence as $L_\infty$-algebras $\bIIA$ and $\bIIAMassive$, respectively. 

\item[In \cref{MCircleReductionOfCharges}]
we recall (from \cite[\S 3.2]{GSS26-KfromM}) the dimensional reduction of cohomotopical M-brane charges along ``topologically small'' M-circle fibers and then strengthen this further to \emph{minimal} small circle reductions.
\end{description}

Below in \cref{Results} we show that this minimal circle reduction of 4-Cohomotopy reproduces and hence flux-quantizes the IIA Borcherds hierarchy, and we identify its natural deformation which analogously reproduces and hence flux-quantizes the massive hierarchy.

\subsection
{
\texorpdfstring
{The IIA Borcherds $L_\infty$-Algebras}
{The IIA Borcherds L-infinity-Algebras}
}
\label
{TheOvercompleteIIAFluxAlgebra}

\paragraph
{The IIA Borcherds Algebra}

Concretely, in the context of hidden exceptional U-duality symmetries of torus-reductions of 11D supergravity (cf. \parencites{CremmerJuliaLuPope1998a}{ObersPioline1999}{deWitNicolai2001}{GSS24-Exceptional}) there is a seminal argument 
\cite{HenryLabordereJuliaPaulot2002,GreitzHowe2011,HenneauxJuliaLevie2012,Palmkvist2012,HowePalmkvist2015} that coefficient algebra characterizing the full tower of higher flux densities and their Bianchi identities in type IIA should be taken to be the \emph{positive IIA Borcherds algebra}, which we denote by%
\footnote{
  Borcherds algebras are often regarded as \emph{super Lie algebras}. In the context of flux algebras this is both insufficient and misleading: Insufficient because the $\mathbb{Z}$-grading matters beyond its parity modulo 2, cf. \cref{SpecializationsOfLInfinityAlgebras}, and 
  misleading because all elements correspond to \emph{bosonic} flux fields of higher form degree, cf. \cref{FluxCoefficients}.
}
\begin{equation}
  \bIIA
  \in 
  \mathrm{dgLieAlg}
  \subset
  L_\infty \mathrm{Alg}
  \mathrlap{\,.}
\end{equation}

\begin{definition}[The IIA Borcherds algebra]
\label[definition]{IIABorcherdsAlgebra}
The \emph{type IIA Borcherds dg-Lie algebra} admits the following presentation, obtained
by specializing the Chevalley--Serre relations
\cite[(A.1)--(A.5)]
{HenneauxJuliaLevie2012}
to the IIA Cartan matrix
\cite[\S~5.2, Eq.~(5.1)]{HenneauxJuliaLevie2012}
(cf. \cite[\S~4.2]{GreitzHowe2011}):
\begin{equation}
\label{TheIIABorcherdsAlgebra}
 \BorcherdsAlgebra_{\mathrm{IIA}}
 :=
 \mathrm{Free}_{\mathrm{dgL}}
 \left(
 \begin{gathered}
   \eta_2,\varphi_1,
   \\
   N^{\varphi}_0,N^{\eta}_0,
   \\
   \bar\varphi_{-1},\bar\eta_{-2}
 \end{gathered}
 \right)
 \begin{tikzpicture}[baseline=(relations.base)]
   \node[inner sep=0pt,outer sep=0pt] (relations)
   {$\displaystyle
     \left(
     \begin{array}{ll}
       [\text{-}]=0
       &
       [N^{\varphi}_0,N^{\eta}_0]=0
       \\[6pt]
       [N^{\varphi}_0,\eta_2] = 0 
       & 
       [N^{\eta}_0,\eta_2] = -\eta_2
       \\
       {[N^{\varphi}_0,\varphi_1]} = -\varphi_1
       &
       [N^{\eta}_0,\varphi_1] = 0
       \\[6pt]
       [N^{\varphi}_0,\bar\eta_{-2}]
       = 0
       &
       [N^{\eta}_0,\bar\eta_{-2}] = \bar\eta_{-2}
       \\
       {[N^{\varphi}_0,\bar\varphi_{-1}]}=\bar\varphi_{-1}
       &
       [N^{\eta}_0,\bar\varphi_{-1}]=0
       \\[6pt]
       {[\eta_2,\bar\eta_{-2}]} = N^{\varphi}_0,
       &
       [\varphi_1,\bar\varphi_{-1}] = N^{\eta}_0
       \\
       {[\eta_2,\bar\varphi_{-1}]} = 0
       &
       [\varphi_1,\bar\eta_{-2}]=0
       \\[6pt]
       {[\varphi_1,\varphi_1]} = 0
       &
       [\bar\varphi_{-1},\bar\varphi_{-1}]=0
     \end{array}
     \right)
     \mathrlap{\,.}
   $};

   \draw[line width=0.35pt,line cap=butt]
     ([xshift=-8mm]relations.south west)
     --
     ([xshift=-2mm]relations.north west);
 \end{tikzpicture}
\end{equation}
The \emph{positive IIA Borcherds algebra} 
is the dg-Lie subalgebra of positive degrees:
\begin{equation}
\label{ThePositiveIIABorcherdsAlgebra}
  \mathrm{Free}_{\mathrm{dgL}}
  \left(
    \begin{aligned}
      & \varphi_1,
      \\
      & \eta_2
    \end{aligned}
  \right)\big/
  \bracket({
    \begin{array}{l}
      [\text{-}] = 0
      \\
      {[\varphi_1, \varphi_1]} = 0
    \end{array}
  })
  \,=:\,
  \bIIA 
  \,\subset\, 
  \BorcherdsAlgebra_{\mathrm{IIA}}
  \mathrlap{\,.}
\end{equation}
Here our generator names are (nonstandard but) suggestively adapted to their role as flux coefficients \cref{UnifiedFluxDensity}. 
\end{definition}
\begin{remark}
The Cartan generators $N^\varphi_0$ and $N^\eta_0$ in \cref{TheIIABorcherdsAlgebra} act as (negative) number operators, in that
for a bracket word $w \in \BorcherdsAlgebra_{\mathrm{IIA}}$ containing
\begin{enumerate}
\item $m_\varphi$ net occurrences of $\varphi_1$ and $\bar \varphi_{-1}$ 
\item $m_\eta$ net occurrences of $\eta_2$ and $\bar \eta_{-2}$
\end{enumerate}
we have
\begin{equation}
  \begin{aligned}
  [N^{\varphi}_0,w] 
    & = -m_\varphi \, w
    \,,
  \\
  [N^{\eta}_0,w]
    & = -m_\eta \, w
  \mathrlap{\,.}
  \end{aligned}
\end{equation}
\end{remark}

\paragraph
{The Borcherds Hierarchy of Higher Bianchi Identities}

\begin{notation}
\label[notation]{FluxCoefficients}
  In the literature, the generators of positive Borcherds algebras are usually denoted $e_i$ (cf. \parencites[\S 1, relations (1)--(5)]{Borcherds1988}[App.~A, (A.1)--(A.5)]{HenneauxJuliaLevie2012}).
  Here, in order to bring out which element of the IIA Borcherds algebra \cref{TheIIABorcherdsAlgebra} is the coefficient of which flux species, we name these elements as follows: 
  \begin{subequations}
  \label{LowDegreeBorcherdsElements}
    \begin{align}
      \varphi_1 
        & := e_1
      \\
      \eta_2 
        & := e_0
      \\
      \varphi_3 
        & := [{\eta_2,\varphi_1}]
      \\
      \varphi_5 
        & := \bracket[{
          \eta_2,
          [{
            \eta_2, \varphi_1
          }]
        }]
      \\
      \eta_6 
       & := 
       \bracket[{
         [\eta_2, \varphi_1],
         [\eta_2, \varphi_1]
       }]
       =
       -\bracket[{
         \varphi_1,
         \bracket[{
           \eta_2,
           [\eta_2, \varphi_1]
         }]
       }]
      \\
      \varphi_7
      & :=
      \bracket[{
        \eta_2,
        \bracket[{
          \eta_2,
          [\eta_2, \varphi_1]
        }]
      }]
      \\
      \eta_8
      & :=
      \bracket[{
        \eta_2,
        \bracket[{
          [\eta_2, \varphi_1],
          [\eta_2, \varphi_1]
        }]
      }]      
      =
      2\bracket[{
        [\eta_2, \varphi_1],
        \bracket[{
          \eta_2,
          [\eta_2, \varphi_1]
        }]
      }]
      =
      -\tfrac{2}{3}
      \bracket[{
        \varphi_1,
        \bracket[{
          \eta_2,
          \bracket[{
            \eta_2,
              [\eta_2, \varphi_1]
          }]
        }]
      }]
      \\
      \varphi_9
      & := 
      \bracket[{
        \eta_2,
        \bracket[{
          \eta_2,
          \bracket[{
            \eta_2,
              [\eta_2, \varphi_1]
          }]
        }]
      }]
      \\
      \label{eta10A}
      \eta^A_{10}
        & :=
        [\eta_2, \eta_8]
        =
        \bracket[{
          \eta_2,
          \bracket[{
            \eta_2,
            \bracket[{
              [\eta_2, \varphi_1],
              [\eta_2, \varphi_1]
            }]
          }]
        }]
        =
        2\bracket[{
          \eta_2,
          \bracket[{
            [\eta_2, \varphi_1],
            \bracket[{
              \eta_2,
              [\eta_2, \varphi_1]
            }]
          }]
        }]
      \\
      \label{eta10B}
      \eta^B_{10}
        & :=
        [\varphi_3, \varphi_7]
        =
        \bracket[{
          [\eta_2, \varphi_1],
          \bracket[{
            \eta_2,
            \bracket[{
              \eta_2,
              [\eta_2, \varphi_1]
            }]
          }]
        }]      
      \\
      \varphi_{11}
      & := 
      \bracket[{
        \eta_2,
          \bracket[{
            \eta_2,
            \bracket[{
              \eta_2,
              \bracket[{
                \eta_2,
                  [\eta_2, \varphi_1]
              }]
            }]
          }]
          }]
      \\
      \vdots
    \end{align}
  \end{subequations}
\end{notation}

This way, the totality of RR/NS-flux densities are the components of a single form with coefficients in $\bIIA$ \cref{ThePositiveIIABorcherdsAlgebra}, by the following expansion: 
\begin{equation}
\label{UnifiedFluxDensity}
  \begin{aligned}
  F_\bullet
  & \in
  \Omega^1_{\mathrm{dR}}\bracket({
    \text{-};
    \bIIA
  })
  \\
  F_\bullet 
  & =
  F_2 \varphi_1 
  +
  F_4 \varphi_3
  +
  F_6 \varphi_5
  +
  F_8 \varphi_7
  +
  F_{10} \varphi_9
  +
  F_{12} \varphi_{11}
  +
  \cdots
  \\
  &
  \;\;\;\;
  +
  H_3 \eta_2
  +
  H_7 \eta_6
  +
  H_9 \eta_8
  +
  H^A_{11} \eta_{10}^A
  +
  H^B_{11} \eta_{10}^B
  +
  \cdots
  \mathrlap{\,,}
 \end{aligned}
\end{equation}
and its closure condition \cref{ClosureSignConvention} gives the IIA Borcherds hierarchy of the IIA Bianchi/Gauss identities:
\begin{equation}
\label
{BorcherdsTowerOfBianchiIdentities}
 \begin{aligned}
    \Omega^1_{\mathrm{cl}}\bracket({
      \text{-}
     ;
      \bIIA
    })
    & 
    \simeq
    \bracketmid\{{
      F_\bullet 
      \in 
      \Omega^1_{\mathrm{dR}}\bracket({
        \text{-};
        \bIIA
      })
    }
    {
      \mathrm{d}\, F_\bullet
      =
      \tfrac{1}{2}
      \,
      \bracket[{
        F_\bullet,
        F_\bullet
      }]
    }
    \}
    \\
    & =
    \left\{
      \begin{aligned}
        \mathcolor{darkgreen}{F_2}
        \\
        \mathcolor{darkgreen}{H_3}
        \\
        \mathcolor{darkgreen}{F_4}
        \\
        \mathcolor{darkgreen}{F_6}
        \\
        \mathcolor{darkorange}{H_7}
        \\
        \mathcolor{darkblue}{F_8}
        \\
        \mathcolor{magenta}{H_9}
        \\
        \mathcolor{darkblue}{F_{10}}
        \\
        \mathcolor{magenta}{H_{11}^A}
        \\[-4pt]
        \mathcolor{purple}{H_{11}^B}
        \\
        \mathcolor{darkblue}{F_{12}}
        \\
        \smash{\vdots}
      \end{aligned}
      \,\middle\vert\,
      \adjustbox{raise=2pt}{$
      \begin{aligned}
        \mathcolor{darkgreen}{\mathrm{d}\, F_2} 
          & 
          \mathcolor{darkgreen}{= 0}
        \\
        \mathcolor{darkgreen}{\mathrm{d}\, H_3} & 
          \mathcolor{darkgreen}{= 0}
        \\
        \mathcolor{darkgreen}{\mathrm{d}\, F_4} & 
          \mathcolor{darkgreen}{= H_3 \, F_2}
        \\
        \mathcolor{darkgreen}{\mathrm{d}\, F_6} 
          & 
          \mathcolor{darkgreen}{= H_3 \, F_4}
        \\
        \mathcolor{darkorange}{\mathrm{d}\, H_7} 
          &
          \mathcolor{darkorange}{
          =
            \sfrac{1}{2}\, 
            F_4\, F_4
            -
            F_2 \, F_6
        }
        \\
        \mathcolor{darkblue}{\mathrm{d}\, F_8} & 
          \mathcolor{darkblue}{= H_3 \, F_6}
        \\
        \mathcolor{magenta}{\mathrm{d}\, H_9} & 
          \mathcolor{magenta}{=  H_3 \, H_7
          + \sfrac{1}{2}\, F_4 \, F_{6}
          - \sfrac{3}{2}\, F_2 \, F_8}
        \\
        \mathcolor{darkblue}{\mathrm{d}\, F_{10}} & 
          \mathcolor{darkblue}{=  H_3 \, F_8}
        \\
        \mathcolor{magenta}{\mathrm{d}\, H^A_{11}}
          & 
        \mathcolor{magenta}{ = H_3 \, H_9 
        - \sfrac{3}{2}\, F_2 \, F_{10}
        + \sfrac{1}{4} \, F_6^2}
        \\
        \mathcolor{purple}{\mathrm{d}\, H^B_{11}} 
         & 
          \mathcolor{purple}{=  
          F_4 \, F_8
          - F_2 \, F_{10}
          - \sfrac{1}{2}\, F_6^2
        }
        \\
        \mathcolor{darkblue}{\mathrm{d}\, F_{12}} & 
          \mathcolor{darkblue}{=  H_3 \, F_{10}}
        \\
        \smash{\vdots}
      \end{aligned}
      $}
    \right\}
    \mathrlap{\,.}
  \end{aligned}
\end{equation}

Of course, on a 9-dimensional manifold $X^9$, the infinite tower \cref{BorcherdsTowerOfBianchiIdentities} truncates to just 
\begin{equation}
\label
{TruncatedBorcherdsTowerOfBianchiIdentities}
    \Omega^1_{\mathrm{cl}}\bracket({
      X^9
      ;
      \bIIA
    })
    =
    \left\{
      \begin{aligned}
        \mathcolor{darkgreen}{F_2}
        \\
        \mathcolor{darkgreen}{H_3}
        \\
        \mathcolor{darkgreen}{F_4}
        \\
        \mathcolor{darkgreen}{F_6}
        \\
        \mathcolor{darkorange}{H_7}
        \\
        \mathcolor{darkblue}{F_8}
        \\
        \mathcolor{magenta}{H_9}
      \end{aligned}
      \,\middle\vert\,
      \adjustbox{raise=2pt}{$
      \begin{aligned}
        \mathcolor{darkgreen}{\mathrm{d}\, F_2} 
          & 
          \mathcolor{darkgreen}{= 0}
        \\
        \mathcolor{darkgreen}{\mathrm{d}\, H_3} & 
          \mathcolor{darkgreen}{= 0}
        \\
        \mathcolor{darkgreen}{\mathrm{d}\, F_4} & 
          \mathcolor{darkgreen}{= H_3 \, F_2}
        \\
        \mathcolor{darkgreen}{\mathrm{d}\, F_6} 
          & 
          \mathcolor{darkgreen}{= H_3 \, F_4}
        \\
        \mathcolor{darkorange}{\mathrm{d}\, H_7} 
          &
          \mathcolor{darkorange}{
          =
            \sfrac{1}{2}\, 
            F_4\, F_4
            -
            F_2 \, F_6
        }
        \\
        \mathcolor{darkblue}{\mathrm{d}\, F_8} & 
          \mathcolor{darkblue}{= H_3 \, F_6}
        \\
        \mathcolor{magenta}{\mathrm{d}\, H_9} & 
          \mathcolor{magenta}{=  H_3 \, H_7
          + \sfrac{1}{2}\, F_4 \, F_{6}
          - \sfrac{3}{2}\, F_2 \, F_8}
      \end{aligned}
      $}
    \right\}
    \mathrlap{\,.}
\end{equation}
Accordingly, the meaning of the higher Bianchi identities in the Borcherds tower has remained unclear in the literature. 
We highlight that the perspective of flux quantization \cref{GeneralFluxQuantizationSchematics} sheds light on this phenomenon.

\paragraph
{Higher Borcherds Stages Control Higher Global Symmetries}

The space of classifying maps $\mathrm{Map}\bracket({ X^9, \mathcal{A} })$ 
encodes not just the topological charge sectors in its connected components, but its higher homotopy groups capture the higher large gauge symmetries, and hence the higher global symmetries of the higher gauge fields sourced by these charges (\cite[\S 4.2.1, pp.~36ff]{SS26-HigherGauge}, cf. \parencites[\S 3.4]{AlfonsiKimLuciani2026}[Rem. 3.6]{PerezLona2026}), see \cref{HigherSymmetries}:

\begin{SCfigure}[.7][htbp]
\adjustbox
  {rndfbox=4pt}
{
\begin{tabular}{l}
$
  \mathrm{Map}\bracket({
    X^d
    ,
    \mathcal{A}
  })
  =
  \left\{
  \begin{tikzcd}[
    sep=70pt
  ]
    X^d
    \ar[
      rr,
      bend left=80,
      "{
        \substack{
          \text{topological} 
          \\
          \text{field configuration}
        }
      }"{description, name=s}
    ]
    \ar[
      rr,
      bend right=80,
      "{
        \substack{
          \text{topological} 
          \\
          \text{field configuration}
        }
      }"{description, name=t}
    ]
    \ar[
      from=s,
      to=t,
      Rightarrow,
      shorten=10pt,
      bend right=70,
      shift left=10pt,
      "{
        \substack{
          \text{large}
          \\
          \text{gauge}
          \\
          \text{transfor}
        }
      }"{description, name=ss}
    ]
    \ar[
      from=s,
      to=t,
      Rightarrow,
      shorten=10pt,
      bend left=70,
      shift right=10pt,
      "{
        \substack{
          \text{large}
          \\
          \text{gauge} 
          \\
          \text{transfor}
        }
      }"{description, name=tt}
    ]
    \ar[
      from=ss,
      to=tt,
      Rightarrow,
      scaling nfold=3,
      bend left=30,
      "{
        \substack{
          \text{gauge-of-gauge}
        }
      }"
    ]
    \ar[
      from=ss,
      to=tt,
      Rightarrow,
      scaling nfold=3,
      bend right=30,
      "{
        \substack{
          \text{gauge-of-gauge}
        }
      }"{swap}
    ]
    &\vdots&
    \mathcal{A}
  \end{tikzcd}
  \right\}
$
\\
\\
$
\begin{array}{cll}
    \chi 
    &
    \in
    \pi_0 
    \bracket({
      \mathrm{Map}\bracket({
        X^d
        ,
        \mathcal{A}
      })
    })
    &
    \substack{
      \text{charges}
    }
    \\[+8pt]
    \ell
    &
    \in
    \pi_1 
    \bracket({
      \mathrm{Map}\bracket({
        X^d
        ,
        \mathcal{A}
      })
      ,
      \chi
    })
    &
    \substack{
      \text{large gauge}
      \\
      \text{monodromies}
    }
    \\[+8pt]
    \ell^{(k)}
    &
    \in
    \pi_k 
    \bracket({
      \mathrm{Map}\bracket({
        X^d
        ,
        \mathcal{A}
      })
      ,
      \chi
    })
    &
    \substack{
      \text{$(1\!<\!k)$th-order}
      \\
      \text{large gauge}
      \\
      \text{monodromies}
    }
  \end{array}
$
\end{tabular}
}
\caption{\label{HigherSymmetries}%
  Where homotopy classes of maps to the classifying space $\mathcal{A}$ correspond to the topological brane charges, the homotopies between these maps correspond to large gauge transformations between these topological field configurations, and higher order homotopies-of-homotopies correspond to higher order gauge-of-gauge transformations.
  This higher gauge structure depends on the homotopy type of $\mathcal{A}$ beyond the dimension $d$ of the domain space and in particular reflects the generators of $\mathfrak{l}\mathcal{A}$ beyond degree $d$. Hence this is where an overcomplete tower of flux densities shows itself.
}
\end{SCfigure}

Hence our goal is to identify a charge classifying space $\mathcal{A}$ that 
\begin{enumerate}
  \item reproduces the full Borcherds hierarchy in that $\mathfrak{l}\mathcal{A} \simeq \bIIA$,
  \item arises via dimensional reduction on small M-circle fibrations from a classifying space for charges in 11D.
\end{enumerate}

\paragraph
{Massive Deformation of the Borcherds Hierarchy}

In order to extend the list of IIA flux species \cref{TruncatedBorcherdsTowerOfBianchiIdentities} to include the \emph{Romans mass} 0-form $F_0$ of massive IIA, with its
Bianchi identity
\begin{equation}
  \massive{
    \mathrm{d}\, F_0
    =
    0
  }
\end{equation}
and its
characteristic deformation of the $F_2$-flux Bianchi identity 
\begin{equation}
\label
{TheMassiveF2Bianchi}
  \mathcolor{darkgreen}{
    \mathrm{d}\, F_2
    =
    0
  }
  \massive{+ F_0 \, H_3}
  \mathrlap{\,,}
\end{equation}
we clearly need to adjoin to $\bIIA$ a corresponding coefficient element $\varphi_{-1}$ satisfying the dual relation
$
  \varphi_{1}
  =
  [\eta_2, \varphi_{-1}]
$ of \cref{TheMassiveF2Bianchi}, in continuation of the pattern $\varphi_{2k+1} = [\eta_2, \varphi_{2k-1}]$ \cref{LowDegreeBorcherdsElements}, as well as the relation 
$
  0
    = 
  [\varphi_{-1}, \varphi_1] 
$
(which reflects the fact that a summand of the form $F_0\, F_2$ does not appear in any massive IIA Bianchi identity).

But since $\bIIA$ \cref{ThePositiveIIABorcherdsAlgebra} is generated from $\varphi_1$ and $\eta_2$ alone, this already fixes the brackets of the new element $\varphi_{-1}$ with all other elements. Hence in variation of  \cref{ThePositiveIIABorcherdsAlgebra} we are led to consider:
\begin{definition}[Massive IIA Borcherds algebra]
\label[definition]{def:massive_Borcherds}
We say that the \emph{massive IIA Borcherds} algebra is:
\begin{equation}
\label{TheMassiveIIABorcherdsAlgebra}
 \bIIAMassive 
   :=
   \FreeDGLie\left(
     \begin{aligned}
     & \varphi_{-1},
     \\
     & \varphi_1,
     \\
     & \eta_2
     \end{aligned}
    \right)
  \bigg/
  \left(
  \begin{array}{rl}
    {[\text{-}]} & = 0
    \\
    {[\varphi_{-1}, \varphi_{-1}]}
     & = 0
    \\
    {[\varphi_{-1},\varphi_1]} 
    & = 0
    \\
    {[\varphi_{-1}, \eta_2]} 
    & = 
    -\varphi_1,
    \\
    {[\varphi_1,\varphi_1]} 
    & = 0
  \end{array}
  \right)
  \in
  L_\infty \mathrm{Alg}
  \mathrlap{\,.}
  \end{equation}
  In other words, this is a deformation of the positive IIA Borcherds algebra \cref{ThePositiveIIABorcherdsAlgebra}
  \begin{equation}
    \begin{tikzcd}
      \bIIA
      \ar[r, hook]
      &
      \bIIAMassive
    \end{tikzcd}
  \end{equation}
  by an element $\varphi_{-1}$, which acts as a degree-$(-1)$ derivation
  \begin{equation}
  \label{TheMassiveDerivation}
  \begin{tikzcd}[sep=0pt]
    \bIIA 
    \ar[
      rr,
      "{ 
        [\varphi_{-1},\, \text{-}]
       }"
    ]
    &\phantom{---}&
    \bIIA
    \\
    \varphi_1 &\mapsto& 0
    \\
    \eta_2 &\mapsto& -\varphi_1
    \mathrlap{\,.}
  \end{tikzcd}
\end{equation}
The sign in $[\varphi_{-1}, \eta_2] = -\varphi_1$, equivalently $[\eta_2, \varphi_{-1}] = \varphi_1$, is the one for which, via \cref{ClosureSignConvention}, the massive deformation of the $F_2$-Bianchi identity comes out as $\mathrm{d}\,F_2 = F_0\,H_3$, in line with $\mathrm{d}\,F_{2k} = H_3\,F_{2k-2} + \cdots$ for all $k \geq 2$.

\end{definition}

\begin{remark}[Relation to the literature]
The actual \cref{def:massive_Borcherds} may not have been stated before, but the idea goes back to
\cite[(5.11)--(5.13)]{LavrinenkoEtAl1999}, where the doubled-field superalgebra of
\cite{CremmerJuliaLuPope1998} was extended to massive type IIA supergravity
via an odd-graded generator there denoted $Y$, recognized as the Lie algebraic
counterpart of the massive Bianchi identity $\mathrm{d}F_2 = m\, H_3$; cf.
\cite[(2.17)]{GreitzHowe2011}.
In the \emph{tensor hierarchy} algebra of \cite{Palmkvist2014}, the embedding tensor is such an element. Correspondingly, in 
\cite{GreitzHowePalmkvist2014,HowePalmkvist2015}, the Maurer--Cartan
differential $\mathrm{d}$ is deformed to $\mathrm{d} + [\,-\,,\Theta]$
\cite[\S 6.2]{HowePalmkvist2015}. That the construction of
\cite{LavrinenkoEtAl1999} is a forerunner, and the simplest example, of
this formalism is pointed out in \parencites[\S 1]{GreitzHowePalmkvist2014}[\S 6]{HowePalmkvist2015}.
\end{remark}

\begin{proposition}[The massive IIA Borcherds hierarchy]
Closed differential forms with coefficients in the massive IIA Borcherds algebra \textup{(\cref{def:massive_Borcherds})} 
constitute the massive Borcherds hierarchy as shown in \cref{OverviewDiagramMassive}.
\end{proposition}
\begin{proof}
We have already noted above \cref{def:massive_Borcherds} that the first three Bianchi identities in the list of \cref{OverviewDiagramMassive} hold by construction \cref{TheMassiveIIABorcherdsAlgebra}.
The other occurrences of $F_0$ in the Bianchi identities in  \cref{OverviewDiagramMassive} are dually, via \cref{ClosureSignConvention}, the following statements:
\begin{align}
  [\varphi_{-1}, \varphi_7] 
   & = \eta_6,
  \\
  [\varphi_{-1}, \varphi_9] 
    & =
  \tfrac{5}{2} \eta_8,
  \\
  [\varphi_{-1},\varphi_{11}] 
   & = 4\eta_{10}^A+\eta_{10}^B
  \mathrlap{\,.}
\end{align}
Recalling from \cref{FluxCoefficients} that  $\eta_{10}^A = [\eta_2,\eta_8]$ \cref{eta10A} and
$\eta_{10}^B = [\varphi_3,\varphi_7]$ \cref{eta10B}, these indeed follow from the previous relations
\cref{LowDegreeBorcherdsElements} 
together with the derivation property of $[\varphi_{-1},\text{-}]$ \cref{TheMassiveDerivation}.
\end{proof}

\subsection
{Circle Reduction of M-Brane Charges}
\label{MCircleReductionOfCharges}

Our aim is to construct a flux/charge-quantization of the massive IIA Borchers hierarchy \cref{TheOvercompleteIIAFluxAlgebra}. 
For this purpose we first consider flux/charge quantization in 11D and then its dimensional reduction on M-circle fibers.

\paragraph
{The Example of M-Brane Charge in Cohomotopy---Hypothesis H}

Towards resolving the above issue \cref{BraneChargesInTwistedKTheory} of twisted K-theory quantizing D2-brane charge (sourcing $F_6$) but not NS1-brane charge (sourcing $H_7$), we highlight that these two are jointly meant to descend from M2-brane charge in 11D (sourcing $G_7$) under dimensional reduction on the M-circle fiber (cf. \cref{MCircleReductionOfCharges}). 
Therefore, to solve this issue one should:
\begin{enumerate}
  \item choose a charge quantization law in 11D that quantizes not just the magnetic M5-brane charge but also the electric M2-brane charge (the \emph{Page charge}),

  \item determine the dimensional reduction \emph{of this charge quantization law} along M-circle fibers.
\end{enumerate}

To this end, notice by \cref{GeneralFluxQuantizationSchematics} that the admissible charge classifying spaces $\mathcal{A}$ in 11D must have Whitehead-bracket $L_\infty$-algebra isomorphic to the \emph{M-theory gauge algebra} (\parencites[(2.6)]{CremmerJuliaLuPope1998}[(3.4)]{LavrinenkoEtAl1999}[(75)]{KalkkinenStelle2003}[(86)]{BandosEtAl2004}[\S 4.9]{Sati2010})
\begin{equation}
\label{MTheoryGaugeAlgebra}
  \mathfrak{l}\mathcal{A}
  \overset{!}{\simeq}
  \mathrm{Free}_{\mathrm{dgL}}
  (\gamma_3, \gamma_6)
  \big/
  \left(
  \begin{aligned}
    [\text{-}] &= 0
    \\
    [\gamma_3, \gamma_3]
    &=
    \gamma_6
  \end{aligned}
  \right)
  \simeq
  \mathfrak{l} S^4
  \,,
\end{equation}
because the latter characterizes the Bianchi/Gauss identities of the C-field flux densities in 11D:
\begin{equation}
\label
{CFieldBianchiIdentities}
  \Omega^1_{\mathrm{cl}}
  \bracket({
    X^{10}
    ;\,
    \mathfrak{l}S^4
  })
  \simeq
  \left\{
  \begin{aligned}
    G_4
    \\
    G_7
  \end{aligned}
  \,\middle\vert\,
  \begin{aligned}
    \mathrm{d}\, G_4 & = 0
    \\
    \mathrm{d}\, G_7 & = 
      \sfrac{1}{2}\, G_4 \, G_4
  \end{aligned}
  \right\}
  \mathrlap{\,.}
\end{equation}
Now, as already indicated by the notation, the M-theory gauge algebra is the Whitehead-bracket $L_\infty$-algebra notably of the 4-sphere $\mathcal{A} \defneq S^4$. As always, this is just one out of an infinitum of other admissible spaces, but it is the \emph{minimal} or (weakly) \emph{universal} choice among these.

Since the nonabelian cohomology theory classified by spheres is known as \emph{Cohomotopy} theory, the hypothesis that $\mathcal{A} := S^4$ is indeed the ``right'' choice for charge quantization of M-brane charges
(\parencites[\S 2.5]{Sati2018}[\S 4]{FSS15-M5WZW}[\S 2]{FSS17-Sphere})
has been called \emph{Hypothesis H} (\parencites{FSS20-H}{FSS21-Hopf}{SS23-Mf}).

Since $\pi_7\bracket({S^4}) \simeq \mathbb{Z} \oplus \mathrm{torsion}$, this hypothesis immediately implies---cf. \cref{BraneChargesInTwistedKTheory}---Page charge quantization of M2-brane charges, and at the same time predicts a variety of non-supersymmetric torsion brane charges:
\begin{equation}
\label{MBraneChargesInCohomotopy}
  \pi_0
  \,
  \mathrm{Map}\bracket({
    \mathbb{R}^{10}
    \setminus
    \mathbb{R}^p
    ,\,
    S^4
  })
  \simeq
  \left\{
  \begin{array}{lll}
    \mathbb{Z}
    & 
    \text{if }
    p = 5
    &
    \substack{
      \text{M5-brane}
    }
    \\
    \mathbb{Z}
    \oplus 
    \mathbb{Z}_{/12}
    &
    \text{if }
    p = 2
    &
    \substack{
      \text{M2-brane}
    }
    \\
    \substack{
      \text{pure}
      \text{torsion}
    }
    &
    \text{otherwise}
    &
    \substack{
      \text{non-supersymmetric}
      \\
      \text{torsion branes}
    }
  \end{array}
  \right\}
  \mathrlap{\,.}
\end{equation}

This \emph{Hypothesis H} is supported by various consistency checks, but it remains a hypothesis (not the least because there is no independent full definition of ``M-brane charges'', much less of M-theory itself). Our main result below (\cref{Results}) is one more such consistency check: We will show that the dimensional reduction of cohomotopical charges in 11D to IIA charges in 10D reproduces the entire Borcherds hierarchy \cref{TheOvercompleteIIAFluxAlgebra} of corresponding flux densities.

\paragraph
{Circle Reduction of Charges}

With $\mathcal{A}$ any classifying space of charges (as in \cref{Motivation}), consider the situation that spacetime and hence the Cauchy surface $X^{d+1}$ is a principal circle bundle of some Chern class $c_1$ over some lower-dimensional $X^{d}$. Then \emph{double dimensional reduction} must be reflected by another classifying space fibered over $B S^1$---to be called $\Cyc \mathcal{A}$---such that the charges classified by $\mathcal{A}$ over $X^{d+1}$ are naturally identified with their \emph{double dimensional reductions} given by charges classified $S^1$-equivariantly by $\Cyc \mathcal{A}$ over $X^d$:
\begin{equation}
\label{TheExtCycAdjunction}
  \left\{
  \begin{tikzcd}
    X^{d+1}
    \ar[
      rr,
      dashed,
      "{
        \substack{
          \text{charges in $d+1$}
        }
      }"
    ]
    \ar[
      d,
      "{
        \mathrm{fib}(c_1)
      }"
    ]
    &&
    \mathcal{A}
    \\
    X^d
    \ar[
      dr,
      "{ c_1 }"{description}
    ]
    \\
    & B S^1
  \end{tikzcd}
  \right\}
  \;\;
  \simeq
  \;\;
  \left\{
  \begin{tikzcd}
    X^{d+1}
    \ar[
      d,
      "{
        \mathrm{fib}(c_1)
      }"
    ]
    \\
    X^d
    \ar[
      dr,
      "{ c_1 }"{description}
    ]
    \ar[
      rr,
      dashed,
      "{
          \text{dimensionally reduced}
      }"{yshift=-1pt},
      "{
        \text{charges}
      }"{swap,yshift=+1pt}
    ]
    &&
    \Cyc \mathcal{A}
    \ar[
      dl
    ]
    \\
    & B S^1
  \end{tikzcd}
  \right\}
  \mathrlap{\,.}
\end{equation}
Such a classifying space for doubly dimensionally reduced charges naturally exists%
\footnote{
The equivalence \cref{TheExtCycAdjunction} is due to \parencites[\S~4]{Schreiber2017}[\S~2.2]{BMSS2019}[\S~2.2]{SS24-Cyc}, following the analogous rational observation in \parencites[\S~3]{FSS17-Sphere}[\S~3]{FSS18-TD}[\S~2]{FSS18-TDR}[\S~8]{FSS2019}, for which see also \parencites[\S~2.4]{SatiVoronov2024}[\S~2.2]{GSS25-TD}. 
The special case when $\mathcal{A} \simeq B G$ is the classifying space of a connected Lie group $G$ was essentially considered in \cite{BergmanVaradarajan2004}, with focus on $G \defneq E_8$; before that, the further specialization to  $c_1 = 0$ was considered in \parencites{MathaiSati04-E8}, cf. also \parencites{Sati2009}{Sati2010-E8}---all this in our context of M/IIA reduction. Much earlier, the special case $\mathcal{A} = B G$, $c_1 = 0$ of \cref{TheExtCycAdjunction} was called the \emph{caloron correspondence} in \parencites{GarlandMurray1988}, cf. \parencites{MurrayVozzo2009}{HekmatiMurrayVozzo2011}.
}
and is given by the homotopy quotient \cref{BorelConstruction} of the free loop space of $\mathcal{A}$ (the space of images of the circle fiber in $\mathcal{A}$) by the rigid rotation $S^1$ action on loops:
\begin{equation}
\label{CyclifiedLoopSpace}
  \Cyc \mathcal{A}
  :=
  \mathrm{Map}\bracket({
    S^1, \mathcal{A}
  })
  \sslash 
  S^1
  \mathrlap{\,.}
\end{equation}

For example, applied to $\mathcal{A} \defneq S^4$ \cref{MBraneChargesInCohomotopy}, this construction reproduces exactly the flux species in IIA which are meant to be descendants (fiber integrals or basic components) of flux densities in 11D, in that
\begin{equation}
\label{TheCycS4Fluxes}
  \Omega^1_{\mathrm{cl}}
  \bracket({
    X^9
    ;\,
    \mathfrak{l}
    \Cyc S^4
  })
  \simeq
  \left\{
  \begin{aligned}
    \mathcolor{darkgreen}{F_2}
    \\
    \mathcolor{darkgreen}{H_3}
    \\
    \mathcolor{darkgreen}{F_4}
    \\
    \mathcolor{darkgreen}{F_6}
    \\
    \mathcolor{darkorange}{H_7}
  \end{aligned}
  \,\middle\vert\,
  \begin{aligned}
    \mathcolor{darkgreen}
      {\mathrm{d}\, F_2} 
     & 
     \mathcolor{darkgreen}{= 0}
    \\
    \mathcolor{darkgreen}
      {\mathrm{d}\, H_3}
      & 
      \mathcolor{darkgreen}{= 0}
    \\
    \mathcolor{darkgreen}
      {\mathrm{d}\, F_4} 
      & 
      \mathcolor{darkgreen}
        {= H_3 \, F_2}
    \\
    \mathcolor{darkgreen}
      {\mathrm{d} \, F_6}
      & 
      \mathcolor{darkgreen}
        {= H_3 \, F_4}
    \\
    \mathcolor{darkorange}
      {\mathrm{d}\, H_7}
      &
      \mathcolor{darkorange}
      {=
      \sfrac{1}{2}\, F_4 \, F_4
      - F_2 \, F_6}
  \end{aligned}
  \right\}
  \mathrlap{\,.}
\end{equation}
Not obtained this way are, however, the further higher-degree fluxes of IIA; notably the electric dual $F_8$ to $F_2$ does not descend from 11D and is hence missing in the list \cref{TheCycS4Fluxes}. 

\paragraph
{Small M-Circle Reduction}

\begin{figure}[htbp]
\centering
\begin{tikzpicture}
  \draw[
    rounded corners,
    line width=1.2pt
  ]
    (-6.1, 2.1) rectangle
    (+6.1, -2.1);
  \node at (0,0) {
    \includegraphics[width=12cm]
      {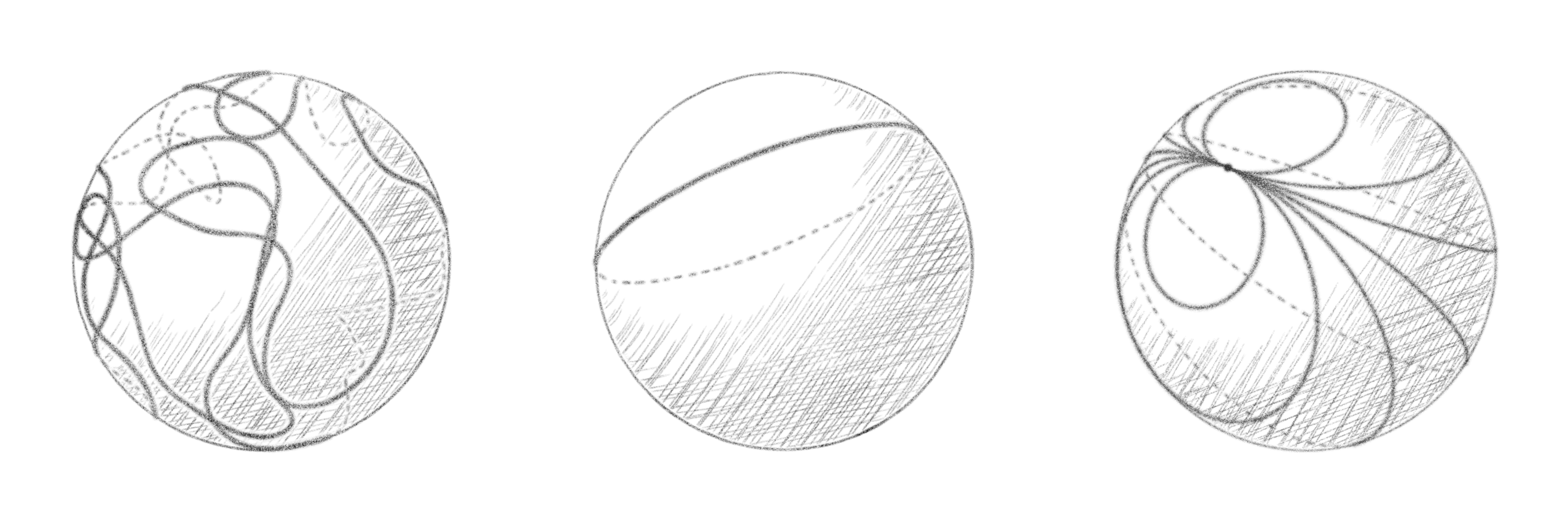}
  };
\draw
  (-2,1.8) --
  (-2,-1.8);
\draw
  (+2,1.8) --
  (+2,-1.8);
 \node at (0,-2.7) {
   \begin{tikzcd}
     \Cyc S^4
     &[1.4cm]
     \SmlCyc S^4
     \ar[
       l,
       hook'
     ]
     &[1.4cm]
     \MinCyc S^4
     \ar[
       l,
       hook'
     ]
   \end{tikzcd}
 };
\end{tikzpicture}
\caption{\label{CircleMaps}%
  Schematically indicated on the left is a generic element of $\Cyc S^4$ \cref{CyclifiedLoopSpace}: a continuous closed curve, understood here as the image under a charge-classifying map of the M-circle fiber $S^1_M$ in the classifying space $S^4$ for M-brane charges according to Hypothesis H \cref{MBraneChargesInCohomotopy}. Similarly indicated in the middle is a generic element of the subspace $\SmlCyc S^4 \subset \Cyc S^4$ of topologically \emph{small circles}, namely such curves that ``have no wiggles and wind at most once'', in that they are just round circles on $S^4$. Indicated on the right is the family of such round circles which is retained in the further minimal subspace $\MinCyc S^4$.
  In each case the degenerate circles are included (a single one on the right) not as plain points $\ast$ but as the shapes $B S^1$ of rotation-orbifold singularities $\ast \sslash S^1$, which we do not attempt to visualize here.
  Cf. also \cref{SystemOfClassifyingSpaces}.
}
\end{figure}

Indeed, type IIA is meant to be the reduction of 11D on a \emph{small} circle fiber. A proposal for what this should mean \emph{in the topological sector} has been made (only) recently in \cite{GSS26-KfromM}, assuming Hypothesis H \cref{MBraneChargesInCohomotopy}. The idea is to constrain the images $\inlinetikzcd{  S^1 \ar[r] \& S^4}$ of the circle fiber in the classifying space to ``have no wiggles and wind at most once'', hence to be \emph{round circles} on $S^4$, including the degenerate circles (constant loops), cf. \cref{CircleMaps}.

This defines the \emph{small cyclification} subspace  (\cite[Def. 3.3, Lem. 3.4]{GSS26-KfromM}, there denoted $\Cyc_1 S^4$)
\begin{equation}
\label{SmallCyclification}
  \begin{tikzcd}
    \SmlCyc S^4
    \ar[r, hook ]
    &
    \Cyc S^4
    \mathrlap{\,.}
  \end{tikzcd}
\end{equation}

Hence requiring that the classifying maps (and their homotopies) of 11D charges seen in 10D factor through the small cyclification
\begin{equation}
  \begin{tikzcd}
    X^{9}
    \ar[
      rr,
      dashed,
      "{
        \text{M-charges reduced on...}
      }"{pos=.6},
      "{ \text{...any M-circle fibers} }"
        {swap, yshift=1pt, pos=.6}
    ]
    \ar[
      drr,
      dashed,
      "{
        \text{...small M-circle fibers}
      }"{swap, sloped}
    ]
    &\phantom{------}&
    \Cyc S^4
    \\[10pt]
    &&
    \SmlCyc S^4
    \ar[u, hook]
  \end{tikzcd}
\end{equation}
is a condition on the charge sector of the theory that reflects a topological notion of ``smallness'' of the M-circle (\cref{CircleMaps}).

Remarkably, this smallness condition is indeed what makes the $F_8$-flux appear that was missing in \cref{TheCycS4Fluxes}, in that \cite[Prop. 3.10]{GSS26-KfromM}:
\begin{equation}
\label{TheSmallCycS4Fluxes}
  \Omega^1_{\mathrm{cl}}
  \bracket({
    X^9
    ;\,
    \mathfrak{l}
    \,
    \SmlCyc S^4
  })
  \simeq
  \left\{
  \begin{aligned}
    \mathcolor{darkgreen}{F_2}
    \\
    \mathcolor{darkgreen}{H_3}
    \\
    \mathcolor{darkgreen}{F_4}
    \\
    \mathcolor{darkgreen}{F_6}
    \\
    \mathcolor{darkorange}{H_7}
    \\
    \mathcolor{darkblue}{F_8}
  \end{aligned}
  \,\middle\vert\,
  \begin{aligned}
    \mathcolor{darkgreen}
      {\mathrm{d}\, F_2} 
     & 
     \mathcolor{darkgreen}{= 0}
    \\
    \mathcolor{darkgreen}
      {\mathrm{d}\, H_3}
      & 
      \mathcolor{darkgreen}{= 0}
    \\
    \mathcolor{darkgreen}
      {\mathrm{d}\, F_4} 
      & 
      \mathcolor{darkgreen}
        {= H_3 \, F_2}
    \\
    \mathcolor{darkgreen}
      {\mathrm{d} \, F_6}
      & 
      \mathcolor{darkgreen}
        {= H_3 \, F_4}
    \\
    \mathcolor{darkorange}
      {\mathrm{d}\, H_7}
      &
      \mathcolor{darkorange}
      {=
      \sfrac{1}{2}\, F_4 \, F_4
      - F_2 \, F_6}
    \\
    \mathcolor{darkblue}
      {\mathrm{d} \, F_8}
      & 
      \mathcolor{darkblue}
        {= H_3 \, F_6}
  \end{aligned}
  \right\}
  \mathrlap{\,.}
\end{equation}

In conclusion so far, 10D IIA supergravity admits flux/charge quantizations
\begin{enumerate}
  
  \item in rank-0 twisted K-theory (cf. \parencites[\S 4.1]{SS25-Flux}), 
  
  which fails to quantize the NS1-brane charge sourcing $\mathcolor{darkorange}{H_7}$-flux,

  \item in cyclic 4-Cohomotopy \cite{GiotopoulosSati2026}, 
  
  which fails to quantize the D0-brane charge sourcing $\mathcolor{darkblue}{F_8}$-flux,

  \item in small cyclic 4-Cohomotopy \cite{GSS26-KfromM}, 
  
  which finally quantizes all the brane charges sourcing the full spectrum of NS/RR-flux densities. 
\end{enumerate}

With this success of the topological smallness condition on the M-circle fiber,
we ask now for a more stringent such smallness condition that would exhibit charges sourcing also the dual dilaton flux $\mathcolor{magenta}{H_9}$ \cref{TheDilatonFlux}, and beyond that the infinite Borcherds tower of higher degree flux densities \cref{BorcherdsTowerOfBianchiIdentities}. 

\paragraph
{Minimal M-Circle Reduction}

Beyond constraining the charge image of each M-circle fiber separately to be ``topologically small'', consider now the stronger condition that also the variation \emph{in families} of these images is ``small'', in fact that only a subset of the small circle images in $S^4$ is admitted, hence that the classifying maps of charges (and their homotopies) factor through a further subspace of $\SmlCyc S^4$. 

Taking this to its logical conclusion, we will consider any \emph{minimal family} of round circles in $S^4$, minimal in that it is just enough to still reflect the fiber integration of M5-brane charge to NS5-brane charge.
\begin{equation}
  \begin{tikzcd}
    X^{9}
    \ar[
      rr,
      dashed,
      "{
        \text{M-charges reduced on...}
      }"{pos=.6},
      "{ \text{...any M-circle fibers} }"
        {swap, yshift=1pt, pos=.6}
    ]
    \ar[
      drr,
      dashed,
      "{
        \text{...small M-circle fibers}
      }"{swap, sloped, pos=.6}
    ]
    \ar[
      ddrr,
      dashed,
      "{
        \substack{
        \text{...minimal families of} 
        \\
        \text{small M-circle fibers}
        }
      }"{swap, sloped}
    ]
    &\phantom{------}&
    \Cyc S^4
    \\[10pt]
    &&
    \SmlCyc S^4
    \ar[u, hook]
    \\[10pt]
    &&
    \MinCyc S^4
    \ar[u, hook]
  \end{tikzcd}
\end{equation}

\paragraph
{The Minimal Cyclification}

The idea is evident from the right panel of \cref{CircleMaps}: We need a 3-parameter family of circles sweeping exactly once over the sphere, while all being tangent at a base point corresponding to a single degenerate circle. One explicit description of such a subspace is the following:
\begin{definition}[Minimal cyclification of $S^4$]
\label[definition]{MinimalSmallCyclificationOfS4}
With $S^4 \subset \mathbb{R}^5$ identified as the unit sphere given by $\sum_{i = 1}^5 (x^i)^2 = 1$, 
and with $\sigma \in \mathbb{R}/\mathbb{Z} \simeq S^1$ the coordinate on the M-circle, consider the subspace of the free loop space of the 4-sphere given by:
\begin{equation}
\label{TheSubspace}
  \begin{tikzcd}[sep=0pt]
    \bracket({
      S^1 \times \mathbb{R}^3
    })_{\cpt}
    \ar[
      rr,
      hook,
      "{ \Gamma }"
    ]
    &&
    \mathrm{Map}\bracket({
      S^1, 
      S^4
    })
    \\
    \bracket({
      \sigma_0, 
      (s^1, s^2, s^3)
    })
    &\longmapsto&
    \left(
    \sigma 
    \mapsto
    \begin{pmatrix}
      \frac
        {\sin 2\pi (\sigma + \sigma_0)}
        {\sqrt{1+|s|^2}}
      \\
      \frac
        {s^1\,
          \scaledbracket({
            1 - \cos 2\pi (\sigma + \sigma_0)
          })
        }
        {1+|s|^2}
      \\
      \frac
        {s^2\,
          \scaledbracket({
            1-\cos 2\pi (\sigma + \sigma_0)
          })
        }
        {1+|s|^2}
      \\
      \frac
        {s^3\,
          \scaledbracket({
            1-\cos 2\pi (\sigma + \sigma_0)
          })
        }
        {1+|s|^2}
      \\
      \frac
        {|s|^2+ \cos 2\pi (\sigma + \sigma_0 )}
        {1+|s|^2}
    \end{pmatrix}
    \right)
    \mathrlap{\,,}
  \end{tikzcd}
\end{equation}
where the point at infinity maps to the loop constant on $(0,0,0,0,1) \in \mathbb{R}^5$.

This map is clearly $S^1$-equivariant.
The corresponding homotopy quotient we call the \emph{minimal cyclification of $S^4$}:
\begin{equation}
\label{TheMinimalCyclificationOfS4}
  \begin{tikzcd}[sep=0pt]
  \MinCyc S^4
  :=
  \bracket({
    \bracket({
      S^1 \times \mathbb{R}^3
    })_{\cpt}
  })
  \sslash S^1
  \ar[
    rr,
    "{ \Gamma \sslash S^1 }"
  ]
  \ar[
    dr,
    "{ 
      \TheInclusion 
    }"{swap}
  ]
  &&
  \mathrm{Map}\bracket({
    S^1,
    S^4
  })
  \sslash 
  S^1
  \defneq
  \Cyc S^4
  \mathrlap{\,.}
  \\
  &
  \SmlCyc S^4
  \ar[
    ur
  ]
  \end{tikzcd}
\end{equation}
\end{definition}
More abstractly, this space is characterized as follows:
\begin{lemma}[Characterization of the minimal cyclification of $S^4$]
\label[lemma]{CharacterizationOfMinimalCyclification}
\begin{enumerate}
\item 
\label{CharacteristicPropertiesOfMinCycS4}
The subspace $\Gamma$ \cref{TheSubspace}
\begin{enumerate}
  \item 
  \label{ContentOfMinimalCyclification}
  contains only
  \begin{enumerate}
  \item
  loops traversing once pairwise distinct round circle submanifolds, 
  \item and a single constant loop assigned to the point at infinity,
  \end{enumerate}
  in particular $\Gamma \sslash S^1$ factors as shown in \cref{TheMinimalCyclificationOfS4}.
  \item
  This family of circles covers the 4-sphere once in that the degree of the map
  \begin{equation}
    \begin{tikzcd}
      \bracket({
        S^1 \times \mathbb{R}^3
      })_{\cpt}
      \ar[
        r, 
        "{ \Gamma }"
      ]
      &
      \mathrm{Map}\bracket({S^1 ,S^4})
      \ar[
        r,
        "{ \mathrm{ev}_0 }",
        "{ \ell \mapsto \ell(0) }"{swap}
      ]
      &
      S^4
    \end{tikzcd}
  \end{equation}
  is unity:
  \begin{equation}
   \deg\bracket({
     \mathrm{ev}_0\circ\Gamma
    }) 
      = 
    1
   \mathrlap{\,.}
  \end{equation}
\end{enumerate}

\item
 The homotopy $S^1$-quotients of all $S^1$-equivariant subspaces $\inlinetikzcd{ \bracket({ S^1 \times \mathbb{R}^3 })_{\cpt} \ar[r, hook] \& \mathrm{Map}\bracket({S^1, S^4}) }$ satisfying \cref{CharacteristicPropertiesOfMinCycS4} are homotopy equivalent over $\SmlCyc S^4$.
\end{enumerate}
\end{lemma}

\begin{proposition}[Homotopy groups of the minimal cyclification]
\label[proposition]{ComparisonMapsOnHomotopyGroups}
The comparison maps \cref{TheMinimalCyclificationOfS4} induce the following homomorphisms on low-degree homotopy groups \textup{(hence of brane charges in 
$\pi_0 \mathrm{Map}\bracket({ \mathbb{R}^9 \setminus \mathbb{R}^p, \mathcal{A} }) \simeq \pi_{8-p}(\mathcal{A})$)} with respect to suitably chosen splittings \textup{(cf. \cite[Tbl. 3]{GSS26-KfromM})}:
\[
\begin{tikzcd}[
  column sep=60pt,
  row sep=0pt,
  /tikz/column 2/.append style={anchor=base west},
  /tikz/column 3/.append style={anchor=base west},
  /tikz/column 4/.append style={anchor=base west},
]
    &[-58pt]
    \MinCyc S^4 \ar[r]
    &   
    \SmlCyc S^4 \ar[r]
    &
    \Cyc S^4
    \\
    \pi_2
    & {\color{darkgreen}\mathbb{Z}}
      \phantom{{}\!\oplus\!\mathbb{Z}_{/12}\!\oplus\!\mathbb{Z}_{/2}}
      \ar[r,"1"]
    & {\color{darkgreen}\mathbb{Z}}
      \phantom{{}\!\oplus\!\mathbb{Z}_{/n}\!\oplus\!\mathbb{Z}_{/n}}
      \ar[r,"1"]
    & {\color{darkgreen}\mathbb{Z}}
    \\
    \pi_3
    & {\color{darkgreen}\mathbb{Z}}
      \phantom{{}\!\oplus\!\mathbb{Z}_{/12}\!\oplus\!\mathbb{Z}_{/2}}
      \ar[r,"1"]
    & {\color{darkgreen}\mathbb{Z}}
      \phantom{{}\!\oplus\!\mathbb{Z}_{/12}\!\oplus\!\mathbb{Z}_{/2}\!}
      \ar[r,"1"]
    & {\color{darkgreen}\mathbb{Z}}
    \\
    \pi_4
    & {\color{darkgreen}\mathbb{Z}}\!\oplus\!\mathbb{Z}_{/2}
      \phantom{{}\!\oplus\!\mathbb{Z}_{/2}\;}
      \ar[r,"{1 \oplus 1}"]
    & {\color{darkgreen}\mathbb{Z}}\!\oplus\!\mathbb{Z}_{/2}
      \phantom{{}\!\oplus\!\mathbb{Z}_{/2}\,}
      \ar[r,"{1 \oplus 1}"]
    & {\color{darkgreen}\mathbb{Z}}\!\oplus\!\mathbb{Z}_{/2}
    \\
    \pi_5
    & \phantom{\mathbb{Z}\!\oplus\!{}}
      \phantom{{}\oplus\mathbb{Z}_{/2}}
      \mathbb{Z}_{/2}^{\oplus2}
      \ar[r,"{1^{\oplus2}}"]
    & \phantom{\mathbb{Z}\!\oplus\!{}}
      \mathbb{Z}_{/2}^{\oplus2}
      \phantom{{}\!\oplus\!\mathbb{Z}_{/2}}
      \ar[r,"{1^{\oplus2}}"]
    & \phantom{\mathbb{Z}\!\oplus\!{}}
      \mathbb{Z}_{/2}^{\oplus2}
    \\
    \pi_6
    & {\color{darkgreen}\mathbb{Z}}
      \!\oplus\!\mathbb{Z}_{/12}\!\oplus\!\mathbb{Z}_{/2}
      \ar[r,"{2\oplus1\oplus1}"]
    & {\color{darkgreen}\mathbb{Z}}
      \!\oplus\!\mathbb{Z}_{/12}\!\oplus\!\mathbb{Z}_{/2}
      \ar[r,"{ 1 \oplus 1 \oplus 1 }"]
    & {\color{darkgreen}\mathbb{Z}}
      \!\oplus\!\mathbb{Z}_{/12}\!\oplus\!\mathbb{Z}_{/2}
    \\
    \pi_7
    & {\color{darkorange}\mathbb{Z}}
      \!\oplus\!\mathbb{Z}_{/12}\!\oplus\!\mathbb{Z}_{/2}^{\oplus2}
      \ar[r,"{1\oplus1\oplus1\oplus0}"]
    & {\color{darkorange}\mathbb{Z}}
      \!\oplus\!\mathbb{Z}_{/12}\!\oplus\!\mathbb{Z}_{/2}^{\oplus2}
      \ar[r,"{ 1 \oplus 1 \oplus 1^{\oplus 2} }"]
    & {\color{darkorange}\mathbb{Z}}
      \!\oplus\!\mathbb{Z}_{/12}\!\oplus\!\mathbb{Z}_{/2}^{\oplus2}
    \\
    \pi_8
    & {\color{darkblue}\mathbb{Z}}
      \!\oplus\!\mathbb{Z}_{/2}^{\oplus4}
      \phantom{{}\!\oplus\! \mathbb{Z}_{/2}^{\oplus 2}}
      \ar[r,"{2\oplus1^{\oplus3}\oplus 0}"]
    & {\color{darkblue}\mathbb{Z}}
      \!\oplus\!\mathbb{Z}_{/2}^{\oplus4}
      \phantom{{}\!\oplus\! \mathbb{Z}_{/2}^{\oplus 2}}
      \ar[r,"0 \oplus 1^{\oplus 4}"]
    & \phantom{\mathbb{Z}\!\oplus\!{}}
      \mathbb{Z}_{/2}^{\oplus4}
    \\
    \pi_9
    & {\color{magenta}\mathbb{Z}}
      \!\oplus\!\mathbb{Z}_{/24}\!\oplus\!\mathbb{Z}_{/3}
      \!\oplus\!\mathbb{Z}_{/2}^{\oplus3}
      \ar[r,"{ 0 \oplus 2 \oplus 1 \oplus 1^{\oplus 2} \oplus 0 }"]
    & \phantom{\mathbb{Z}\oplus{}}
      \mathbb{Z}_{/24}\!\oplus\!\mathbb{Z}_{/3}
      \!\oplus\!\mathbb{Z}_{/2}^{\oplus3}
      \ar[r,"{ 1 \oplus 1 \oplus 1 \oplus 1 \oplus 0 }"]
    & \phantom{\mathbb{Z}\oplus{}}
      \mathbb{Z}_{/24}\!\oplus\!\mathbb{Z}_{/3}
      \!\oplus\!\mathbb{Z}_{/2}^{\oplus2}
      \mathrlap{\,.}
  \end{tikzcd}
\]  
\end{proposition}

\section
{Results}
\label{Results}

\subsection
{Obtaining Borcherds IIA}
\label
{ObtainingBorcherdsIIA}

Our first main observation now is that flux quantization in minimal cyclic 4-Cohomotopy (\cref{MCircleReductionOfCharges}) implies the IIA Borcherds hierarchy:
\begin{theorem}
\label[theorem]{prop:RationalModelOfMinCyc}
\begin{enumerate}
\item
\label{RationalMinCycIsPositiveBorcherds}
The Whitehead-bracket $L_\infty$-algebra $\mathfrak{l}(\text{-})$ 
\textup{(\cref{WHBracketLinfinityAlgebras})} of $\MinCyc S^4$ \textup{(\cref{MinimalSmallCyclificationOfS4})}
is the positive type IIA Borcherds algebra $\bIIA$ \textup{(\cref{IIABorcherdsAlgebra})}.

\item The inclusion $\TheInclusion$ \cref{TheMinimalCyclificationOfS4} into $\SmlCyc S^4$ induces the projection of the positive IIA Borcherds algebra onto its dg-Lie quotient by the coefficient $\eta_8$ \cref{LowDegreeBorcherdsElements}, 
\end{enumerate}
in that the following square of $L_\infty$-algebras commutes:
\begin{equation}\label{eq:two-models}
  \begin{tikzcd}[
    column sep=30pt,
    row sep=15pt
  ]
    \mathfrak{l}\bracket({\MinCyc S^4})
      \ar[r, "{\mathfrak{l}(\TheInclusion)}"]
      \ar[d, "\alpha"', "\sim"{sloped}]
    &
    \mathfrak{l}\bracket({\SmlCyc S^4})
      \ar[d, "\beta", "\sim"{sloped,swap}]
    \\
    \bIIA
      \ar[r, two heads, "p"]
    &
    \bIIA/\langle\eta_8\rangle.
  \end{tikzcd}
\end{equation}
\end{theorem}
\begin{proof}[Proof outlines]
\begin{enumerate}
  \item 
  First observe that we have a homotopy equivalence
  \begin{equation}
  \label{MinCycEquivalentToWedgeSum}
    \MinCyc S^4
    \defneq
    \bracket({
      S^1 \times \mathbb{R}^3
    })_{\cpt}
    \sslash S^1
    \;\simeq\;
    B S^1 \vee S^3
  \end{equation}
  between the minimal cyclification and the wedge sum of the delooping of the circle with the 3-sphere: This is because one-point compactification turns Cartesian products into smash products \cref{CompactificationOfProductSpace},
  \begin{equation}
  \label
  {OnePointCompactificationOfProductAsWedgeProduct}
    \bracket({
      S^1 \times \mathbb{R}^3
    })_{\cpt}
    \simeq
    S^1_{\plus}
    \wedge
    \mathbb{R}^3_{\cpt}
    \mathrlap{\,,}
  \end{equation}
  and with \cref{CharacterizationOfMinimalCyclification} \cref{CharacteristicPropertiesOfMinCycS4} \cref{ContentOfMinimalCyclification} we have that
  \begin{enumerate}
  \item
    away from $\infty \in \mathbb{R}^3_{\cpt}$ the $S^1$-action is free, yielding a contribution 
    $\bracket({S^1 \times \mathbb{R}^3}) \sslash S^1 \simeq \mathbb{R}^3$,
  \item while at $\infty$, where the $S^1$-factor collapses in the smash product \cref{OnePointCompactificationOfProductAsWedgeProduct}, it is trivial, yielding
  a contribution $\{\infty\} \sslash S^1 \simeq B S^1$.
  \end{enumerate}
  Gluing these two contributions at $\infty$ yields the claim \cref{MinCycEquivalentToWedgeSum}.

  Now, the minimal Sullivan models of the summands on the right of \cref{MinCycEquivalentToWedgeSum} are well-known to be (cf. \parencites{Menichi2015}[p. 21]{SS25-Flux}):
  \begin{equation}
  \label{RationalModelsOfBS1AndS3}
    \begin{aligned}
      \mathrm{CE}\bracket({
        \mathfrak{l} B S^1
      })
      &
      \simeq
      \mathrm{Free}_{\mathrm{dgcA}}
      ({ f_2 })
      /
      \bracket({
      \begin{aligned}
        \mathrm{d}\, f_2
        & 
        = 0
      \end{aligned}
      })
      \simeq
      \mathrm{CE}\bracket({
        \mathrm{Free}_{\mathrm{dgL}}
        (\varphi_1)
        \big/
        \left(
        \substack{
          [\text{-}] = 0
          \\
          [\varphi_1, \varphi_1] = 0
        }
        \right)
      })
      \\
      \mathrm{CE}\bracket({
        \mathfrak{l} S^3
      })
      &
      \simeq
      \mathrm{Free}_{\mathrm{dgcA}}
      ({ h_3 })
      /
      \bracket({
      \begin{aligned}
        \mathrm{d}\, h_3
        & 
        = 0
      \end{aligned}
      })
      \simeq
      \mathrm{CE}\bracket({
        \mathrm{Free}_{\mathrm{dgL}}(\eta_2)
        \big/
        \left(
        \substack{
          [\text{-}] = 0
        }
        \right)
      })
      \mathrlap{\,,}
    \end{aligned}
  \end{equation}
  which follows with \cref{UnderlyingChOfWhiteheadLInfinity,CohomologyOfCEOfWhiteheadLinfinity}. With the Whitehead $L_\infty$-algebras readily identified on the right of \cref{RationalModelsOfBS1AndS3}, their wedge sum (their cofiber product under the terminal object) is
  \begin{equation}
    \def\arraystretch{1.8}
    \begin{array}{lll}
    \mathfrak{l} \bracket({
      \MinCyc S^4
    })
    & \simeq
    \mathfrak{l}\bracket({
      B S^1 \vee S^3
    })
    &
    \substack{
      \text{by \cref{MinCycEquivalentToWedgeSum}}
    }
    \\
    & \simeq
    \bracket({
      \mathfrak{l} BS^1 
    })
    \sqcup
    \bracket({
      \mathfrak{l} S^3
    })
    &
    \substack{
      \text{cf. \cite[\S A.4]{LazarevMarkl2015}}
    }
    \\
    &
    \simeq
    \bracket({
    \mathrm{Free}_{\mathrm{dgL}}
    (\varphi_1)
    \big/\!
    \left(
    \substack{
      [\text{-}] = 0
      \\
      [\varphi_1, \varphi_1] = 0
    }
    \right)
    })
    \sqcup
    \bracket({
    \mathrm{Free}_{\mathrm{dgL}}
    (\eta_2)
    \big/\!
    \left(
    \substack{
      [\text{-}] = 0
    }
    \right)
    })
    &
    \substack{
      \text{by \cref{RationalModelsOfBS1AndS3}}
    }
    \\
    &
    \simeq
    \mathrm{Free}_{\mathrm{dgL}}
    (\varphi_1, \eta_2)
    \big/
    \left(
      \substack{
        [\text{-}] = 0
        \\
        [\varphi_1, \varphi_1] = 0
      }
    \right)
    \\
    &
    \simeq
    \bIIA
    &
    \substack{
      \text{by \cref{ThePositiveIIABorcherdsAlgebra}.}
    }
    \end{array}
  \end{equation}
This determines $\alpha$ in \cref{eq:two-models}, which sends $\varphi_1,\eta_2$ to the corresponding generators of $\bIIA$.
  \item
  The proof that the inclusion induces the quotient projection by $\eta_8$ is more technical; we relegate it to
\cref{lem:minimal_cyclification_and_borcherds} in \cref{app:proofs}. 
\qedhere
\end{enumerate}
\end{proof}

In view of \cref{Motivation} this result means that:\\
\begin{minipage}{7.2cm}
\begin{standout}
  Minimal cyclic 4-Cohomotopy is an admissible flux-quantization law for type IIA that captures the entire Borcherds hierarchy of higher gauge fields and their higher symmetries.
\end{standout}
\end{minipage}
\adjustbox{rndfbox=4pt}{
$
  \def\arraystretch{1.9}
  \begin{array}{cc}
    \substack{
      \text{minimal cyclic}
      \\
      \text{4-Cohomotopy moduli}
    }
    & 
    \mathrm{Map}\bracket({
      X^9
      ,
      \MinCyc S^4
    })
    \\[-5pt]
    \pi_0
    =
    &
    \left\{
    \substack{
      \text{complete type IIA charges}
      \\
      \text{including NS1- \& D0-brane charge}
      \\
      \text{and including dual dilaton charge}
    }
    \right\}
    \\
    \pi_{\geq 1}
    =
    &
    \left\{
    \substack{
      \text{higher global symmetries}
      \\
      \text{reflecting the entire}
      \\
      \text{Borcherds hierarchy}
    }
    \right\}
  \end{array}
$}

\begin{remark}[Integral lattice for dual-dilaton]
\label[remark]{IntegralLatticeForDualDilaton}
Minimal cyclic 4-Cohomotopy quantizes (discretizes) the
dual dilaton flux $H_9$. Indeed, by \cref{prop:RationalModelOfMinCyc} \cref{RationalMinCycIsPositiveBorcherds}, the realification of $\pi_9\bracket({\MinCyc S^4})\cong \mathbb Z\oplus \pi_9\bracket({\MinCyc S^4})_{\text{tor}}$ (from \cref{ComparisonMapsOnHomotopyGroups}) is $\R\langle\eta_8\rangle$, where $\eta_8$ is the coefficient of $H_9$ in \cref{UnifiedFluxDensity}. Thus on $S^9$ with the lower fluxes set to zero, a lift to $\MinCyc S^4$ restricts $\int_{S^9} H_9$ to a discrete set of values. More generally, in view of \cref{TheDilatonFlux}, the integral lift is a condition on the full cocycle 
\begin{equation}
  \widehat H_9
  =
  H_9
    -
  B_2 \, H_7
    -
  \tfrac{1}{2} C_3\widetilde F_6
    +
  \tfrac{3}{2} C_1 \widetilde F_8
  \mathrlap{\,,}
\end{equation}
where 
$
  \widetilde F_{2k}
  :=
  \bracket({
    e^{-B_2} \sum_i F_{2i}
  })_{2k}
$ are the untwisted RR-flux densities, and where $B_2$, $C_1$, $C_3$ are local potentials with $\mathrm{d}\, B_2 = H_3$, $\mathrm{d}\, C_1 = \widetilde F_2 = F_2$, and $\mathrm{d}\, C_3 = \widetilde F_4 = F_4 - B_2\, F_2$. With these conventions, $\mathrm{d}\, \widehat H_9 = 0$ follows from the Bianchi identities \cref{TruncatedBorcherdsTowerOfBianchiIdentities}, using that $\mathrm{d}\,\widetilde F_{2k} = 0$.

This quantization law for the dual dilaton flux is a consequence (hence a prediction) of the charge quantization of type IIA theory in minimal cyclic 4-Cohomotopy.

A familiar analogue is the dual RR axion flux in type IIB. The period of the axion field strength $F_1$ around a linking circle measures the magnetic D7-brane charge, and the period of its dual $F_9$ over a linking nine-sphere measures the electric D-instanton charge. In type IIB this electric--magnetic pairing supplies a physical argument for quantization.
\end{remark}

\begin{remark}
  Indeed, \cref{prop:RationalModelOfMinCyc} has a IIB analogue, in view of \cref{MinCycEquivalentToWedgeSum}: The IIB Borcherds algebra is isomorphic to $\mathfrak{l}\bracket({ S^3 \vee S^3 })$. We will discuss this elsewhere. 
\end{remark}

\subsection
{Massive Borcherds IIA}
\label{MassiveBorcherds}

The relation of massive type IIA to 11D, if any, has remained somewhat obscure. Bergshoeff, Lozano and Ort\'{\i}n obtained the bosonic Romans theory by reducing a deformation of eleven-dimensional supergravity with a prescribed Killing vector \cite{BergshoeffLozanoOrtin1998}. Hull proposed an M-theory compactification on a $T^2$-bundle over $S^1$ in a limit in which the volume of the total space tends to zero, with the Romans mass encoded by the monodromy \cite{Hull1998Massive}. A closer precedent for the construction below is the proposal of Adams and Evslin to identify the quantized Romans mass with the level of a central  extension of the loop group of $E_8$ \cite{AdamsEvslin2003}. Bergman and Varadarajan showed how the low-degree characteristic classes of the resulting bundles encode the massive IIA Bianchi identities, including $\mathrm dF_2=F_0H_3$ \cite[\S 3.2]{BergmanVaradarajan2004}.

We now make the following suggestive observations:

\paragraph
{The Minimal Family of Circles Near the Collapsed Circle}

The wedge sum realization of the minimal cyclification  \cref{MinCycEquivalentToWedgeSum} reveals equivalently a realization as a homotopy cofiber:
\begin{equation}
\label{MinCycAsHomotopyCofiber}
  \MinCyc S^4
  \simeq
  B S^1 \vee S^3
  \simeq
  \mathrm{hocofib}\bracket({
    0 :
    \inlinetikzcd{
      S^2 \ar[r]
      \&
      B S^1
    }
  })
  \mathrlap{\,,}
\end{equation}
namely as the cell attachment via the zero-map:
\begin{equation}
  \begin{tikzcd}
    S^2 
    \ar[r]
    \ar[
      rr,
      uphordown,
      "{ 0 }"{description}
    ]
    \ar[
      d,
      hook
    ]
    \ar[
      dr,
      phantom,
      "{ \ulcorner }"{pos=.9}
    ]
    &
    \ast
    \ar[r]
    \ar[d]
    \ar[
      dr,
      phantom,
      "{ \ulcorner }"{pos=.9}
    ]
    &
    B S^1
    \ar[d]
    \\
    D^3
    \ar[r]
    &
    S^3
    \ar[r, hook]
    &
    B S^1 \vee S^3
    \mathrlap{\,.}
  \end{tikzcd}
\end{equation}

In general, the homotopy classes of maps from $S^2$ to $B S^1$ are labeled by the integers, these being the Chern classes of the principal circle bundles $P$ over $S^2$ that are classified thereby:
\begin{equation}
  \pi_0 
  \mathrm{Map}\bracket({
    S^2, B S^1
  })
  \simeq
  \mathbb{Z}
  \,,
  \;\;\;\;\;\;\;
  \begin{tikzcd}
    S^1
    \ar[
      r,
      hook
    ]
    \ar[
      d
    ]
    \ar[
      dr,
      phantom,
      "{ \lrcorner }"{pos=.1}
    ]
    &
    P
    \ar[r]
    \ar[
      d,
      ->>
    ]
    \ar[
      dr,
      phantom,
      "{ \lrcorner }"{pos=.1}
    ]
    &
    \ast
    \ar[d]
    \\
    \ast
    \ar[r]
    &
    S^2
    \ar[
      r,
      "{ m }"
    ]
    &
    B S^1
    \mathrlap{\,.}
  \end{tikzcd}
\end{equation}

In our case $\MinCyc S^4$ \cref{MinCycAsHomotopyCofiber}:
\begin{enumerate}
\item
the 2-sphere is the boundary of the minimal 3-ball of round circles that cover the 4-sphere away from the base point in \cref{MinimalSmallCyclificationOfS4,CircleMaps}, 
\item
the principal circle bundle $P$ has as fiber over a point the actual small circle labeled by that point,
\item the triviality of that bundle (classified by the 0-map) reflects that this trivial bundle extends over a 3-ball neighborhood of the base point, reflecting the collapse of the family of circles to the constant one.
\end{enumerate}

\paragraph
{Removing the Collapsed M-Circle}

This analysis reveals a natural ``twisting'' of this situation: If we don't admit and hence delete the collapsed circle from the family, then the remaining family of non-degenerate circles may exhibit a nontrivial bundle over the asymptotic 2-sphere around the excised point. The resulting modified spaces of round circles on the 4-sphere are therefore the homotopy cofibers of the classifying maps $\inlinetikzcd{ m : S^2 \ar[r] \& B S^1 }$ of these asymptotic bundles:
\begin{equation}
\label{MinCycN}
  \MinCyc_m S^4
  :=
  \mathrm{hocofib}\bracket({
    m : 
    \inlinetikzcd{
      S^2 
      \ar[r]
      \&
      B S^1
    }
  })
  \mathrlap{\,,}
\end{equation}
hence are these cell attachments:
\begin{equation}
\label{MinCycNCellAttachment}
  \begin{tikzcd}
    S^2
    \ar[
      r,
      "{ m }"
    ]
    \ar[
      d,
      hook
    ]
    \ar[
      dr,
      phantom,
      "{ \ulcorner }"{pos=.9}
    ]
    &
    B S^1
    \ar[
      d
    ]
    \\
    D^3
    \ar[
      r,
      hook
    ]
    &
    \AttachingSpace{m}
    \mathrlap{
      \,
      \simeq
      \MinCyc_m S^4
      \,.
    }
  \end{tikzcd}
\end{equation}

\begin{definition}
\label[definition]{MassiveMinimalCyclicCohomotopy}
We denote the disjoint union of all these deformations \cref{MinCycNCellAttachment} by
\begin{equation}
\label{MassMinCyc}
  \MassMinCyc S^4
  :=
  \textstyle{\coprod_{m \in \mathbb{Z}}}
    \MinCyc_m S^4
\end{equation}
and speak of the \emph{mass-deformed} or just  \emph{massive minimal cyclification} of the 4-sphere. 

Accordingly, the nonabelian cohomology theory \cref{NonabelianCohomologyFunctor} represented by this space we call \emph{massive minimal cyclic 4-Cohomotopy} $H^0\bracket({\text{-}; \MassMinCyc S^4})$.
\end{definition}

\paragraph
{The Whitehead dg-Lie Algebra of Massive Minimal Cyclified $S^4$}

Our second main observation now is the following massive generalization of \cref{prop:RationalModelOfMinCyc}:
\begin{theorem}
\label[theorem]{prop:RationalModelOfMassiveMinCyc}
The Whitehead dg-Lie algebra \textup{(\cref{WHBracketLinfinityAlgebras})} of
\begin{enumerate}
\item the massive minimal cyclification of $S^4$ \cref{MassMinCyc} 
\item 
with respect \cref{TheComponentLabels} to the evident component labeling 
$\ell : \MinCyc_m S^4 \mapsto m$, 
\end{enumerate}
is the massive deformation \cref{TheMassiveIIABorcherdsAlgebra} of the positive IIA Borcherds algebra:
\begin{equation}
\label{MassiveMinCycWhiteheadAlgebra}
  \mathfrak{l}\bracket({
    \MassMinCyc S^4
  })
  \simeq
  \bIIAMassive
  \mathrlap{\,.}
\end{equation}
\end{theorem}
We relegate the proof of this statement to \cref{ProofOfSecondMainTheorem}.

By \cref{GeneralFluxQuantizationSchematics}, this (\cref{prop:RationalModelOfMassiveMinCyc}) shows that (we expand on this in \cref{Conclusion} below):
\begin{standout}
  Charge quantization in massive minimal cyclic 4-Cohomotopy $H^0\bracket({\text{-}; \MassMinCyc S^4})$ constitutes a global completion of massive IIA supergravity that quantizes the entire massive IIA Borcherds hierarchy of charges and higher global symmetries.
\end{standout}

We next show that 
our nonabelian cohomology theory \cref{MassMinCyc} \emph{massive minimal cyclic 4-Cohomotopy}, $H^0\bracket({\text{-}; \MassMinCyc S^4})$, has an approximation within twisted abelian cohomology theories
by twisted K-theory \cref{BraneChargesInTwistedKTheory}, when ignoring the NS1-charge, the dual dilaton and the rest of the massive IIA Borcherds hierarchy.

\paragraph{Comparison to Twisted K-Theory}

Recall the comparison from small cyclic 4-Cohomotopy:

\begin{proposition}[{\cite[Thms.~3.17 and 3.21, Rem.~3.18]{GSS26-KfromM}}]
\label[proposition]{ComparisonMapOnSmlCyc}
  The comparison map $\Phi$ from \textup{\cite[Rem.~3.18]{GSS26-KfromM}} \textup{(here at $n = 2$)}, from the small cyclification of the 4-sphere \cref{SmallCyclification} to the classifying space for twisted K-theory \cref{BraneChargesInTwistedKTheory},
  \begin{equation}
    \label{TheComparisonMapOnSmlCyc}
    \begin{tikzcd}[
      column sep=12pt,
      row sep=4pt
    ]
      \SmlCyc S^4 
        \ar[
          rr,
          "\Phi"
        ] 
        \ar[
          dr,
          "h_3^C"'
        ]
        && 
        B\mathrm{U}\sslash B\mathrm{U}(1) 
        \arrow[
          dl,
          "h_3^K"
        ] 
        \\
        & 
        B^3\mathbb{Z} 
        \mathrlap{\,,}
    \end{tikzcd}
  \end{equation}
  is such that:
  \begin{enumerate}
  \item
  \label
  {ChargeIdentificationUpToFactorOf2}
  it preserves the universal NS5-brane charge $h_3$ and multiplies D6-brane charge $f_2$ and D4-brane
  charge $f_4$ by 2,%
  \footnote{
    The D4 comparison is taken on the homotopy fiber of $f_2^C$, cf. \cite[(76c)]{GSS26-KfromM}.
  }

  \item
  while the shifted restriction
  $\Phi_{\vert f_2^C,h_3^C = 0} - \beta$
  is the identity on the brane charges,

  \item
  whose Chern character is
  \begin{equation}
    \mathrm{ch}_{\leq 10}
      \bracket({
        \Phi_{\vert f_2^C, h_3^C = 0}
          -
        \beta
      })
    = 
    \widetilde F_4 
      + 
    2\widetilde F_6 
      + 
    2\widetilde F_8
      + 
    2\widetilde F_{10}
      + 
    \kappa\,\widetilde F_4\,\widetilde F_6,
    \;\;\; \kappa\in\mathbb{Q}.
  \end{equation}
  \end{enumerate}
\end{proposition}
Here the \emph{Bott generator}
\begin{equation}
  \label{BottGenerator}
  \beta\in\widetilde{\mathrm{KU}}^0(S^4)
\end{equation}
is oriented by $\int_{S^4}\mathrm{ch}_2(\beta)=1$,
and tilded
RR-flux densities are the untwisted components
\begin{equation}
  \label{UntwistedRRFluxDensities}
  \widetilde F_{2k}
    := 
  \bracket({
    e^{-B_2} \textstyle{\sum_i} F_{2i}
  })_{2k}
  \mathrlap{\,,}
\end{equation}
with $\mathrm{d}\, B_2 = H_3$ the assumed global trivialization of the flux sourced by NS5-brane charge.

This means that small cyclic 4-Cohomotopy is an accurate approximation of twisted K-theory only up to factors of 2.

We next show that passing to the minimal cyclification gives a genuine integral half of the massless comparison and extends it to every Romans mass.
\begin{theorem}
\label[theorem]{TheMassiveComparisonMap}
  The classifying space for massive minimal cyclic 4-Cohomotopy
  of \cref{MassMinCyc} admits a comparison to NS5-charge twisted K-theory \cref{BraneChargesInTwistedKTheory}
  \begin{equation}
    \label{MassiveComparisonMap}
    \begin{tikzcd}[
      column sep=13pt,
      row sep=3pt
    ]
      \MassMinCyc S^4 \arrow[rr,"\Psi"] \arrow[dr,"h_3^C"']
        && (B\mathrm{U}\times\mathbb{Z})\sslash B\mathrm{U}(1)
             \arrow[dl,"h_3^K"] \\
        & B^3\mathbb{Z} \mathrlap{\,,} &
    \end{tikzcd}
  \end{equation}
  such that:
  \begin{enumerate}
    \item
    \label{RankOfPhi}
      $
        \forall_{ m \in \mathbb{Z} } \;:\; \mathrm{rk}\bracket({ 
          \Psi_{\vert f_0^C = m} 
        }) = m
      $,
    \item 
    \label{PhiMasslessSector}
     in the massless sector, $f_0^C = 0 = m$, 
     this divides the restriction of $\Phi$ \cref{TheComparisonMapOnSmlCyc}
     along the inclusion $\inlinetikzcd{ \MinCyc S^4 \ar[r, hook, "{ \iota_{\mathrm{min}} }"] \& \SmlCyc S^4 }$ \cref{TheMinimalCyclificationOfS4}, in that
     \begin{equation}
     \label{DivisionBy2}
        2[\Psi_{\vert f_0^C = 0}]
          =
        \iota_{\mathrm{min}}^*[\Phi],
     \end{equation}
      making $\Psi_{\vert f_0^C = 0}$ the identity on D4- and D6-brane charges \textup{(cf. \cref{ChargeIdentificationUpToFactorOf2} in \cref{ComparisonMapOnSmlCyc})},
    \item 
    \label{ChernCharacterOnMinCycS4}
      its universal Chern character satisfies
      \begin{equation}
        \label{ChernCharacterOfPsi}
        \mathrm{ch}
          \bracket({
            \Psi_{
              \vert
              f_0^C, f_2^C, h_3^C = 0
            }
          })
        = 
        \widetilde F_4
          +
        \widetilde F_6
          +
        \widetilde F_8
          +
        \widetilde F_{10}
          +
        \widetilde F_{12}
          +
        \cdots
        \mathrlap{\,.}
      \end{equation}
  \end{enumerate}
\end{theorem}

\begin{proof}[Proof outline \textup{(full proof relegated to \cref{proof:TheMassiveComparisonMap})}]
  As in \cite{GSS26-KfromM}, we use lowercase versions of the symbols for flux densities to denote the corresponding universal cohomology classes on $\MassMinCyc S^4$ and on the twisted K-theory classifying space, with a superscript ``$C$'' for the former and ``$K$'' for the latter. For instance $f_2^C$ denotes the universal D6-brane charge on $\MassMinCyc S^4$ and $f_2^K$ that on $\bracket({ B \mathrm{U} \times \mathbb{Z} }) \sslash B \mathrm{U}(1)$. 

  This particular class is identified with the Chern class of the universal complex line bundle $L$ over $B S^1 \simeq B \mathrm{U}(1)$, (which is associated to the avatar of the M-circle bundle):
  \begin{equation}
  \label{ClassOfL}
    [L] = f_2^C
    \mathrlap{\,.}
  \end{equation}
  By \cite[(91)]{GSS26-KfromM}, on the
  constant-loop locus $S^4 \times B S^1$ of $\SmlCyc S^4$, the map $\Phi$ classifies the virtual vector bundle
  $2(L-1) + \beta\bracket({[L]+1})$.
  But under restriction to the minimal cyclification, 
  which retains a single constant loop $p\in S^4$,
  the Bott generator $\beta$ \cref{BottGenerator} clearly vanishes on $\{p\}\times B S^1$, leaving the virtual bundle $2(L-1)$.

  To see that this can be divided by 2 on the whole space  
  $\MinCyc S^4\simeq B S^1\vee S^3$, observe that the twist is trivialized on $B S^1$
  and restricts to the generator $h_3^C$ on $S^3$. But 
  $\mathrm{KU}^0_{h_3^C}\bracket({S^3})=0$. Therefore the Mayer--Vietoris sequence for the wedge sum
  gives an isomorphism
  \begin{equation}
    \label{PsiAsLMinusOne}
    \begin{tikzcd}[sep=0pt]
      \mathrm{KU}^0_{h_3^C}\bracket({
        B S^1\vee S^3
      })
      \ar[
        rr,
        "{ \sim }"
      ]
      &&
      \widetilde{\mathrm{KU}}^0\bracket({
        B S^1
      })
      &
      = 
      [L-1]
        \cdot
      \mathbb{Z}\bracket\llbracket{L-1}\rrbracket
      \\
      \bracket[{
        \Psi_{\smash{\vert f_0^C = 0}}
      }]
      &\longmapsto& 
      \bracket[{
         L - 1
      }]
      \mathrlap{\,.}
    \end{tikzcd}
  \end{equation}
  This proves the unique halving \cref{DivisionBy2} and its unit D6-brane charge. The D4-brane charge
  is additive on the $f_2^C$-fiber $S^3 \vee S^4$, so the unit D4-charge follows from \cite[(76c)]{GSS26-KfromM} by canceling 2
  in $H^4\bracket({S^3\vee S^4;\mathbb{Z}}) \simeq \mathbb{Z}$.
\end{proof}

\begin{remark}[D6-brane charge as the root of the hierarchy]
\label[remark]{D6BraneChargeAsRoot}
  The complex line bundle $L$ in \cref{ClassOfL} represents the universal D6-brane charge in minimal cyclic 4-Cohomotopy. Hence \cref{PsiAsLMinusOne} shows that in some sense the entire IIA Borcherds hierarchy of brane charges in minimal cyclic 4-Cohomotopy is rooted in D6-brane charge. We obtain a more geometric (cobordism) perspective on this phenomenon in \cref{AsD6CobordismTheory} below.
\end{remark}

\begin{remark}
\label[remark]
{OriginalMotivationForHypothesisK}
\Cref{ChernCharacterOnMinCycS4} in \cref{TheMassiveComparisonMap}
says that 
the universal Chern character \cref{ChernCharacterOfPsi} of minimal cyclic 4-Cohomotopy agrees with that of K-theory. 
But the original motivation for the \emph{Hypothesis K} that D-brane charge is quantized in K-theory \cref{KTheoryBraneCharges} rested on the observation \parencites[(2.11)]{GreenHarveyMoore1996}[(5.1)]{CheungYin1997}[(1.1)]{MinasianMoore1997}
that a common formula for rational D-brane charge looks just like the K-theoretic Chern character (times a conventional rescaling%
\footnote{%
  The traditional inclusion of the factor $\sqrt{\widehat{A}}$  in the expression of rational D-brane charge is just a historical normalization convention \parencites[ftn. 12]{FreedHopkins2000}[(2.8)]{Freed2002}[(1.8)]{BrodzkiMathaiRosenbergSzabo2008}[ftn. 8]{DistlerFreedMoore2011} that may be absorbed \cite[(1.11)]{BrodzkiMathaiRosenbergSzabo2008} into the cohomological pairing, which plays no role here.
}
by $\widehat{A}^{\sfrac{1}{2}}$).

Hence \cref{TheMassiveComparisonMap} \cref{ChernCharacterOnMinCycS4} entails that:
\begin{standout}
  The original motivation for D-brane charge quantization in K-theory cannot distinguish K-theory from minimal cyclic 4-Cohomotopy.
\end{standout}
\end{remark}
Beyond this original motivation, actual checks of Hypothesis K, available only in simple examples, have remained inconclusive.%
\footnote{
  \cite{FredenhagenQuella2005}: ``It might surprise that despite all the progress that has been made in understanding branes on group manifolds, there are usually not enough D-branes known to explain the whole charge group predicted by (twisted) K-theory. [...] it is fair to say that a satisfactory answer is still missing.''
}
It is therefore fair to say that the question how to globally complete (massive) IIA theory by brane charge quantization has remained open. This brings us to our conclusion.

\begin{SCtable}[.66][htb]
\caption{\label{SystemOfClassifyingSpaces}%
 The system of classifying spaces $\mathcal{A}$ considered here, and the nonabelian cohomology theories
 $H^0({\text{-}; \mathcal{A}})$ which they classify.
 Here $\Psi$ is the identity on the Chern character \cref{ChernCharacterOfPsi}, so that the original motivation for IIA charge quantization in K-theory applies verbatim also to massive minimal cyclic 4-Cohomotopy (\cref{OriginalMotivationForHypothesisK}).
 }
\adjustbox{rndfbox=4pt}{
$
  \begin{tikzcd}[
    column sep=1pt
  ]
    \substack{
      \text{cyclic 4-Cohomotopy}
      \\
      \text{(\cref{CircleMaps,TheCycS4Fluxes})}
    }
    &
    \Cyc S^4
    &[20pt]
    \\
    \substack{
      \text{small}
      \\
      \text{cyclic 4-Cohomotopy}
      \\
      \text{(\cref{CircleMaps,SmallCyclification})}
    }
    &
    \SmlCyc S^4
    \ar[
      u,
      hook
    ]
    & 
    \\
    \substack{
      \text{minimal}
      \\
      \text{cyclic 4-Cohomotopy}
      \\
      \text{(\cref{CircleMaps,MinimalSmallCyclificationOfS4})}
    }
    &
    \MinCyc S^4
    \ar[
      u, hook
    ]
    \ar[
      d, hook
    ]
    \ar[
      r
    ]
    & 
    B \mathrm{U} \sslash B \mathrm{U}(1)
    \ar[
      d,
      hook
    ]
    &
    \substack{
      \text{rank-0}
      \\
      \text{twisted K-theory}
      \\
      \text{\cref{KTheoryBraneCharges}}
    }
    \\
    \substack{
      \text{massive minimal}
      \\
      \text{cyclic 4-Cohomotopy}
      \\
      \text{(\cref{MassiveMinimalCyclicCohomotopy})}
    }
    &
    \MassMinCyc S^4
    \ar[
      r,
      "{ \Psi }",
      "{
          \substack{\text{%
     (\cref{TheMassiveComparisonMap})
          }}
      }"{swap, yshift=-4pt}
    ]
    & 
    \bracket({
      B \mathrm{U} 
      \!\times\!
      \mathbb{Z}
    })
      \sslash B \mathrm{U}(1)
    &
    \substack{
      \text{twisted K-theory}
      \\
      \text{\cref{BraneChargesInTwistedKTheory}}
    }
  \end{tikzcd}
$
}
\end{SCtable}

\section
{Conclusion}
\label{Conclusion}

\subsection
{Globally Completed Massive Borcherds IIA Theory}
\label{GloballyCompletedMassiveIIA}

By the general relation \cref{GeneralFluxQuantizationSchematics} between 
\begin{enumerate}
\item
brane charge cohomology theories $H^0(\text{-};\mathcal{A}) \simeq \pi_0\, \mathrm{Map}\bracket({\text{-}, \mathcal{A}})$ and 
\item
sets $\Omega^1_{\mathrm{cl}}\bracket({\text{-}; \mathfrak{l}\mathcal{A}})$ of fluxes satisfying Bianchi identities,
\end{enumerate}
our main results (\cref{Results}) yield:
\begin{enumerate}
\item
a global completion of massive type IIA supergravity, 
\item
compatible with M/IIA duality in the massless sector (and beyond, cf. \cref{AsD6CobordismTheory}),
\item
quantizing all brane charges/fluxes and higher global symmetries in the massive IIA Borcherds hierarchy (\cref{OverviewDiagram,OverviewDiagramMassive}),

\item
in the nonabelian cohomology theory \emph{massive minimal cyclic 4-Cohomotopy} $H^0(\text{-};\mathcal{A})$, 

whose classifying space \cref{MassMinCyc} is
\begin{equation}
\label{TheClassifyingSpaceAsAttachingSpace}
  \MassMinCyc S^4
  \simeq
  \coprod_m
  \bracket({
    \AttachingSpace{m}
  })
  \mathrlap{\,,}
\end{equation}

\item
which is a nonabelian refinement of twisted K-theory, as such exhibited by a comparison cohomology operation $\Psi_\ast$ \cref{MassiveComparisonMap}
\begin{equation}
  \begin{tikzcd}
    \substack{
      \text{massive minimal}
      \\
      \text{cyclic 4-Cohomotopy}
    }
    \;\;
    H^0\bracket({
      \text{-};
      \MassMinCyc S^4
    })
    \ar[
      r,
      "{ \Psi_\ast }"
    ]
    &
    H^0\bracket({
      \text{-};
      \bracket({
        B\mathrm{U}
          \!\times\! 
        \mathbb{Z}
      })
      \sslash
      B \mathrm{U}(1)
    })    
    \;\;
    \substack{
      \text{twisted}
      \\
      \text{K-theory}
    }
    \mathrlap{\,,}
  \end{tikzcd}
\end{equation}
which reproduces the brane charges expected in K-theory (\cref{TheMassiveComparisonMap,OriginalMotivationForHypothesisK}).
\end{enumerate}

To put this into context, we recall what this means for the global completion of 10D IIA supergravity and highlight why it is relevant:

\paragraph
{Global Completion of Higher Gauge Theories}

It is familiar since \cite{Dirac1931} that an electromagnetic field configuration is a gauge potential 1-form $A_1$ \emph{only locally}, namely on a contractible chart of spacetime. The global definition of the electromagnetic field, sensitive to the global topology of spacetime, requires further choices, involves more data, and quantizes (discretizes) the magnetic and possibly also the electric charges. 
A popular choice is the original \emph{Dirac charge quantization}, which says that the magnetic charges are in integral 2-cohomology \cref{Ordinary2Cohomology} and electromagnetic field configurations are globally cocycles in \emph{differential} integral 2-cohomology, as such equivalently modeled for instance as $\mathrm{U}(1)$-principal bundles with connection.

\begin{table}[htbp]
\centering
\adjustbox{
  rndfbox=4pt
}{
$
  \def\arraystretch{2}
  \begin{array}{ccccc}
    \substack{
      \textbf{higher}
      \\
      \textbf{gauge}
      \\
      \textbf{theory\rule[-5pt]{0pt}{0pt}}
    }
    &
    \substack{
      \textbf{local}
      \\
      \textbf{higher gauge}
      \\
      \textbf{potential forms\rule[-5pt]{0pt}{0pt}}
    }
    &
    \substack{
      \textbf{global} 
      \\
      \textbf{completion in}
      \\
      \textbf{differential...\rule[-5pt]{0pt}{0pt}}
    }
    &
    \substack{
      \textbf{classifying space}
      \\
      \mathcal{A}
    }
    &
    \substack{
      \textbf{discussed}
      \\
      \textbf{in}
    }
    \\
    \hline\hline
    \substack{
      \text{Maxwell}
      \\
      \text{magnetic}
      \\
      \text{field}
    }
    &
    \substack{
      A_1^{\mathrm{mg}}
    }
    &
    \substack{
      \text{integral}
      \\    
      \text{2-cohomology}
    }
    &
    \substack{
      B \mathrm{U}(1)
      \simeq
      B^2 \mathbb{Z}
    }
    &
    \substack{
      \text{\parencites{Alvarez1985}[\S 7.1]{Brylinski1993}}
      \\
      \text{\parencites[\S 2]{Freed2002}}
    }
    \\
    \hline
    \substack{
      \text{Maxwell}
      \\
      \text{electromagnetic}
      \\
      \text{field}
    }
    &
    \substack{
      A_1^{\mathrm{mg}},
      A_1^{\mathrm{el}}
    }
    &
    \substack{
      \mathbb{Z}^2
      \\
      \text{2-cohomology}
    }
    &
    \substack{
      B \mathrm{U}(1)^2
      \simeq
      B^2 \mathbb{Z}^2
    }
    &
    \substack{
      \text{\parencites[(1.26)]{FreedMooreSegal2007}[(3.27)]{FreedMooreSegal2007b}}
      \\
      \text{\parencites[Rem. 2.3]{BeckerBeniniSchenkelSzabo2017}{LazaroiuShahbazi2018}}
    }
    \\
    \hline
    \substack{
      \text{magnetic}
      \\
      \text{Kalb--Ramond}
      \\
      \text{B-field}
    }
    &
    \substack{
      B_2^{\mathrm{mg}}
    }
    &
    \substack{
      \text{integral}
      \\
      \text{3-cohomology}
    }
    &
    \substack{
      B^2 \mathrm{U}(1)
      \simeq
      B^3 \mathbb{Z}
    }
    &
    \substack{
      \text{\parencites{Gawedzki1988}{FreedWitten1999}}
      \\
      \text{\parencites{CareyJohnsonMurray2004}{BonoraFerrariSavelli2008}}
    }
    \\
    \hline
    \substack{
      \text{massive}
      \\
      \text{IIA RR-fields}
      \\
      \text{\& magn. B-field}
    }
    &
    \substack{
      \massive{A_{-1}^{\mathrm{mg}}},
      \mathcolor{darkgreen}{A_1^{\mathrm{mg}}}, 
      \mathcolor{darkgreen}{A_3^{\mathrm{mg}}}, 
      \mathcolor{darkgreen}{A_5^{\mathrm{el}}}, 
      \mathcolor{darkblue}{A_7^{\mathrm{el}}},
      \mathcolor{darkblue}{A_9^{\mathrm{el}}}
      \\
      \mathcolor{darkgreen}{B_2^{\mathrm{mg}}}
    }
    &
    \substack{
      \text{degree-0}
      \\
      \text{twisted K-theory}
    }
    &
    \scalebox{.66}{$
      \big(
        B \mathrm{U}
        \times
        \mathbb{Z}
      \big) 
      \sslash B \mathrm{U}(1)
    $}
    &
    \substack{
      \text{\parencites{MinasianMoore1997}{Witten1998}...}
      \\
      \text{(for more cf. p. \pageref{HypothesisK})}
    }
    \\
    \hline
    \substack{
      \text{electromagnetic}
      \\
      \text{C-field}
    }
    &
    \substack{
      C_3^{\mathrm{mg}},
      C_6^{\mathrm{el}}
    }
    &
    \substack{
      \text{4-Cohomotopy}
    }
    &
    \substack{
      S^4
    }
    &
    \substack{
      \text{\parencites[\S 2.5]{Sati2018}{FSS20-H}{FSS21-Hopf}}...
      \\
      \text{cf. \parencites[\S 4]{GSS26-SuGra}}
    }
    \\
    \hline
    \substack{
      \text{$S^1$-reduction of}
      \\
      \text{electromagnetic}
      \\
      \text{C-field}
    }
    &
    \substack{
      \mathcolor{darkgreen}{A_1^{\mathrm{mg}}},
      \mathcolor{darkgreen}{A_3^{\mathrm{mg}}},
      \mathcolor{darkgreen}{A_5^{\mathrm{el}}}
      \\
      \mathcolor{darkgreen}{B_2^{\mathrm{mg}}},
      \mathcolor{darkorange}{B_6^{\mathrm{el}}}
    }
    &
    \substack{
      \text{cyclic}
      \\
      \text{4-Cohomotopy}
    }
    &
    \substack{
      \Cyc S^4
    }
    &
    \substack{
      \text{\parencites{GiotopoulosSati2026}[\S 3.2]{GSS26-SuGra}}
    }
    \\
    \hline
    \substack{
      \text{IIA RR-fields}
      \\
      \text{\& EM B-field}
    }
    &
    \substack{
      \mathcolor{darkgreen}{A_1^{\mathrm{mg}}}, 
      \mathcolor{darkgreen}{A_3^{\mathrm{mg}}}, 
      \mathcolor{darkgreen}{A_5^{\mathrm{el}}}, 
      \mathcolor{darkblue}{A_7^{\mathrm{el}}}
      \\
      \mathcolor{darkgreen}{B_2^{\mathrm{mg}}},
      \mathcolor{darkorange}{B_6^{\mathrm{el}}}
    }
    &
    \substack{
      \text{small cyclic}
      \\
      \text{4-Cohomotopy}
    }
    &
    \substack{
      \SmlCyc S^4
    }
    &
    \substack{
      \text{\cite{GSS26-KfromM}}
    }
    \\
    \hline
    \substack{
      \text{IIA Borcherds}
      \\
      \text{hierarchy fields}
    }
    &
    \substack{
      \mathcolor{darkgreen}{A_1^{\mathrm{mg}}}, 
      \mathcolor{darkgreen}{A_3^{\mathrm{mg}}}, 
      \mathcolor{darkgreen}{A_5^{\mathrm{el}}}, 
      \mathcolor{darkblue}{A_7^{\mathrm{el}}}
      \\
      \mathcolor{darkgreen}{B_2^{\mathrm{mg}}},
      \mathcolor{darkorange}{B_6^{\mathrm{el}}}
      \\
      \mathcolor{magenta}{D_8}, \cdots
    }
    &
    \substack{
      \text{minimal cyclic}
      \\
      \text{4-Cohomotopy}
    }
    &
    \substack{
      \MinCyc S^4
    }
    &
    \substack{
      \text{\cref{ObtainingBorcherdsIIA}}
    }
    \\
    \hline
    \substack{
      \text{massive}
      \\
      \text{IIA Borcherds}
      \\
      \text{hierarchy fields}
    }
    &
    \substack{
      \massive{A_{-1}^{\mathrm{mg}}},
      \mathcolor{darkgreen}{A_1^{\mathrm{mg}}}, 
      \mathcolor{darkgreen}{A_3^{\mathrm{mg}}}, 
      \mathcolor{darkgreen}{A_5^{\mathrm{el}}}, 
      \mathcolor{darkblue}{A_7^{\mathrm{el}}},
      \mathcolor{darkblue}{A_9^{\mathrm{el}}}
      \\
      \mathcolor{darkgreen}{B_2^{\mathrm{mg}}},
      \mathcolor{darkorange}{B_6^{\mathrm{el}}}
      \\
      \mathcolor{magenta}{D_8}, \cdots
    }
    &
    \substack{
      \text{massive}
      \\
      \text{minimal cyclic}
      \\
      \text{4-Cohomotopy}
    }
    &
    \substack{
      \coprod_m
      \MinCyc_m S^4
    }
    &
    \substack{
      \text{\cref{MassiveBorcherds,Conclusion}}
    }
  \end{array}
$
}
\caption{\label{HigherGaugeFields}
  Higher gauge fields are only locally, on a single chart of spacetime, expressed by higher gauge potential forms (for which cf. \parencites[\S 4.1]{GSS26-SuGra}{Giotopoulos26}). Globally completed gauge field configurations are instead cocycles in differential cohomology theories (cf. \parencites{Freed2002}{Szabo2013}{MooreSaxena2025}{SS26-HigherGauge}). These in turn are encoded by classifying spaces $\mathcal{A}$ (cf. \parencites{SS24-Phase}[\S 3.2]{SS25-Flux} following \cite[\S 9]{FSS23-Char}): The torsion-free homotopy groups of $\mathcal{A}$ encode the local gauge potentials, the Whitehead-bracket $\mathfrak{l}\mathcal{A}$ on these encodes the Bianchi identities on the flux densities, homotopy classes of maps to $\mathcal{A}$ encode the topological brane charges, and the higher homotopies between these maps encode the higher global symmetries of the globally completed higher gauge theory. 
}
\end{table}

In more abstract language that generalizes (cf. \cref{HigherGaugeFields}): 
\begin{enumerate}
\item
$\mathrm{U}(1)$-principal bundles with connection, and hence globally completed electromagnetic field configurations, are equivalently cocycles in \emph{differential integral 2-cohomology},
\item
$\mathrm{U}(1)$-bundle gerbes with connections, the most popular global completion of the Kalb--Ramond B-field, are equivalently cocycles in \emph{differential integral 3-cohomology},

\item 
\label{RRFieldInDifferentialKTheory}
cocycles in \emph{differential twisted K-theory} are the globally completed RR- and magnetic B-fields according to the popular Hypothesis K, cf. \cref{KTheoryBraneCharges,BraneChargesInTwistedKTheory}.
\end{enumerate}

The choice of global completion of the type IIA field in \cref{RRFieldInDifferentialKTheory} has received much attention, but it fails to describe some of the fluxes/charges of IIA theory (cf. \cref{OverviewDiagram}): Notably it misses the $H_7$-flux and hence the NS1-brane charges sourcing it. Beyond that it also misses the dual dilaton flux $H_9$ \cref{TheDilatonFlux} and the corresponding charge (cf. \cref{IntegralLatticeForDualDilaton}) that is the first higher stage in the IIA Borcherds hierarchy.

This shortcoming is not a coincidence: A no-go theorem \cite[Prop. 2.2]{GSS26-KfromM} shows that:
\begin{standout}
  No generalized abelian (stable) cohomology theory can quantize charges whose fluxes satisfy non-linear Bianchi identities.
\end{standout}

Instead, global completion of general higher gauge fields (of ``higher Maxwell type'', as appearing in 11D SuGra and its descendants) has to take place in \emph{differential nonabelian cohomology} theories \parencites[\S 9]{FSS23-Char}{SS26-HigherGauge}, whose classifying spaces $\mathcal{A}$ are not necessarily (bundles of) infinite-loop spaces.

The first globalization of the full field content of type IIA was achieved in \cite{GSS26-KfromM}, via the nonabelian cohomology theory classified by the \emph{small cyclification of $S^4$} (cf. \cref{CircleMaps}), shown there to be a nonabelian (unstable) deformation of twisted K-theory.

Here we asked for a further enhancement of this result that would capture also the rest of the IIA Borcherds hierarchy (\cref{OverviewDiagram}) as well as its massive deformation (\cref{OverviewDiagramMassive}), while inheriting the comparison map to twisted K-theory. Remarkably, the relevant nonabelian cohomology theory---massive minimal cyclic 4-Cohomotopy \cref{TheClassifyingSpaceAsAttachingSpace}---not only exists, but reflects expected fine-print of massive IIA theory even beyond its design criteria.  This is what we turn to in \cref{AsD6CobordismTheory} below.

\subsection
{As a D6-Brane/KK-Monopole Cobordism Theory}
\label{AsD6CobordismTheory}

\paragraph
{A 6-Brane Cobordism Cohomology Theory}
Interestingly, by the specific identification \cref{MinCycNCellAttachment} of our classifying space \cref{TheClassifyingSpaceAsAttachingSpace}, 
the evident variant of the classical (unstable) \emph{Pontryagin construction}
(\cite{Pontrjagin1959}, cf. \parencites[\S IX]{Kosinski1993}[\S IX]{Bredon1993}, for exposition in our context see \parencites[\S 2.1]{SS20-Tad}[\S 2.2 \& \S 3.2]{SS23-Mf}) identifies
our cohomology theory of massive Borcherds IIA charges with a form of unstable \emph{cobordism cohomology} for 6-brane worldvolumes coupled to the M-circle bundle, as follows.

Namely, over a connected space $X^9$ (just for notational simplicity now), the classifying map \cref{MassMinCyc} picks an integer component $m \in \mathbb{Z}$, factoring as
\begin{equation}
\label{FactoredClassifyingMap}
\begin{tikzcd}
  X^9 
    \ar[r, "{ c }"] 
  &
  \MinCyc_m S^4
    \simeq
  \AttachingSpace{m}
  \mathrlap{\,,}
\end{tikzcd}
\end{equation}
thus encoding the following data:
\begin{enumerate}
\item a codimension-3 brane locus
\begin{equation}
\label{The6Brane}
  Q^6 
  :=
  c^{-1}\bracket({ \{0\} \subset D^3 })
  \subset 
  X^9
  \mathrlap{\,,}
\end{equation}

\item with a framed tubular neighborhood
\begin{equation}
  U^9
  :=
  c^{-1}\bracket({ D^3 })
  \simeq
  Q^6 \times D^3
  \subset 
  X^9
  \mathrlap{\,,}
\end{equation}
and linking sphere bundle
\begin{equation}
\label{TheLinkingSphere}
  Q^6 \times S^2_L
  :=
  \partial U^{9}
  =
  c^{-1}\bracket({ S^2 })
  \subset X^9
  \mathrlap{\,,}
\end{equation}

\item 
away from that, over the closed complement space
\begin{equation}
\label{TheComplementSpace}
  X^9 \setminus \mathrm{int}(U^9)
  \;\subset\;
  X^9
  \mathrlap{\,,}
\end{equation}
away from the brane locus $Q^6$, an M-circle bundle $X^{10}$,
whose restriction to the linking spheres
\cref{TheLinkingSphere}
around the 6-brane $Q^6$ \cref{The6Brane} has Chern class $m$:
\begin{equation}
\label{ThePartialMCircleBundle}
  \begin{tikzcd}
     S^1 
       \ar[
         r,
         hook
       ]
     &[-10pt]
     X^{10}_{\vert \{q\} \times S^2_L}
     \ar[
       r,
       hook
     ]
     \ar[
       d,
       ->>
     ]
     \ar[
       dr,
       phantom,
       "{ \lrcorner }"{pos=.1}
     ]
     &[-10pt]
     X^{10}
     \ar[
       d,
       ->>
      ]
     \ar[r]
     \ar[
       dr,
       phantom,
       "{ \lrcorner }"{pos=.1}
     ]
     &[+13pt]
     E S^1
     \ar[
       d,
       ->>
      ]
     \\
     &
     \{q\} \times S^2_L
     \ar[r, hook]
     \ar[
       rr,
       downhorup,
       "{ m }"{description}
     ]
     & 
     \bracket({
       X^9 \!\setminus\! \mathrm{int}(U^9)
     })
     \ar[
       r,
       "{
         c\vert_{X^9 \setminus \mathrm{int}(U^9)}
       }"
     ]
     &
     B S^1
     \mathrlap{\,.}
  \end{tikzcd}
\end{equation}
\end{enumerate}
Finally, homotopy classes of classifying maps \cref{FactoredClassifyingMap} correspond to suitable cobordism classes of the above data, again in direct variation of the Pontryagin theorem.

By this correspondence we have that:
\begin{standout}
  This 6-brane cobordism cohomology theory is equivalently the home of the brane charges and higher global symmetries of the global completion of massive Borcherds IIA theory from \cref{GloballyCompletedMassiveIIA}. 
\end{standout}

\paragraph
{Relation to D6-Branes \& 11D KK-Monopoles}

While we did not ask for it, the above data \cref{ThePartialMCircleBundle} reproduces just the geometry of \emph{D6-branes lifted to KK-monopoles} (cf. \parencites[p. 6]{Townsend1995-revisited}):
\begin{enumerate}
  \item 
  the bottom row of \cref{ThePartialMCircleBundle} encodes exactly the charge $m$ of a black/singular D6-brane in IIA seen on its linking sphere,

  \item
  the top row of \cref{ThePartialMCircleBundle} encodes exactly its lift to the \emph{Kaluza--Klein monopole} of 11D SuGra as a principal circle bundle away from the brane locus $Q^6$,

  \item while this brane locus $Q^6$ \cref{The6Brane} \emph{remains included} in spacetime, being a submanifold of a smooth 11D spacetime.
\end{enumerate}

\begin{SCfigure}[.57][htbp]
\caption{%
\label{TaubNUTRadialSchematicMono}
  (following \cite[Fig. 5]{AcharyaGukov04})
  The \emph{Kaluza--Klein monopole} is famously a smooth (non-singular) solution of 11D supergravity. Away from a 6-brane locus $Q^6$ \cref{The6Brane} it is topologically a principal bundle of M-circles, whose Chern class $m \in \mathbb{Z}$ over the linking sphere $\partial U^9$ \cref{TheLinkingSphere} is the charge of the corresponding D6-brane. 
  Right at the brane locus $Q^6$ the M-circles collapse and $X^{10}$ fails to be a principal fiber bundle over $X^9$.
}
\includegraphics[width=.6\textwidth]
  {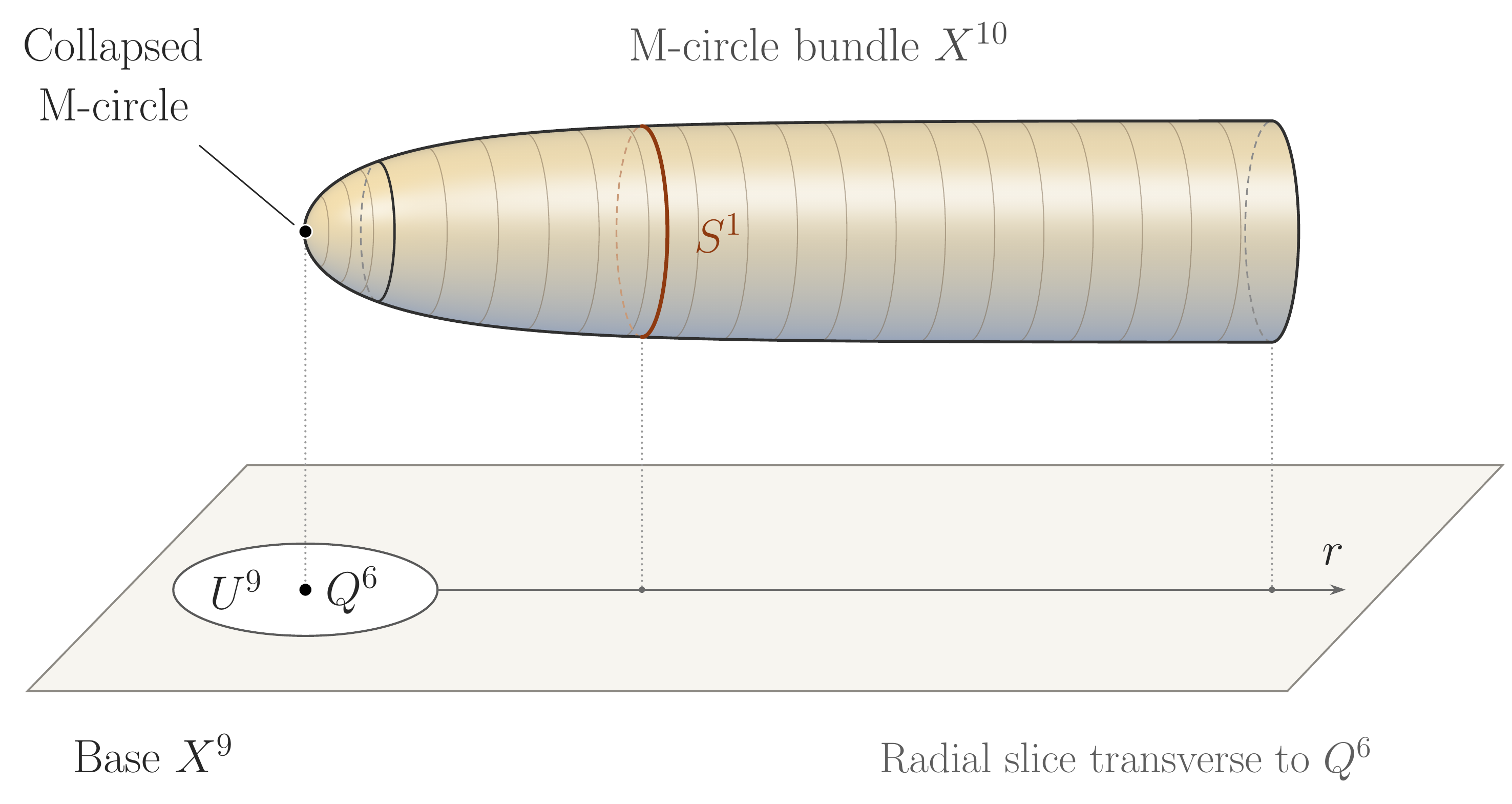}
\end{SCfigure}

Indeed, the 11D KK-monopole spacetime is well-known to be smooth (non-singular) (cf. \parencites[p. 364]{EguchiGilkeyHanson1980}[p. 6]{Townsend1995-revisited}), with the M-circle collapsing over the 6-brane/monopole singularity (cf. \parencites[\S 2]{Sen1997}[\S 9]{ItzhakiMaldacenaSonnenscheinYankielowicz1998}), as indicated in \cref{TaubNUTRadialSchematicMono}.

\begin{remark}[Relation to D-brane charge in K-theory]
Thereby the classifying space \cref{TheClassifyingSpaceAsAttachingSpace} automatically resolves the issue of how to measure D6-brane charge on linking spheres without deleting the brane locus from spacetime, which in the K-theory classification
\cref{KTheoryBraneCharges} with its classifying space
$B S^1 \subset B \mathrm{U}$ requires some conceptual contortions
(\parencites
[(1.6)]{MaldacenaMooreSeiberg2001}
[(2.2)]{BergmanGimonKol2001}
[(2.10)]{MooreWitten2000}, cf. discussion in \cite[\S A.1]{GSS26-KfromM}). 

Here the cell attachment \cref{MinCycNCellAttachment} from 
$B S^1$ to $\AttachingSpace{m}$ is exactly what reflects the presence of the D6-brane worldvolume $Q^6$ \cref{The6Brane} inside its linking sphere \cref{TheLinkingSphere}.
\end{remark}

\paragraph
{Outlook}

Thereby we obtain the remarkable statement that 
\begin{standout}
The global completion (\cref{GloballyCompletedMassiveIIA}) of massive IIA theory compatible with cohomotopically completed 11D supergravity, despite or by virtue of its infinite Borcherds hierarchy of flux/brane/symmetry species (\cref{OverviewDiagram,OverviewDiagramMassive}), is fundamentally a theory (just) of D6-branes/KK-monopoles subject to cobordism.
\end{standout}
This is clearly another incarnation of how D6-brane charge emerged above (\cref{D6BraneChargeAsRoot}) as the root of the cohomotopically quantized IIA Borcherds hierarchy, in the proof of \cref{TheMassiveComparisonMap}. 

Further discussion of this theory and the following further aspects is left for elsewhere:
\begin{enumerate}

\item 
\textbf{Tangential Couplings.}
Not taken into account above is coupling of charges to spacetime tangential structures, such as in the half-integral shift of $G_4$ (and correspondingly in the Freed-Witten condition on $H_3$, cf. \cite[(5.10)]{Witten2000}). In terms of flux/charge-quantization these correspond to \emph{tangential twistings} of the coefficient cohomology theory (as in \parencites[Prop. 3.13]{FSS20-H}[\S 12]{FSS23-Char}), reflecting corrections to the $G_7$-Bianchi identity \cref{CFieldBianchiIdentities} in curvature-corrected 11D supergravity (cf. \cite{CederwallGranNilssonTsimpis2005}). We will discuss this elsewhere.%
\footnote{
  Cf. \href{http://ncatlab.org/schreiber/show/The+I8+Project}{\texttt{ncatlab.org/schreiber/show/The+I8+Project}}.
}

\item
\textbf{9D Borcherds Hierarchy.}
  The construction presented here is clearly just the beginning of a tower of topological enhancements of hidden exceptional structures in $n$-torus-reductions of 11D supergravity (refining the unrestricted rational analysis in \parencites{SatiVoronov2024}).

  In the next step, for classifying charge quantization of the reduction of 11D SuGra on a 2-torus $T^2$, one finds a minimal subspace of the \emph{2-toroidification} of the 4-sphere,
  \begin{equation}
  \label{MinimalToroidification}
    \begin{tikzcd}
      \MinTor S^4
      \ar[r]
      \ar[
        d,
        phantom,
        "{:=}"{rotate=-90}
      ]
      &
      \operatorname{Tor} S^4
      \ar[
        d,
        phantom,
        "{:=}"{rotate=-90}
      ]
      \\
      B T^2
        \vee
      \Omega S^3
      \ar[
        r,
        hook,
        "{ \iota^{(2)}_{\mathrm{min}} }"
      ]
      &
      \mathrm{Map}\bracket({
        T^2, S^4
      })\sslash T^2
      \mathrlap{\,,}
    \end{tikzcd}
  \end{equation}
  whose Whitehead bracket $L_\infty$-algebra is the positive 9D Borcherds algebra (cf. \cite[\S 4.2]{KleinschmidtPalmkvist2013}):
  \begin{equation}
  \label
  {WhiteheadAlgebraOfMinTorS4}
    \mathfrak{l}\bracket({
      \MinTor S^4
    })
    \simeq
    \mathfrak{bor}^+_{\mathrm{9D}}
    \mathrlap{\,.}
  \end{equation}

  This subspace \cref{MinimalToroidification} is homotopy invariant under the shear subgroup of the modular mapping class group of the torus,
  \begin{equation}
    \begin{tikzcd}
      \bracketmid\{{
        \begin{pmatrix}
          1 & k
          \\
          0 & 1
        \end{pmatrix}
      }{
        k \in \mathbb{Z}
      }\}
      =:
      \mathbb{Z}_{\mathrm{shr}} 
        \ar[r, hook] 
        &
      \mathrm{SL}(2,\mathbb{Z})
      \mathrlap{\,,}
    \end{tikzcd}
  \end{equation}
  and its corresponding homotopy quotient maps to the cyclification of the minimal cyclification of the 4-sphere, as shown on the right here:
  \begin{equation}
    \begin{tikzcd}[
      column sep=40pt
    ]
      &
      \bracket({
        \MinTor S^4
      }) 
        \sslash 
      \mathbb{Z}_{\mathrm{shr}}
      \ar[
        r,
        "{
          \scaledbracket({
            \iota^{(2)}_{\mathrm{min}}
          })
          \sslash
          \mathbb{Z}_{\mathrm{shr}}
        }"
      ]
      \ar[
        d
      ]
      &
      \bracket({
        \operatorname{Tor} S^4
      }) 
        \sslash 
      \mathbb{Z}_{\mathrm{shr}}
      \ar[
        d,
        "{ \sim }"{sloped}
      ]
      \\
      X^{8}
      \ar[
        r
      ]
      \ar[
        ur, 
        dashed,
        end anchor=south west
      ]
      &
      \Cyc\MinCyc S^4
      \ar[
        r, 
        hook,
        "{
          \Cyc (\iota_{\mathrm{min}})
        }"
      ]
      &
      \Cyc\Cyc S^4
      \mathrlap{\,.}
    \end{tikzcd}
  \end{equation}
  Homotopy classes of dashed maps in the above diagram hence constitute cohomology classes quantizing all charges in the Borcherds hierarchy of 9D SuGra, compatible with dimensional reduction of the global completion of IIA Borcherds SuGra established above in \cref{Conclusion}. (Here $B \mathbb{Z}_{\mathrm{shr}}$ classifies the charges of wrapped D7-branes seen in 9D, cf. \parencites[Fig. 5]{GSS26-KfromM}).
  In fact, the shear action on $\MinTor S^4$ extends to an $\mathrm{SL}(2,\mathbb{Z})$-action in a way that implements S-duality on the 9D Borcherds hierarchy of flux species \cref{WhiteheadAlgebraOfMinTorS4}. 
  
  We relegate details and proofs of these statements to \cite{GSS26-9D}. 
\end{enumerate}

\appendix

\section
{Background}

For reference in the main text, we briefly review some mathematical background material.

\subsection{Topological Homotopy Theory}
\label{HomotopyTheory}

We just need some basics, cf. \parencites{Schwarz1994}{Eschrig2011}{FomenkoFuchs2016}.

As usual in topological homotopy theory, by a \emph{space} we mean a \emph{topological space} and by that we really mean an object in a \emph{convenient} category $\Top$ of topological spaces (cf. \parencites[\S 0]{FHT2000}).
Similarly, by a \emph{map} between spaces we mean a \emph{continuous map}, a morphism $\inlinetikzcd{ X \ar[r] \& Y }$ in $\Top$.  We display dashed arrows $\inlinetikzcd{ X \ar[r, dashed] \& Y }$ to highlight unspecified such maps.

\paragraph
{Mapping Spaces and Nonabelian Cohomology}

For spaces $X,Y \in \Top$, the set of continuous maps between them,
\begin{equation}
\label{MappingSpace}
  \mathrm{Map}\bracket({
    X, Y
  })
  \defneq
  \big\{
  \inlinetikzcd{
    X \ar[r, dashed]
    \&
    Y
  }
  \big\}
  \mathrlap{\,,}
\end{equation}
carries a topology such that maps into this \emph{mapping space} are in natural bijection with maps out of the product space of the two domains:
\begin{equation}
  \big\{
  \inlinetikzcd{
    Z
    \ar[r, dashed]
    \&
    \mathrm{Map}(X, Y)
  }
  \big\}
  \;\simeq\;
  \big\{
  \inlinetikzcd{
    Z \times X
    \ar[r, dashed]
    \&
    Y
  }
  \big\}
  \mathrlap{\,.}
\end{equation}
A path in $\mathrm{Map}(X,Y)$ is called a \emph{homotopy} 
and the path-connected components of the mapping space are the \emph{homotopy classes} of maps:
\begin{equation}
\label{HomotopyClasses}
  \pi_0\, \mathrm{Map}(X,Y)
  \defneq
  \big\{
  \inlinetikzcd{
    Z \times X
    \ar[r, dashed]
    \&
    Y
  }
  \big\}
  \big/
  \left\{
  \begin{tikzcd}
    X
    \ar[
     r,
      bend left=30,
      "{\ }"{name=s,swap}
    ]
    \ar[
      r,
      bend right=30,
      "{\ }"{name=t}
    ]
    \ar[
      from=s,
      to=t,
      dashed,
      shorten=-3pt,
      Rightarrow
    ]
    &
    Y
  \end{tikzcd}
  \right\}
  \mathrlap{\,.}
\end{equation}

For fixed codomain $\mathcal{A} \in \Top$ this construction is a contravariant functor
\begin{equation}
\label{NonabelianCohomologyFunctor}
  \inlinetikzcd{
  H^0(\text{-}; \mathcal{A})
  :=
  \pi_0 \mathrm{Map}\bracket({
    \text{-},
    \mathcal{A}
  })
  :
  \Top^{\mathrm{op}}
  \ar[r]
  \&
  \mathrm{Set}
  \mathrlap{\,.}
  }
\end{equation}
This is the \emph{nonabelian cohomology theory} (\parencites[Def. 6.0.6]{Toen2002}[Def. 2.3]{Schreiber2009OWR}[Def. 6]{Lurie2014}[\S~2]{FSS23-Char}[\S~1]{SS25-TEC}; under suitable conditions the \emph{nonabelian Poincar{\'e} dual} of \emph{factorization homology} \parencites{Lurie2014}[\S~4]{AyalaFrancis2015})
with coefficients in the \emph{classifying space} $\mathcal{A}$.

Notice that \emph{cohomology operations} $\phi_\ast$ between nonabelian cohomology theories \cref{NonabelianCohomologyFunctor}, namely the natural transformations between them, are equivalently homotopy classes of maps $\phi$ between their classifying spaces (by the Yoneda lemma):
\begin{equation}
  \begin{tikzcd}[
    row sep=0pt
  ]
    \mathcal{A}
    \ar[
      r,
      "{ \phi }"
    ]
    &
    \mathcal{A}'
    \\
    H^0\bracket({
      \text{-};
      \mathcal{A}
    })
    \ar[
      r,
      "{ \phi_\ast }"
    ]
    &
    H^0\bracket({
      \text{-};
      \mathcal{A}'
    })
    \mathrlap{\,.}
  \end{tikzcd}
\end{equation}

\paragraph
{Principal Bundles}

A \emph{topological group} $\mathcal{G} \in \mathrm{Grp}(\Top)$ is a topological space equipped with continuous group structure. 

A \emph{$\mathcal{G}$-principal bundle} is a map $\inlinetikzcd{P \ar[r] \& X}$ which over charts $\bracket\{{ U_i \subset X }\}_{i \in I}$ of an open cover may be identified with $U_i \times \mathcal{G}$ such that the transition maps 
$g_{ij} \acts \bracket({(U_i \cap U_j) \times \mathcal{G}})$ are given by fiberwise group multiplication.

The \emph{delooping} of a topological group is a space 
\begin{equation}
  \label{TopologicalDelooping}
  B \mathcal{G}
  \in
  \Top
\end{equation}
which classifies \cref{NonabelianCohomologyFunctor} isomorphism classes of such $\mathcal{G}$-principal bundles (cf. \parencites[\S 3]{RudolphSchmidt2017}[\S 5.2.1]{SS26-Bun}):
\begin{equation}
\label{NonabelianCohomologyInDegree1}
  H^1\bracket({
    \text{-};
    \mathcal{G}
  })
  :=
  H^0\bracket({
    \text{-};
    B \mathcal{G}
  })
  \simeq
  \mathcal{G}\mathrm{PrncBndl}(\text{-})_{/\sim}
  \mathrlap{\,.}
\end{equation}

\paragraph
{Homotopy Types}

A \emph{pointed space} is a space $X \in \Top$ equipped with a choice of element $x_0 \in X$, the \emph{basepoint}. We write $\Top^\ast \subset \Top$ for the full subcategory of pointed spaces with basepoint-preserving maps between them.
Accordingly, for $X, Y \in \Top^\ast$, we write
\begin{equation}
\label{PointedMappingSpace}
  \mathrm{Map}^\ast(X,Y)
  \subset
  \mathrm{Map}(X,Y)
\end{equation}
for the subspace of the mapping space \cref{MappingSpace} on the basepoint-preserving maps.

For example, with the circle $S^1$ regarded with any choice of basepoint,
\begin{equation}
  \Omega X
  :=
  \mathrm{Map}^\ast\bracket({
    S^1, 
    X
  })
\end{equation}
is the \emph{based loop space} of $X$. Concatenation of such loops induces a group structure on homotopy classes \cref{HomotopyClasses}, called the \emph{fundamental group} of $X \in \Top^\ast$
\begin{equation}
  \pi_1(X)
  :=
  \pi_0 \bracket({
    \Omega X
  })
  \defneq
  \pi_0\, \mathrm{Map}^\ast\bracket({
    S^1, X
  })
  \mathrlap{\,.}
\end{equation}
Similarly, concatenation of images of $n$-spheres $S^n \simeq [0,1]^n/ \partial [0,1]^n$ gives the \emph{higher homotopy groups} of $X \in \Top^\ast$
\begin{equation}
  \pi_n(X)
  :=
  \pi_0\, 
  \mathrm{Map}^\ast\bracket({
    S^n, X
  })
  \mathrlap{\,,}
\end{equation}
which are all abelian for $n \geq 2$.

For $X, Y \in \Top$, a map between them is a \emph{weak homotopy equivalence},  or just \emph{equivalence}, for short, denoted
\begin{equation}
  \inlinetikzcd{
    X \ar[r, "{ \sim }"]
    \&
    Y
    \mathrlap{\,,}
  }
\end{equation}
iff it induces isomorphisms on all homotopy sets/groups $\pi_n$, $n \in \mathbb{N}$, for all basepoints.

The corresponding equivalence classes are the \emph{homotopy types} of spaces. These are the actual objects of interest of homotopy theory, where topological spaces serve just as representatives.

For example, every topological group is equivalent to the loop space of its delooping \cref{NonabelianCohomologyInDegree1}, whence the terminology:
\begin{equation}
  \mathcal{G}
  \simeq
  \Omega B \mathcal{G}
  \mathrlap{\,.}
\end{equation}
Moreover, every connected space is a delooping \cref{NonabelianCohomologyInDegree1} up to equivalence, hence for every $\mathcal{A} \in \Top$ we have
\begin{equation}
  \mathcal{A}
  \;\simeq\;
  \coprod
    _{\mathclap{s \in \pi_0 \mathcal{A}}}
  B \mathcal{G}_s
  \,,
  \;\;\;\;\;\;\;
  H^0\bracket({
    \text{-}; \mathcal{A}
  })
    \simeq
  \coprod
   _{\mathclap{s \in \pi_0 \mathcal{A}}}
  H^1\bracket({
    \text{-};
    \mathcal{G}_s
  })
  \mathrlap{\,.}
\end{equation}

\paragraph
{Homotopy Quotients and Borel
Construction}

For $G$ a topological group (such as the circle group $S^1 = \mathrm{U}(1)$), there is a contractible space 
$E G$ (all $\pi_\bullet(E G)$ are trivial) carrying a \emph{free} continuous $G$-action. 
For $X \in \Top$ equipped with a continuous $G$-action, its
\emph{homotopy quotient},
or \emph{Borel construction}, is
\begin{equation}
\label{BorelConstruction}
  X \sslash G
    :=
  \bracket({
    X \times E G
  })/G
  \mathrlap{\,,}
\end{equation}
where on the right $G$ acts diagonally.

For example, with $X = \ast$ the point space, its homotopy $G$-quotient is equivalent to the delooping classifying space 
 \cref{TopologicalDelooping} of $G$:
\begin{equation}
  \inlinetikzcd{
    \ast \sslash G
    \defneq
    \bracket({ E G })
    /
    G
    \ar[
      r,
      "{ \sim }"
    ]
    \&
    B G
    \mathrlap{\,.}
  }
\end{equation}
The quotient projection $\inlinetikzcd{ E G \ar[r, ->>] \& B G }$ is the \emph{universal} $G$-principal bundle \cref{NonabelianCohomologyInDegree1}.

\paragraph
{Cell attachments and Cofiber Coproducts}

Given a map $\inlinetikzcd{f : S^{n-1} \ar[r] \& X}$,
its \emph{attaching space} $D^{n} \coprod_{S^{n-1}}^f X$ is the result of gluing an $n$-disk $D^n$ to $X$ by identifying its boundary $\partial D^n \cong S^{n-1}$ with $f(\partial D^n)$.

In particular, this makes a commuting square of maps of the form
\begin{equation}
  \begin{tikzcd}
    S^{n-1}
    \ar[r, "{ f }"]
    \ar[
      d,
      hook
    ]
    \ar[
      dr,
      phantom,
      "{ \ulcorner }"{pos=.9}
    ]
    &
    X
    \ar[
      d,
      hook
    ]
    \\
    D^n
    \ar[r, hook]
    &
    D^n 
    \underset
      {\mathclap{S^{n-1}}}
      {\coprod}^f 
    X
    \mathrlap{\,,}
  \end{tikzcd}
\end{equation}
where the little hook indicates that this is in fact the \emph{universal} such square (given $f$), called the (homotopy) \emph{pushout} square characterizing a (homotopy) \emph{cofiber coproduct}. 

For example, $D^n \displaystyle{\coprod_{S^{n-1}}} \ast \cong S^n$.

Another kind of cofiber coproduct is the \emph{wedge sum} $X\vee Y$ of two pointed spaces, the result of gluing them at their basepoints:
\begin{equation}
  \begin{tikzcd}
    \ast
    \ar[r]
    \ar[d]
    \ar[
      dr,
      phantom,
      "{ \ulcorner }"{pos=.9}
    ]
    &
    Y
    \ar[
      d,
      hook
    ]
    \\
    X 
      \ar[
        r,
        hook
      ]
    &
    X \vee Y
    \mathrlap{\,.}
  \end{tikzcd}
\end{equation}

\paragraph
{One-Point Compactification}

For well-behaved spaces $X$ (``locally compact Hausdorff''), their
\emph{one-point compactification} $X_{\cpt}$ (cf. \cite[(17)]{SS23-Mf}) is obtained by adjoining a point $\infty$,  whose open neighborhoods are the complements of compact subsets of $X$.

For example:
\begin{enumerate}
\item
$\mathbb R^n_{\cpt}\cong S^n$ (cf. \parencites[(1)]{SS23-Mf}).

\item
If $X$ is already compact, then $X_{\cpt} \cong X_{\plus} := X \sqcup \{\infty\}$, hence with the point a infinity being disjoint. 

\item
The one-point compactification of a product space is the \emph{smash product} of the compactification of the factor spaces:
\begin{equation}
\label
{CompactificationOfProductSpace}
  \bracket({
    X \times Y
  })_{\cpt}
  \cong 
  X_{\cpt} \wedge Y_{\cpt}
  :=
  \frac
    { \rule[-5pt]{0pt}{0pt}
      X_{\cpt} \times Y_{\cpt} 
    }
    { \rule{0pt}{11pt}
      \bracket({
        \{\infty_X\} \times Y 
      })
        \cup  
      \bracket({
        X \times \{ \infty_Y \}
      })
    }
  \mathrlap{\,.}
\end{equation}
\end{enumerate}

\subsection{Homotopy Lie Algebra}
\label{app:HigherLieAlgebras}

We mostly just need basics (following \parencites{LadaStasheff1992})    , though the literature on the points relevant to us remains scattered. For review in our context see \parencites{FSS2019}[\S 4]{FSS23-Char}[\S 4]{Voronov2025}{SS25-Flux}{CastellaniDAuria2025}[\S 2.1]{GSS25-TD}. The discussion of polynomial disconnected Sullivan models in \cref{MinimalSullivanModel} \cref{MinimalSullivanModelOfDisconnected}) is new.

Our ground field is the real numbers $\mathbb R$.

\paragraph
{$L_\infty$-Algebras}

An \emph{$L_\infty$-algebra} structure on a $\mathbb{Z}$-graded vector space 
\begin{equation}
  \mathfrak{a} 
    = 
  \bigoplus_{k\in\mathbb Z} \mathfrak{a}_k
\end{equation}
has (in the unshifted homological convention of
\cite[Def.~2.1]{LadaMarkl1995}) graded-antisymmetric brackets $[\text{-}, \cdots, \text{-}]$ of positive arities $n \in \mathbb{N}_{\geq 1}$, of degree $n - 2$,
\begin{equation}
  \bracket|{
    \bracket[{
      x^{(1)}, \ldots, x^{(n)}
    }]
  }|
    = 
  n - 2
  +
  \textstyle{\sum_{i=1}^n}
    \bracket|{x^{(i)}}| 
  \mathrlap{\,,}
  \;\;\;
  x^{(i)} \in \mathfrak{a}
\end{equation}
and satisfying higher Jacobi identities 
\cite[(2)]{LadaMarkl1995}, cf. \cref{DualJacobiIdentity} below.

In particular:
\begin{enumerate}
\item
the unary bracket is a differential $\partial := [\text{-}]$ of degree  $\bracket\vert{[\text{-}]}\vert = -1$,

\item 
the binary bracket has degree $\bracket\vert{[\text{-},\text{-}]}\vert = 0$.
\end{enumerate}

We denote homogeneous degrees by subscripts,
\[
  x_k \in \mathfrak{a}_k
  \mathrlap{\,,}
\]
whence the graded-commutativity of the binary bracket of an $L_\infty$-algebra reads:
\begin{equation}
  [x_n, x'_{n'}]
  =
  -(-1)^{n n'} 
  [x'_{n'}, x_{n}]
  \mathrlap{\,.}
\end{equation}

\paragraph
{Chevalley--Eilenberg Algebras}

For $L_\infty$-algebras $\mathfrak{a}$ that are \emph{of finite type}, in that they are degreewise finite-dimensional,
let $\mathfrak{a}^{\vee}$ denote their degreewise dual graded vector space
\begin{equation}
  \bracket({\mathfrak{a}^\vee})_k
  :=
  ({
    \mathfrak{a}_k
  })^\ast
  \mathrlap{\,.}
\end{equation}
Then the linear duals of the $L_\infty$-brackets induce a degree-1 derivation $\mathrm{d}$ on the graded-commutative algebra 
\begin{equation}
\label{GradedGrassmannAlgebra}
  \wedge^\bullet \mathfrak{a}^\vee 
    := 
  \mathrm{Sym}\bracket({
    \mathfrak{a}^\vee[1]
  })
  \mathrlap{\,,}
\end{equation} 
defined by
\begin{equation}
\label
{TheDifferential}
  \begin{tikzcd}[
    row sep=15pt,
    column sep=90pt
  ]
    \wedge^\bullet \mathfrak{a}^\vee
    \ar[
      r,
      "{ \mathrm{d} }"
    ]
    &
    \wedge^\bullet \mathfrak{a}^\vee
    \mathrlap{\,,}
    \\
    \wedge^1 \mathfrak{a}^\vee
    \ar[
       ur,
       "{
         [\text{-}]^\ast
         +
         [\text{-},\text{-}]^\ast
         +
         [\text{-},\text{-},\text{-}]^\ast
         +
         \cdots
       }"{swap, sloped}
    ]
    \ar[
      u,
      hook
    ]
  \end{tikzcd}
\end{equation}
whose nilpotency is equivalently the higher Jacobi identities:
\begin{equation}
\label{DualJacobiIdentity}
  \text{higher Jacobi identities}
  \;\;\;
  \Leftrightarrow
  \;\;\;
  \mathrm{d}^2 = 0
  \mathrlap{\,.}
\end{equation}

The resulting dgc-algebra is the \emph{Chevalley--Eilenberg algebra}
\begin{equation}
\label{TheCEAlgebra}
  \mathrm{CE}(\mathfrak{a})
  :=
  \bracket({
    \wedge^\bullet \mathfrak{a}^{\vee}
    ,
    \mathrm{d}
  })
  \;\;\in
  \mathrm{dgcAlg}
  \mathrlap{\,.}
\end{equation}

\paragraph
{The Category of $L_\infty$-Algebras}

Our cases of interest are $L_\infty$-algebras of finite type and \emph{(-1)-connective} in that
  \begin{equation}
  \label{ConnectiveFiniteType}
    \mathrm{dim}(\mathfrak{a}_n)
    \begin{cases}
      < \infty & \text{if } n \geq -1
      \\
      = 0 & \text{otherwise,}
    \end{cases}
    \;\;\;
    \Leftrightarrow
    \;\;\;
    \mathfrak{a}
    \in L_\infty\mathrm{Alg}^{\mathrm{ft}}_{\geq -1}
  \end{equation}
  which entails that
  $\wedge^\bullet \mathfrak{a}^\vee  \defneq \mathrm{Sym}\bracket({\mathfrak{a}^\vee[1]})$ \cref{GradedGrassmannAlgebra} is generated in non-negative degrees.

  We take the category of these $(-1)$-connective finite-type $L_\infty$-algebras to be the full subcategory of the opposite of dgc-algebras on their CE-algebras \cref{TheCEAlgebra}:
    \begin{equation}
    \label
    {LInfinityAsSubcategory}
      \begin{tikzcd}
        L_\infty\mathrm{Alg}
          ^{\mathrm{ft}}
          _{\mathrm{\geq -1}}
        \ar[
          r,
          hook,
          "{ \mathrm{CE}(\text{-}) }"
        ]
        &[30pt]
        \mathrm{dgcAlg}^{\mathrm{op}}
        \mathrlap{\,.}
      \end{tikzcd}
    \end{equation}

  For $\mathfrak{b}_{-1} = 0$ the resulting morphisms $\inlinetikzcd{\mathfrak{a} \ar[r] \& \mathfrak{b}}$ are the usual (non-strict, sh-map) homomorphisms of $L_\infty$-algebras (\parencites[p. 8]{LadaStasheff1992}[Thm.~2.3]{LadaMarkl1995}, cf. \parencites[Def. 13]{SatiSchreiberStasheff2009}[Def. 4.13]{FSS23-Char}), while for $\mathfrak{b}_{-1} \neq 0$ this includes \emph{curved} such morphisms (\parencites{ChuangLazarev2011}[Def. 2.1]{FSS18-TD}), where the dual CE-algebra homomorphisms may shift degree-0 generators by scalars.

  Conversely, the non-curved homomorphisms are precisely those that respect the canonical \emph{augmentations} of our CE-algebras, given by the homomorphism
    \begin{equation}
    \label{CanonicalAugmentation}
    \begin{tikzcd}
      \mathrm{CE}(\mathfrak{a})
      \simeq
      \wedge^\bullet \mathfrak{a}^\vee
      \ar[r, ->>, "{ \epsilon_0 }"]
      &
      \wedge^0 \mathfrak{a}^\vee
      \simeq
      \mathbb{R}
    \end{tikzcd}
    \end{equation}
    which discards all generators and retains just the ground field in degree zero.

\paragraph
{Closed $L_\infty$-Valued Forms}

For $X$ a smooth manifold, denote its de Rham dgc-algebra of smooth differential forms by 
\begin{equation}
  \Omega^\bullet_{\mathrm{dR}}(X)
  \in
  \mathrm{dgcAlg}
  \mathrlap{\,.}
\end{equation}

The set of \emph{differential 1-forms} on $X$ \emph{with coefficients in} $\mathfrak{a} \in L_\infty\mathrm{Alg}^{\mathrm{ft}}_{\geq -1}$ \cref{LInfinityAsSubcategory} is
\begin{equation}
  \Omega^1_{\mathrm{dR}}\bracket({
    X
    ;
    \mathfrak{a}
  })
  \simeq
  \mathrm{Hom}_{\mathrm{gcAlg}}
  \bracket({
    \wedge^\bullet \mathfrak{a}^\vee
    ,
    \Omega^\bullet_{\mathrm{dR}}(X)
  })
  \mathrlap{\,,}
\end{equation}
and its subset of \emph{closed} forms is
\begin{equation}
  \label{ClosedFormsAsDgcHoms}
  \Omega^1_{\mathrm{cl}}\bracket({
    X
    ;
    \mathfrak{a}
  })
  \simeq
  \mathrm{Hom}_{\mathrm{dgcAlg}}
  \bracket({
    \mathrm{CE}(\mathfrak{a})
    ,
    \Omega^\bullet_{\mathrm{dR}}(X)
  })
  \mathrlap{\,.}
\end{equation}
\emph{In terms of algebra-valued forms.}
For $\mathfrak{a}$ a graded Lie algebra (a dg-Lie algebra with vanishing differential, \cref{SpecializationsOfLInfinityAlgebras}), such as the (massive) IIA Borcherds algebras of \cref{TheOvercompleteIIAFluxAlgebra}, the closed $\mathfrak{a}$-valued forms \cref{ClosedFormsAsDgcHoms} have the following equivalent description in terms of brackets of differential forms tensored with elements of $\mathfrak{a}$.

Choose a homogeneous linear basis $(t_a)_{a}$ of $\mathfrak{a}$, with structure constants $[t_b, t_c] = f^a{}_{bc}\, t_a$, and write $(t^a)_a$ for the dual basis of generators of $\mathrm{CE}(\mathfrak{a})$, of degree $\vert t^a \vert = \vert t_a \vert + 1$. With the signs implicit in \cref{TheDifferential} made explicit, the Chevalley--Eilenberg differential reads
\begin{equation}
  \label{CEDifferentialInComponents}
  \mathrm{d}\, t^a
  =
  \tfrac{1}{2}
  \textstyle{\sum_{b,c}}
  (-1)^{\vert t_b \vert (\vert t_c \vert + 1)}
  \,
  f^a{}_{bc}
  \;
  t^b\, t^c
  \mathrlap{\,,}
\end{equation}
where the summand is symmetric under the exchange of $b$ and $c$.
On the other hand, regard $\Omega^\bullet_{\mathrm{dR}}(X) \otimes \mathfrak{a}$ as a graded Lie algebra, graded by form degree minus algebra degree, with the de Rham differential acting on the form factor and with bracket
\begin{equation}
  \label{KoszulSignRule}
  \bracket[{
    \alpha \otimes x
    ,\,
    \beta \otimes y
  }]
  :=
  (-1)^{\vert x \vert \, \vert \beta \vert}
  \,
  (\alpha \wedge \beta) \otimes [x, y]
  \mathrlap{\,.}
\end{equation}
Since a graded-commutative algebra homomorphism $\phi$ out of $\wedge^\bullet \mathfrak{a}^\vee$ is fixed by its values on generators, the assignment
\begin{equation}
  \phi
  \;\mapsto\;
  F_\phi
  :=
  \textstyle{\sum_a}\,
  \phi(t^a) \otimes t_a
\end{equation}
(a finite sum, since $\Omega^k_{\mathrm{dR}}(X) = 0$ for $k > \mathrm{dim}(X)$ and $\mathfrak{a}$ is of finite type) identifies $\Omega^1_{\mathrm{dR}}(X;\mathfrak{a})$ with the elements of total degree 1 in $\Omega^\bullet_{\mathrm{dR}}(X) \otimes \mathfrak{a}$. Under this identification, $\phi$ respects the differentials, $\mathrm{d}\, \phi(t^a) = \phi(\mathrm{d}\, t^a)$, iff
\begin{equation}
  \label{ClosureSignConvention}
  \mathrm{d}\, F_\phi
  =
  \tfrac{1}{2}
  \bracket[{
    F_\phi, F_\phi
  }]
  \mathrlap{\,.}
\end{equation}
Indeed, the $t_a$-component of the left-hand side is $\mathrm{d}\,\phi(t^a)$, while by \cref{KoszulSignRule}, with $\vert \phi(t^c) \vert = \vert t_c \vert + 1$, the $t_a$-component of the right-hand side is $\tfrac{1}{2}\sum_{b,c} (-1)^{\vert t_b \vert (\vert t_c \vert + 1)} f^a{}_{bc}\, \phi(t^b)\, \phi(t^c) = \phi(\mathrm{d}\, t^a)$, by \cref{CEDifferentialInComponents}. Hence both definitions of closed $\mathfrak{a}$-valued forms agree:
\begin{equation}
  \Omega^1_{\mathrm{cl}}(X;\mathfrak{a})
  \;\simeq\;
  \bracketmid\{{
    F \in \Omega^1_{\mathrm{dR}}(X;\mathfrak{a})
  }{
    \mathrm{d}\, F = \tfrac{1}{2}[F,F]
  }\}
  \mathrlap{\,.}
\end{equation}
(Under $F \mapsto -F$, this is the familiar \emph{Maurer--Cartan equation} $\mathrm{d}\,F + \tfrac{1}{2}[F,F] = 0$.) For example, with \cref{CEDifferentialInComponents} the relation $\varphi_3 = [\eta_2, \varphi_1]$ in $\bIIA$ \cref{LowDegreeBorcherdsElements} gives $\mathrm{d}\, f_4 = h_3\, f_2$ in $\mathrm{CE}(\bIIA)$, as in \cref{SullivanModelOfTwistedKTheory}, and generally the brackets of the (massive) IIA Borcherds algebra yield the Bianchi identities in the form $\mathrm{d}\, F_{2k} = H_3 \, F_{2k-2} + \cdots$, as in \cref{BorcherdsTowerOfBianchiIdentities} and \cref{OverviewDiagramMassive}.

\paragraph
{Dg-Lie Algebras}

\emph{Dg-Lie algebras} are the $L_\infty$-algebras whose $(k > 2)$-ary brackets vanish (cf. \cite[Remark~5.3]{LadaMarkl1995}), forming a full subcategory of \cref{LInfinityAsSubcategory}
\begin{equation}
\label{SubcategoryOfDGLieAlgebras}
  \begin{tikzcd}
    \mathrm{dgLieAlg}
      _{\geq -1}
      ^{\mathrm{ft}}
    \ar[
      r,
      hook
    ]
    &
    L_\infty \mathrm{Alg}
      _{\geq -1}
      ^{\mathrm{ft}}
    \mathrlap{\,.}
  \end{tikzcd}
\end{equation}

For dg-Lie algebras, the higher Jacobi identities \cref{DualJacobiIdentity} simplify to the graded Jacobi identity of the binary bracket and the nilpotent graded Lie derivation property of the unary bracket $\partial := [\text{-}]$:
\begin{equation}
\label{DGLConditions}
  \begin{aligned}
   \bracket[{
     x_n, [x'_{n'}, x''_{n''}]
   }]
   &
    =
    \bracket[{
      [x_n,x'_{n'}], x''_{n''}
    }]
    +(-1)^{n n'}
  \bracket[{
    x'_{n'},[x_n,x''_{n''}]
  }]
  \\
  \partial [x_n, x'_{n'}]
    &
    = [\partial x_n, x'_{n'}]
      +
    (-1)^{n}[x_n, \partial x'_{n'}],
  \\
  \partial \partial & = 0.
 \end{aligned}
\end{equation}

There is the evident sense of dg-Lie algebras \emph{freely generated} from a $\mathbb{Z}$-graded vector space, and hence from (the linear span of) a set of graded generators. We denote this construction by 
\begin{equation}
  \begin{tikzcd}
    \mathrm{dgLieAlg}
    \ar[
      rr,
      shift right=6pt,
      "{
         \text{underlying}
      }"{swap}
    ]
    \ar[
      rr,
      phantom,
      "{ \bot }"{scale=.6}
    ]
    &&
    \mathrm{Vect}^{\mathbb{Z}}
    \mathrlap{\,.}
    \ar[
      ll,
      shift right=6pt,
      "{
        \mathrm{Free}_{\mathrm{dgL}}
      }"'
    ]
  \end{tikzcd}
\end{equation}

These conditions \cref{DGLConditions} make manifest that, furthermore, dg-Lie algebras with vanishing differential $\partial := [\text{-}] = 0$ are super Lie algebras whose $\mathbb{Z}_{/2}$-grading is lifted to a $\mathbb{Z}$-grading. 

In conclusion we have the following sequence of specializations of $L_\infty$-algebras:
\begin{equation}
\label{SpecializationsOfLInfinityAlgebras}
  \begin{tikzcd}
    L_\infty\mathrm{Alg}
    &
    \underset
    {
      \substack{
        \\
        \text{higher}
        \\
        \text{brackets vanish}
      }
    }
    {
      \mathrm{dgLieAlg}
    }
    \ar[
      l,
      hook'
    ]
    &
    \underset
    {
      \substack{
        \text{also unary}
        \\
        \text{bracket vanishes}
      }
    }
    {
      \mathbb{Z}\mathrm{sLieAlg}
    }
    \ar[
      l,
      hook'
    ]
    &
    \underset
    {
      \substack{
        {}
        \\
        \text{moreover}
        \\
        \text{concentrated}
        \\
        \text{in degree $0$}
      }
    }
    {
      \mathrm{LieAlg}
    }
    \mathrlap{\,.}
    \ar[
      l,
      hook'
    ]
  \end{tikzcd}
\end{equation}

\paragraph
{Whitehead-Bracket $L_\infty$-Algebras}

A key class of examples of (-1)-connective $L_\infty$-algebras \cref{LInfinityAsSubcategory} are real \emph{Whitehead-bracket $L_\infty$-algebras} $\mathfrak{l}\mathcal{A}$ (cf. \parencites[Prop. 5.11, Rem. 5.4]{FSS23-Char}) of topological spaces $\mathcal{A}$, which are those whose CE-algebras are the spaces' \emph{minimal Sullivan models}. We recall this with emphasis on the rarely discussed generality of possibly disconnected spaces (that we need in \cref{MassiveBorcherds}).

The following is a standard notion  
extended to include negative-degree generators in $\mathfrak{a}$:
\begin{definition}
[Minimal CE-algebra]
\label[definition]{MinimalCEAlgebra}
Consider a CE-algebra $\mathrm{CE}(\mathfrak{a})$ \cref{TheCEAlgebra} of a $(-1)$-connective finite-type $L_\infty$-algebra $\mathfrak{a}$ \cref{ConnectiveFiniteType}.
\begin{enumerate}
  \item 
  \label{OrdinaryMinimalCEAlgebra}
  If $\mathfrak{a}_{-1} \cong 0$
  then $\mathrm{CE}(\mathfrak{a})$ is called \emph{minimal} iff 
  (cf. \parencites[Def. 7.2]{BousfieldGugenheim1976}) 
  
  there exists a well-ordered degree-homogeneous linear basis $\{ \alpha^{(i)} \}_{i \in I}$ of $\mathfrak{a}^\vee$ such that
  \begin{equation}
    \begin{aligned}
    \forall
    \bracket({
      i \in I
    })
      \;:\;\;
      &
      \mathrm{d}\, \alpha^{(i)}
      \in
      \wedge^{\bullet}
      \bracketmid\langle
      {
        \alpha^{(j)}
      }{
        j < i
      }
      \rangle
      \mathrlap{\,,}
      \\
      &
      \mathrm{d}
      \bracket({
        \wedge^1 \mathfrak{a}^\vee
      })
      \subset
        \wedge^{\geq 2}
        \mathfrak{a}^\vee
      \mathrlap{\,.}
    \end{aligned}
  \end{equation}

  \item 
  \label{MinimalCEAlgebraOverPolynomialDeg0}
  If $\mathfrak{a}_{-1} \cong \mathbb{R}\langle \varphi_{-1} \rangle$ then we call
  $\mathrm{CE}(\mathfrak{a})$ \emph{minimal over $\wedge^\bullet \mathfrak{a}_{-1}^{\vee} \cong \mathbb{R}[f_0]$ at $f_0 = 0$} 
  iff 
  
  there exists a well-ordered degree-homogeneous linear basis $\{ \alpha^{(i)} \}_{i \in I}$ of $\mathfrak{a}_{\geq 0}^\vee$ such that
  \begin{equation}
  \label{ConditionsForMinimalityOverPolynomialDegree0}
    \begin{aligned}
    \forall
    \bracket({
      i \in I
    })
      \;:\;\;
      &
      \mathrm{d}\, \alpha^{(i)}
      \in
      \bracket({
        \wedge^\bullet \mathfrak{a}_{-1}^\vee
      })
      \otimes
      \wedge^{\bullet}
      \bracketmid\langle
      {
        \alpha^{(j)}
      }{
        j < i
      }
      \rangle
      \mathrlap{\,,}
      \\
      & 
      \mathrm{d}\bracket({
        \wedge^1 \mathfrak{a}^\vee
      })
      \subset
      \wedge^{\geq 2}
      \mathfrak{a}^\vee
      \mathrlap{\,,}
      \\
      &
      \mathrm{d}
      \bracket({
        \wedge^1 \mathfrak{a}_{-1}^\vee
      })
        = 
      0
      \mathrlap{\,.}
    \end{aligned}
  \end{equation}

\end{enumerate}
\end{definition}

\begin{notation}[The piecewise smooth de Rham algebra of a space]
In rational homotopy theory, for a topological space $\mathcal{A}$ (not assuming or using the structure of a differentiable manifold) one defines a dgc-algebra 
\begin{equation}
  \Omega^\bullet_{\mathrm{pdR}}\bracket({
    \mathcal{A}
  })
  \in
  \mathrm{dgcAlg}
\end{equation}
of \emph{piecewise} smooth differential forms,
namely of ordinary smooth differential forms on $n$-simplices
\begin{equation}
  \Delta^n 
   \;:=\; 
   \bracketmid\{{
     (x^0, \cdots, x^n) 
       \in 
      \mathbb{R}^{n+1}_{\geq 0}
  }{ 
    \sum_i x^i = 1 
  }\} 
    \subset 
  \mathbb{R}^{n+1}
\end{equation}
for each $n \in \mathbb{N}$ and for each continuous map $\inlinetikzcd{ \Delta^n \ar[r] \& \mathcal{A} }$, glued together in the evident way (for technical details cf. \parencites[pp. 1--7]{BousfieldGugenheim1976}[\S 9.1]{GriffithsMorgan2013}[\S 4.2]{Voronov2025}, denoted (cf. \cite[Def. 5.3]{FSS23-Char}):
\begin{equation}
\label
{EndFormulaForPiecewiseDeRhamAlgebra}
  \Omega^\bullet_{\mathrm{pdR}}
  \bracket({
    \mathcal{A}
  })
  :=
  \int_{
    n
  }
  \prod_{
    \Delta^n \to \mathcal{A}
  }
  \Omega_{\mathrm{dR}}^\bullet
  \bracket({
    \Delta^n
  })
  \mathrlap{\,.}
\end{equation}
\end{notation}

\begin{definition}[Minimal Sullivan models beyond the connected cases]
\label[definition]{MinimalSullivanModel}
Consider a space $\mathcal{A}$ whose connected components are nilpotent and of finite rational type (\parencites[Def. 5.1]{FSS23-Char}) and a dgc-algebra homomorphism
\begin{equation}   \label{MinimalModelComparisonMap}
  \begin{tikzcd}
    \mathrm{CE}(\mathfrak{a})
    \ar[
      r,
      "{ \rho }"
    ]
    &
    \Omega
      ^\bullet
      _{\mathrm{pdR}}
    \bracket({
      \mathcal{A}
    })
  \end{tikzcd}
\end{equation}
to its piecewise smooth de Rham algebra \cref{EndFormulaForPiecewiseDeRhamAlgebra}, from some CE-algebra. 
\begin{enumerate}
\item
\label{MinimalSullivanModelOfConnected}
  For $\pi_0\bracket({\mathcal{A}}) \cong \ast$   

  one says that \cref{MinimalModelComparisonMap} is a \emph{minimal Sullivan model} iff
  \begin{enumerate}
  \item
  $\mathrm{CE}(\mathfrak{a})$ is minimal in the sense of \cref{MinimalCEAlgebra} \cref{OrdinaryMinimalCEAlgebra}, 

  \item
  $\rho$ \cref{MinimalModelComparisonMap} is a quasi-isomorphism.
  \end{enumerate}

  \item
  \label{MinimalSullivanModelOfDisconnected}
  Given $\inlinetikzcd{
    \pi_0(\mathcal{A})
    \ar[r, "{ \sim }"]
    \&
    \mathbb{Z}
  }$
  and hence a function
  \begin{equation}
  \label{TheComponentLabels}
    \begin{tikzcd}
      \mathcal{A}
      \ar[
        r,
        ->>
      ]
      \ar[
        rrr,
        uphordown,
        "{ \ell }"
      ]
      &
      \pi_0(\mathcal{A})
      \ar[
        r,
        "{ \sim }"
      ]
      &
      \mathbb{Z}
      \ar[
        r,
        hook
      ]
      &
      \mathbb{R}
      \mathrlap{\,,}
    \end{tikzcd}
  \end{equation}
  we say that $\rho$ \cref{MinimalModelComparisonMap} is a \emph{minimal Sullivan model over $\mathbb{Z}$} iff 
  \begin{enumerate}
    \item 
    $\wedge^1 \mathfrak{a}_{-1}^\vee \simeq \mathbb{R}\langle f_0\rangle$
    and $\rho(f_0) = \ell$ \cref{TheComponentLabels},

    \item
    $\mathrm{CE}(\mathfrak{a})$ is \emph{minimal over $\mathbb{R}[f_0]$ at $f_0 = 0$} in the sense of \cref{MinimalCEAlgebra} \cref{MinimalCEAlgebraOverPolynomialDeg0},

    \item 
    $\rho$ \cref{MinimalModelComparisonMap} induces quasi-isomorphisms%
    \begin{equation}
    \label
    {FiberwiseQuasiIsos}
      \begin{tikzcd}
        \mathllap{
          \forall_{m \in \mathbb{Z}}
          :
          \;
        }
        \mathrm{CE}
        \bracket({\mathfrak{a}})
        /(f_0 - m)
        \ar[
          r,
          "{ \sim }"
        ]
        &
        \Omega^\bullet_{\mathrm{pdR}}
        \bracket({\mathcal{A}_m})
        \mathrlap{\,,}
      \end{tikzcd}
    \end{equation}
    equivalently
    \begin{equation}
      \begin{tikzcd}
        \mathbb{R}^{\mathbb{Z}}
        \otimes_{\mathbb{R}[f_0]}
        \mathrm{CE}
        ({ \mathfrak{a} })
        \ar[
          r,
          "{ \sim }"
        ]
        &
        \Omega^\bullet_{\mathrm{pdR}}
        (\mathcal{A})
        \mathrlap{\,.}
      \end{tikzcd}
    \end{equation}
  \end{enumerate}
\end{enumerate}
\end{definition}
\begin{remark}
\label[remark]{MinimalityInZeroComponent}
 The quasi-isomorphism \cref{FiberwiseQuasiIsos} exhibits a minimal model for the connected component $\mathcal{A}_m$ in the sense of \cref{MinimalSullivanModelOfConnected} if $m = 0$, but not in general.%
 \footnote{
  To see the claim in \cref{MinimalityInZeroComponent}:
  Reducing the conditions \cref{ConditionsForMinimalityOverPolynomialDegree0} modulo $f_0$, the first gives
  $
    \mathrm{d}\,\alpha^{(i)}
    \in
    \wedge^\bullet
    \bracketmid\langle{\alpha^{(j)}}{j < i}\rangle
  $,
  and the second survives, since the only summands of $\mathrm{d}\bracket({\wedge^1 \mathfrak{a}^\vee})$ that could become linear are those of the form $f_0^k\, \beta$, with $k \geq 1$ and $\beta \in \wedge^1 \mathfrak{a}^\vee_{\geq 0}$, and these vanish. Hence $\mathrm{CE}(\mathfrak{a})/(f_0) \cong \mathrm{CE}(\mathfrak{a}_{\geq 0})$, with $\mathfrak{a}_{\geq 0}$ carrying the brackets that do not involve $\varphi_{-1}$ (these close on $\mathfrak{a}_{\geq 0}$ since $\mathrm{d}\, f_0 = 0$), is minimal in the sense of \cref{MinimalCEAlgebra} \cref{OrdinaryMinimalCEAlgebra}, and \cref{FiberwiseQuasiIsos} for $m = 0$ exhibits it as a minimal Sullivan model of $\mathcal{A}_0$.
  For $m \neq 0$, the same summands become linear, so that the fiber is in general not minimal; for instance, $\mathrm{d}\, f_2 = m\, h_3$ in $\mathrm{CE}\bigl(\bIIAMassive\bigr)/(f_0 - m)$, cf. \cref{OverviewDiagramMassive}.
}
\end{remark}

\begin{proposition}[{cf. \cite[Props.~5.11--5.13 \& Rem.~5.4]{FSS23-Char}}]
\label[proposition]{WhiteheadAlgebraInConnectedCase}
For connected $\mathcal{A}$ \textup{(\cref{MinimalSullivanModelOfConnected})}, minimal Sullivan models exist uniquely up to isomorphism, and the corresponding $L_\infty$-algebra 
$
  \mathfrak{l}\mathcal{A}
    \defneq 
  \mathfrak{a} 
$ \cref{MinimalModelComparisonMap} has:
\begin{enumerate}
 \item
 as elements the $\mathbb{R}$-rationalized homotopy groups of $\Omega \mathcal{A}$:
 \begin{equation}
   \bracket({
     \mathfrak{l}
     \mathcal{A}
   })_\bullet
   \simeq
   \pi_\bullet\bracket({
     \Omega \mathcal{A}
   })
   \otimes
   \mathbb{R}
   \mathrlap{\,,}
 \end{equation}
 
 \item
 as brackets the $\mathbb{R}$-rationalized \emph{higher Whitehead brackets} on these homotopy groups \textup{(\parencites{AndrewsArkowitz1978}{BelchiEtAl2017}{BuijsMorenoFernandez2019})}.
\end{enumerate}
\end{proposition}

\begin{notation}[Whitehead-bracket $L_\infty$- and dg-Lie algebras]
\label[notation]{WHBracketLinfinityAlgebras}
 In the connected case (\cref{MinimalSullivanModelOfConnected}), \cref{WhiteheadAlgebraInConnectedCase} says that $\mathfrak{l}\mathcal{A}$ is the \emph{real Whitehead-bracket $L_\infty$-algebra} of $\mathcal{A}$.

In the case $\inlinetikzcd{
    \pi_0(\mathcal{A})
    \ar[r, "{ \sim }"]
    \&
    \mathbb{Z}
  }$ (\cref{MinimalSullivanModelOfDisconnected})   
  a minimal model need not exist and when it does exist it need not be unique (by which we mean: up to isomorphism fixing $f_0$). It is only guaranteed to be unique when specialized to the 0-components, by \cref{MinimalityInZeroComponent,WhiteheadAlgebraInConnectedCase}.
    
  But when it exists uniquely, then that again determines an $L_\infty$-algebra $\mathfrak{a}$ \cref{MinimalModelComparisonMap}, which we denote again by $\mathfrak{l}\mathcal{A} \defneq \mathfrak{a}$.

  In this vein, when the minimal model $\mathrm{CE}(\mathfrak{a})$ becomes unique under the further condition that $\mathfrak{a}$ is a dg-Lie algebra $\inlinetikzcd{\mathfrak{a} \in \mathrm{dgLieAlg} \ar[r, hook] \&  L_\infty \mathrm{Alg}}$ \cref{SubcategoryOfDGLieAlgebras} then we may accordingly speak of \emph{the Whitehead-bracket dg-Lie algebra}, which may be denoted $\mathfrak{l}_{\mathrm{dgL}} \mathcal{A}$, but which we still denote by $\mathfrak{l}\mathcal{A} \defneq \mathfrak{a}$ when the context is clear (\cref{prop:RationalModelOfMassiveMinCyc}).
\end{notation}

\section{Proofs}
\label{app:proofs}

This appendix contains the proofs deferred from the main text.

\subsection
{
  \texorpdfstring
  {For \cref{ObtainingBorcherdsIIA}}
  {For the IIA Hierarchy}
}
\label{ProofOfFirstMainTheorem}

\begin{lemma}\label[lemma]{lem:minimal_cyclification_and_borcherds}
The minimal Sullivan model of \(\SmlCyc S^4\) is equivalent to 
\(
\mathrm{CE}\!\left(
\mathfrak{bor}^{+}_{\mathrm{IIA}}/\langle\eta_8\rangle
\right).
\)
The isomorphism of $L_{\infty}$-algebras
\begin{equation}
  \beta:
  \mathfrak l(\SmlCyc S^4)
  \simeq
  \mathfrak{bor}^{+}_{\mathrm{IIA}}/\langle\eta_8\rangle
\end{equation}
may be chosen compatibly with the Borel projection $\SmlCyc S^4\to BS^1$ with the following normalizations:
\begin{enumerate}
\item
 The constant-loop section at a fixed point of $S^4$ \textup{(cf.~\cref{CircleMaps})} is represented by
 \begin{equation}
   \begin{tikzcd}
   \mathrm{CE}\bracket({
     \mathfrak{bor}^{+}_{\mathrm{IIA}}
       /
      \langle\eta_8\rangle
   })
   \ar[r]
   &
   \mathrm{Free}_{\mathrm{dgcA}}(f_2)
     /
   \bracket({\mathrm{d} f_2 = 0})
   \mathrlap{\,,}
   \end{tikzcd}
  \end{equation}
   sending \(f_2\) to itself and every other generator to zero.
\item 
\label{InducingQuotient}
An $S^3$-family of small loops, 
\[
  \inlinetikzcd{
    S^3 
      \cong 
    \R^3_{\cpt} 
      \ar[r, hook]
    \&
    (S^1\times \R^3)_{\cpt} 
    \ar[
      r, 
      ->>
    ] 
    \&
    (S^1\times \R^3)_{\cpt} 
      \sslash 
    S^1
    \defneq
    \MinCyc S^4
    \ar[r, hook]
    \&
    \SmlCyc S^4
  }
\]
induces
  \begin{equation}
   \begin{tikzcd}
   \mathrm{CE}\bracket({
     \bIIA
       /
     \langle\eta_8\rangle
   })
   \ar[r]
   &
   \mathrm{Free}_{\mathrm{dgcA}}(h_3)
     /
    (\mathrm{d} h_3 = 0)
    \mathrlap{\,,}
    \end{tikzcd}
  \end{equation}
   sending $h_3$ to itself and every other generator to zero.

  Dually, these are the strict maps generated by $p(\varphi_1)$ and $p(\eta_2)$, respectively.
\end{enumerate}
\end{lemma}
\begin{proof}
Throughout, we 
\begin{itemize}
\item
abbreviate
$
  \mathfrak{g}
  \;:=\;
  \bIIA/\langle\eta_8\rangle
  \mathrlap{\,,}
$
\item denote the generators of $\mathrm{CE}(\bIIA)$ dual to the elements $\varphi_1,\eta_2,\varphi_3,\varphi_5,\eta_6,\varphi_7,\eta_8,\dots$ of \cref{LowDegreeBorcherdsElements} by $f_2,h_3,f_4,f_6,h_7,f_8,h_9,\dots$, as in \cref{BorcherdsTowerOfBianchiIdentities}, 
\item 
write $H^\bullet(\mathfrak{h}) := H^\bullet\bracket({\mathrm{CE}(\mathfrak{h})})$
for a dg-Lie algebra $\mathfrak{h}$. 
\end{itemize}

\begin{enumerate}
\item 
\label{Step1StructureOfg}
\textbf{The structure of $\mathfrak{g}$.}
In $\bIIA$ we have $[\varphi_1,\varphi_3] = 0$, by the Jacobi identity
$
  \bracket[{\varphi_1,[\eta_2,\varphi_1]}]
  =
  \bracket[{[\varphi_1,\eta_2],\varphi_1}]
  +
  \bracket[{\eta_2,[\varphi_1,\varphi_1]}]
  =
  -\bracket[{\varphi_1,[\eta_2,\varphi_1]}]
$.
Let $\delta$ be the derivation of degree $1$ of the free graded Lie algebra $\LL(\eta_2,\varphi_3)$ with $\delta(\eta_2) = -\varphi_3$ and $\delta(\varphi_3) = 0$. Then $\delta^2 = \tfrac12[\delta,\delta]$ is a derivation vanishing on the generators, so $\delta^2 = 0$, and therefore setting $[\varphi_1,\varphi_1] := 0$ and $[\varphi_1, x] := \delta(x)$ makes $\R\varphi_1 \oplus \LL(\eta_2,\varphi_3)$ a graded Lie algebra $\R\varphi_1 \ltimes_\delta \LL(\eta_2,\varphi_3)$. By the presentation \cref{ThePositiveIIABorcherdsAlgebra}, there is a Lie map
\begin{equation}
\label{BorcherdsAsSemidirectProduct}
  \begin{tikzcd}[sep=0pt]
    \bIIA
    \ar[rr, "{ \sim }"]
    &&
    \R\varphi_1 \ltimes_\delta \LL(\eta_2,\varphi_3)
    \\
    \varphi_1 &\longmapsto& \varphi_1
    \\
    \eta_2 &\longmapsto& \eta_2
    \mathrlap{\,,}
  \end{tikzcd}
\end{equation}
and it is an isomorphism: its inverse is $\varphi_1 \mapsto \varphi_1$ together with the Lie map $\psi : \LL(\eta_2,\varphi_3) \to \bIIA$ given by $\eta_2 \mapsto \eta_2$, $\varphi_3 \mapsto [\eta_2,\varphi_1]$. This respects the brackets with $\varphi_1$, because $x \mapsto \psi(\delta x)$ and $x \mapsto [\varphi_1,\psi(x)]$ are derivations along $\psi$ which agree on the generators, by $[\varphi_1,\eta_2] = -[\eta_2,\varphi_1]$ and $[\varphi_1,[\eta_2,\varphi_1]] = 0$; and both composites are the identity on generators. Under \cref{BorcherdsAsSemidirectProduct} we have
$\varphi_3 = [\eta_2,\varphi_1]$, $\eta_6 = [\varphi_3,\varphi_3]$ and $\eta_8 = [\eta_2,[\varphi_3,\varphi_3]]$.
Since $\delta(\eta_8) = -[\varphi_3,[\varphi_3,\varphi_3]] = 0$, $\varphi_3$ being odd, the ideal $(\eta_8)$ of $\LL(\eta_2,\varphi_3)$ is $\delta$-stable, hence coincides with the ideal $\langle\eta_8\rangle$ of $\bIIA$, and so
\begin{equation}
\label{QuotientAsSemidirectProduct}
  \mathfrak{g}
  \;\cong\;
  \R\varphi_1 \ltimes \mathfrak{k}
  \mathrlap{\,,}
  \qquad
  \mathfrak{k}
  \;:=\;
  \LL(\eta_2,\varphi_3)/(\eta_8)
  \mathrlap{\,.}
\end{equation}
In degree $8$, $\LL(\eta_2,\varphi_3)$ is spanned by $[\eta_2,[\varphi_3,\varphi_3]]$ and $[\varphi_3,[\eta_2,\varphi_3]]$, which are proportional by the Jacobi identity; hence $(\bIIA)_8 = \R\eta_8$ and $\mathfrak{g}_8 = 0$. Dualizing the quotient map exhibits $\mathrm{CE}(\mathfrak{g})$ as a sub-dgc-algebra of $\mathrm{CE}(\bIIA)$, which therefore has, in degrees $\leq 9$, exactly the generators $f_2,h_3,f_4,f_6,h_7,f_8$, with differentials as in \cref{BorcherdsTowerOfBianchiIdentities}:
\begin{equation}
\label{LowDegreeCEDifferentials}
  \begin{aligned}
  \mathrm{d}\, f_2 &= 0
    \mathrlap{\,,}
  \\
  \mathrm{d}\, h_3 &= 0
    \mathrlap{\,,}
  \\
  \mathrm{d}\, f_4 & = h_3 f_2
    \mathrlap{\,,}
  \end{aligned}
  \;\;\;\;\;\;\;
  \begin{aligned}
  \mathrm{d}\, f_6 & = h_3 f_4
    \mathrlap{\,,}
  \\
  \mathrm{d}\, h_7 
    & = \tfrac12 f_4 f_4 - f_2 f_6
    \mathrlap{\,,}
  \\
  \mathrm{d}\, f_8 & = h_3 f_6
    \mathrlap{\,.}
  \end{aligned}
\end{equation}
Finally, by \cref{QuotientAsSemidirectProduct}, $\mathrm{CE}(\mathfrak{g}) \cong \R[f_2] \otimes \mathrm{CE}(\mathfrak{k})$ as graded algebras, $f_2$ is a cocycle ($\varphi_1$ not being a bracket), and the quotient map $\mathrm{CE}(\mathfrak{g}) \to \mathrm{CE}(\mathfrak{g})/(f_2) \cong \mathrm{CE}(\mathfrak{k})$ is the dual of the inclusion $\mathfrak{k} \hookrightarrow \mathfrak{g}$.

\item 
\label{Step2CohomologyOfk}
\textbf{The cohomology of $\mathfrak{k}$.}
We claim that
\begin{equation}
\label{CohomologyOfTheIdeal}
  H^n(\mathfrak{k})
  \;\cong\;
  \left\{\!\!
  \begin{array}{ll}
    \R & \text{for } n \in \{0,3,4,10\}\,,
    \\
    0 & \text{otherwise}\,,
  \end{array}
  \right.
  \qquad
  H^3(\mathfrak{k}) = \R\, h_3
  \,,\quad
  H^4(\mathfrak{k}) = \R\, f_4
  \mathrlap{\,.}
\end{equation}
The last two statements hold because $\mathrm{CE}(\mathfrak{k})$ vanishes in degrees $1$ and $2$, equals $\R h_3$ and $\R f_4$ in degrees $3$ and $4$, and $\mathrm{d}h_3 = 0 = \mathrm{d}f_4$ there, by \cref{LowDegreeCEDifferentials} modulo $f_2$.
For the first statement, let $V := \R\langle\eta_2,\varphi_3\rangle$, let $T := T(V)$ be the free graded associative algebra, $T^+ = T\cdot V$ its augmentation ideal, and
\[
  r
  \;:=\;
  \eta_8
  \;=\;
  [\eta_2,[\varphi_3,\varphi_3]]
  \;=\;
  2\bracket({
    \eta_2\varphi_3\varphi_3
    -
    \varphi_3\varphi_3\eta_2
  })
  \;\in\;
  T
  \mathrlap{\,,}
\]
so that the universal enveloping algebra of $\mathfrak{k}$ is $R := T/(r)$ \parencite[\S 21]{FHT2000}.
\begin{enumerate}
\item
Call a word in the letters $\eta_2,\varphi_3$ \emph{normal} if it does not contain the subword $\varphi_3\varphi_3\eta_2$. The normal words span $R$: modulo $r$, an occurrence of $\varphi_3\varphi_3\eta_2$ may be replaced by $\eta_2\varphi_3\varphi_3$, which makes the word lexicographically smaller (for $\eta_2 < \varphi_3$) among the finitely many words of the same length and degree.

\item
Let $N(z) := \sum_n \#\{\text{normal words of degree } n\}\, z^n$. Every nonempty normal word is uniquely of the form $w x$ with $w$ normal and $x$ a letter. Conversely, if $w$ is normal and $x$ is a letter, then $wx$ is either normal or of the form $u\,\varphi_3\varphi_3\eta_2$ with $u$ normal; and every such $u\,\varphi_3\varphi_3\eta_2$ arises this way, with $w = u\,\varphi_3\varphi_3$, which is normal because $\varphi_3\varphi_3\eta_2$ ends with $\eta_2$. Hence
$(z^2+z^3)\,N(z) = N(z) - 1 + z^8 N(z)$, that is
\begin{equation}
\label{HilbertSeriesOfNormalWords}
  N(z)
  \;=\;
  \frac{1}{1 - z^2 - z^3 + z^8}
  \mathrlap{\,.}
\end{equation}

\item
Consider the sequence of left $R$-modules
\begin{equation}
\label{ResolutionOfTheGroundField}
  \begin{tikzcd}
    0
    \ar[r]
    &
    R \otimes \R r
    \ar[r, "{ \partial_2 }"]
    &
    R \otimes V
    \ar[r, "{ \partial_1 }"]
    &
    R
    \ar[r, "{ \epsilon }"]
    &
    \R
    \ar[r]
    &
    0
    \mathrlap{\,,}
  \end{tikzcd}
\end{equation}
with $\epsilon$ the augmentation, $\partial_1(y \otimes v) := y v$ and $\partial_2(y\otimes r) := -2y\varphi_3\varphi_3 \otimes \eta_2 + 2y\eta_2\varphi_3 \otimes \varphi_3$, so that $\partial_1\partial_2(y\otimes r) = y r = 0$. It is exact at $\R$, at $R$, and at $R \otimes V$: writing $I := (r) \subset T$, we have $R\otimes V \cong T^+/(I\cdot V)$ via $y\otimes v \mapsto yv$, under which $\partial_1$ becomes the projection $T^+/(I \cdot V) \to T^+/I$, whose kernel $I/(I\cdot V)$ is the image of $\partial_2$, since $I = T r T = T r + I\cdot V$, by $T = \R \oplus T\cdot V$. Therefore the kernel $K$ of $\partial_2$ has Hilbert series $H_K(z) = H_R(z)\,(1 - z^2 - z^3 + z^8) - 1$. By (a) we have $H_R = N - E$ for a power series $E$ with nonnegative coefficients, and so, by \cref{HilbertSeriesOfNormalWords}, $H_K = -E\cdot(1-z^2-z^3+z^8)$. If $E \neq 0$, its lowest nonzero coefficient would reappear in $H_K$ with negative sign, which is absurd. Hence $E = 0$ and $K = 0$, so that \cref{ResolutionOfTheGroundField} is a free resolution of $\R$.
\end{enumerate}
Since $\partial_2$ and $\partial_1$ take values in $R^+ \otimes V$ and $R^+$, respectively, applying $\R \otimes_R (\text{-})$ to \cref{ResolutionOfTheGroundField} gives zero differentials, whence $\mathrm{Tor}^R_\bullet(\R,\R)$ is $\R$, $V$ and $\R r$ in homological degrees $0$, $1$ and $2$, respectively. The Chevalley--Eilenberg chains of $\mathfrak{k}$ compute $\mathrm{Tor}^{R}(\R,\R)$, in total degree equal to homological degree plus internal degree \parencite[\S 22]{FHT2000}, and $\mathrm{CE}(\mathfrak{k})$ is their degreewise linear dual. Since $\vert\eta_2\vert = 2$, $\vert\varphi_3\vert = 3$ and $\vert r \vert = 8$, this gives the claim \cref{CohomologyOfTheIdeal}.

\item 
\label{Step3AnUpperBound}
\textbf{An upper bound for $H^\bullet(\mathfrak{g})$.}
By \cref{Step1StructureOfg}, multiplication by $f_2$ is an injective cochain map of degree 2 on $\mathrm{CE}(\mathfrak{g})$ with cokernel $\mathrm{CE}(\mathfrak{k})$. The resulting long exact sequence
\[
  \begin{tikzcd}
    \cdots
    \ar[r]
    &
    H^{n-2}(\mathfrak{g})
    \ar[r, "{ f_2 \cdot }"]
    &
    H^{n}(\mathfrak{g})
    \ar[r]
    &
    H^{n}(\mathfrak{k})
    \ar[r, "{ \Delta }"]
    &
    H^{n-1}(\mathfrak{g})
    \ar[r, "{ f_2 \cdot }"]
    &
    H^{n+1}(\mathfrak{g})
    \ar[r]
    &
    \cdots
  \end{tikzcd}
\]
shows that $\dim H^n(\mathfrak{g}) \leq \dim f_2 \cdot H^{n-2}(\mathfrak{g}) + \dim \ker\bracket({\Delta\vert_{H^n(\mathfrak{k})}})$. Here $[h_3] \neq 0$ in $H^3(\mathfrak{g})$, since $\mathrm{CE}(\mathfrak{g})$ is $\R f_2$ in degree 2 and $\mathrm{d}f_2 = 0$; and since $\mathrm{d}f_4 = f_2 h_3$, the connecting map sends $[f_4] \mapsto [h_3]$, hence is injective on $H^4(\mathfrak{k})$, while $f_2\cdot [h_3] = [\mathrm{d} f_4] = 0$. With \cref{CohomologyOfTheIdeal} it follows, by induction on $n$, that
\begin{equation}
\label{UpperBoundOnCohomology}
  \dim H^n(\mathfrak{g})
  \;\leq\;
  \mu_n
  \;:=\;
  \left\{\!\!
  \begin{array}{ll}
    1 & \text{for } n \in \{0,2,3,4,6,8\}\,,
    \\
    2 & \text{for } n \text{ even and } n \geq 10\,,
    \\
    0 & \text{otherwise}\,.
  \end{array}
  \right.
\end{equation}

\item 
\label{Step4AModelOfSmlCyS4}
\textbf{A model of $\SmlCyc S^4$.}
By \cite[Lem.~3.4]{GSS26-KfromM}, the space of round loops modulo rotations is the homotopy pushout
\begin{equation}
\label{eq:homotopy_pushout}
  \SmlCyc S^4
  \;\simeq\;
  \operatorname{hocolim}\left(
  \begin{tikzcd}
    \widetilde{\operatorname{Gr}}_2(\R^5)
    &
    F
    \ar[l, "{ q_A }"{swap}]
    \ar[r, "{ q_C }"]
    &
    S^4 \times BS^1
  \end{tikzcd}
  \right)
  \mathrlap{\,,}
\end{equation}
where
\[
  F
  \;:=\;
  \frac{\mathrm{SO}(5)}{\mathrm{SO}(2)\times\mathrm{SO}(2)}
  \;\cong\;
  \bigl\{
    (P,w)
    \,\big\vert\,
    P\in\widetilde{\operatorname{Gr}}_2(\R^5),\;
    w\in S^4,\;
    w\perp P
  \bigr\}
  \mathrlap{\,,}
\]
$\widetilde{\operatorname{Gr}}_2(\R^5)$ is the space of oriented great circles $S^4 \cap P$, the map $q_A$ forgets the pole $w$ towards which the great circle may be shrunk, and $q_C$ sends $(P,w)$ to the point $w$ to which it shrinks and to the classifying map of the circle bundle of rigid parametrizations of $S^4 \cap P$; the constant loops correspond to $S^4 \times BS^1$. By van Kampen, $\SmlCyc S^4$ is simply connected.

On $F$, write $f_2 := e(P)$ and $e_2 := e(Q)$ for the Euler classes of the tautological oriented plane bundles $P$ and $Q := P^\perp \cap w^\perp$, with $Q$ oriented such that $P \oplus Q = w^\perp = T_w S^4$ as oriented bundles. The circle bundle of rigid parametrizations of $S^4\cap P$ is the unit circle bundle of $P$, so that $q_C^\ast$ sends the generator $f_2 \in H^2(BS^1)$ to $e(P) = f_2$; on $\widetilde{\operatorname{Gr}}_2(\R^5)$ we also write $f_2 := e(P)$. Since $\widetilde{\operatorname{Gr}}_2(\R^5) = \mathrm{SO}(5)/\bracket({\mathrm{SO}(2)\times\mathrm{SO}(3)})$ and $F$ are homogeneous spaces of equal rank, their Cartan models \parencite[Thm.~2.71]{FelixOpreaTanre2008} are Koszul complexes on the regular sequences given by restricting the Pontryagin classes $p_1, p_2 \in H^\bullet(B\mathrm{SO}(5))$, so that these spaces are formal with cohomology $H^\bullet(BH)/\bracket({H^+(B\mathrm{SO}(5))})$. With the Whitney sum formulas $p(P)\,p(P^\perp) = 1 = p(P)\,p(Q)$ (from $\R^5 = P \oplus P^\perp = P \oplus \R w \oplus Q$), and eliminating $p_1(P^\perp) = -f_2^2$, this gives the algebras $A$ and $B$ in
\begin{equation}
\label{eq:homotopy_pullback}
  \begin{tikzcd}[
    sep=0pt
  ]
    \overbrace
      {
      \frac{\R[f_2]}{(f_2^4)}
      }^{A}
    \ar[rr, "{ \iota_A }"]
    &&
    \overbrace{
      \frac{\R[f_2,e_2]}{(f_2^2+e_2^2,\;f_2^2e_2^2)}
     }^{B}
    &&
    \overbrace{
      \frac{\R[f_2,x_4]}{(x_4^2)}
    }^{C}
    \ar[ll, "{ \iota_C }"{swap}]
    \\
    f_2 &\mapsto& f_2
    &\mapsfrom& f_2
    \\
    \\
    &&
    \tfrac{1}{2} f_2 e_2
    &\mapsfrom& x_4
    \mathrlap{\,,}
  \end{tikzcd}
\end{equation}
where $C = H^\bullet(S^4 \times BS^1)$ with $\bracket\langle{x_4,[S^4]}\rangle = 1$, and where $\iota_A = q_A^\ast$ and $\iota_C = q_C^\ast$, the latter because $w^\ast e(TS^4) = e(P\oplus Q) = f_2 e_2$ while $e(TS^4) = \chi(S^4)\,x_4 = 2x_4$. In $B$ we have $f_2^4 = -f_2^2 e_2^2 = 0$, so that $B$ has the basis $f_2^i e_2^j$, $0 \leq i \leq 3$, $0 \leq j \leq 1$, with top class $f_2^3 e_2$.

The diagram \cref{eq:homotopy_pushout} is formal, in that its rational model is \cref{eq:homotopy_pullback}: The spaces $\widetilde{\operatorname{Gr}}_2(\R^5)$ and $S^4 \times BS^1$ have the minimal models $\Lambda(f_2, a_7)$ with $\mathrm{d}a_7 = f_2^4$, and $\Lambda(f_2, x_4, c_7)$ with $\mathrm{d}c_7 = x_4^2$, whose projections to $A$ and $C$ are quasi-isomorphisms, and $F$ is formal. A dgc-algebra map from either of these minimal models to $B$ sends the odd generator to zero, $B$ being concentrated in even degrees, and the even generators, being cocycles, to their cohomology classes; hence it is determined by the induced map on cohomology. Therefore Sullivan representatives of $q_A$ and $q_C$ \parencite[\S 12]{FHT2000}, followed by a formality quasi-isomorphism of $F$ inducing the above identification $H^\bullet(F;\R) \cong B$, are exactly the maps $\iota_A$ and $\iota_C$ precomposed with these projections. It follows \parencite[Prop.~13.6]{FHT2000} that $\SmlCyc S^4$ is modeled by the homotopy pullback of \cref{eq:homotopy_pullback}, which is realized by the pullback
\[
  \mathcal{P}
  \;:=\;
  (A\times C)\times_{B\times B}B[s,\mathrm ds]
  \mathrlap{\,,}
  \qquad
  B[s,\mathrm ds]
  \;:=\;
  B\otimes\R[s]\otimes\Lambda(\mathrm ds)
  \mathrlap{\,,}
\]
of $\iota_A \times \iota_C$ along evaluation at $s = 0, 1$ (a surjection), consisting of the triples $(a, c, b)$ with $b\vert_{s=0} = \iota_A(a)$ and $b\vert_{s=1} = \iota_C(c)$. The projections of $\mathcal{P}$ to $A$ and to $C$ model the two structure maps of the homotopy pushout.

The projection $\mathcal{P} \to A \times C$ is surjective, with kernel the triples $(0,0,b)$ with $b$ vanishing at $s = 0,1$, whose cohomology is $B\cdot\mathrm{d}s$, that is $B^{n-1}$ in degree $n$. The connecting map of the resulting long exact sequence is, up to sign, $\rho := \iota_C - \iota_A : A \oplus C \to B$, so that we have short exact sequences
\begin{equation}
\label{MayerVietorisSequence}
  \begin{tikzcd}
    0
    \ar[r]
    &
    \operatorname{coker}\bracket({\rho^{n-1}})
    \ar[r]
    &
    H^n(\mathcal{P})
    \ar[r]
    &
    \ker\bracket({\rho^n})
    \ar[r]
    &
    0
    \mathrlap{\,,}
  \end{tikzcd}
\end{equation}
the first map sending the class of $b \in B^{n-1}$ to $\bigl[(0,0,b\,\mathrm{d}s)\bigr]$, the second induced by the projection. Since $\iota_C(f_2^k) = f_2^k$ and $\iota_C(f_2^{k-2}x_4) = \tfrac12 f_2^{k-1}e_2$, the map $\rho$ is surjective in all degrees except $2$, where $\operatorname{coker}(\rho^2) \cong \R\, e_2$. Counting dimensions, we find
\begin{equation}
\label{CohomologyOfSmlCyc}
  \dim H^n(\mathcal{P})
  \;=\;
  \mu_n
  \qquad
  \text{for all } n
  \mathrlap{\,,}
\end{equation}
with $\mu_n$ from \cref{UpperBoundOnCohomology}.

\item
\label{Step5TheComparisonMap}
\textbf{The comparison map.}
Define the algebra homomorphism $\theta:\mathrm{CE}(\mathfrak{g})\to\mathcal{P}$ by
\[
\begin{aligned}
f_2&\longmapsto (f_2,\, f_2,\, f_2),\\
h_3&\longmapsto(0,\,0,\,\tfrac12e_2\,\mathrm ds),\\
f_4&\longmapsto(0,\,x_4,\,\tfrac12s f_2e_2),\\
f_6&\longmapsto
\bigl(\tfrac18f_2^3,\,0,\,\tfrac18(1-s^2)f_2^3\bigr),\\
h_7&\longmapsto0,\\
f_8&\longmapsto
\bigl(0,\,\tfrac1{12}f_2^2x_4,\,
\tfrac1{16}(s-\tfrac13 s^3)f_2^3e_2\bigr),
\end{aligned}
\]
and by sending all generators of degree $\geq 10$ to zero (there are none in degree 9, by \cref{Step1StructureOfg}). These triples satisfy the endpoint conditions; e.g., the last component of $\theta(f_8)$ restricts at $s = 1$ to $\tfrac{1}{24}f_2^3 e_2 = \iota_C\bracket({\tfrac{1}{12}f_2^2x_4})$. Using only $e_2^2 = -f_2^2$ and $f_2^4 = 0 = x_4^2$, one checks that $\theta$ respects the differentials \cref{LowDegreeCEDifferentials} on the generators of degree $\leq 8$:
\[
  \begin{aligned}
    \mathrm{d}\,\theta(f_4)
    & =
    \bigl(0,0,\tfrac12 f_2 e_2\,\mathrm{d}s\bigr)
    =
    \theta(h_3 f_2)
    \mathrlap{\,,}
    \\
    \mathrm{d}\,\theta(f_6)
    & =
    \bigl(0,0,-\tfrac14 s f_2^3\,\mathrm{d}s\bigr)
    =
    \bigl(0,0,\tfrac14 s f_2 e_2^2\,\mathrm{d}s\bigr)
    =
    \theta(h_3 f_4)
    \mathrlap{\,,}
    \\
    \theta\bracket({
      \tfrac12 f_4 f_4 - f_2 f_6
    })
    & =
    \bigl(
      -\tfrac18 f_2^4,\,
      \tfrac12 x_4^2,\,
      \tfrac18 s^2 f_2^2 e_2^2
      -
      \tfrac18(1-s^2) f_2^4
    \bigr)
    = 0 =
    \mathrm{d}\,\theta(h_7)
    \mathrlap{\,,}
    \\
    \mathrm{d}\,\theta(f_8)
    & =
    \bigl(0,0,\tfrac1{16}(1-s^2) f_2^3 e_2\,\mathrm{d}s\bigr)
    =
    \theta(h_3 f_6)
    \mathrlap{\,,}
  \end{aligned}
\]
and trivially on $f_2$ and $h_3$. For a generator $g$ of degree $\geq 10$, $\mathrm{d}g$ is a quadratic polynomial in generators, of degree $\geq 11$. In degrees $\geq 10$, $\mathcal{P}$ coincides with its $C$-component, which vanishes in odd degrees. The $C$-components of the images of $f_2, f_4, f_8$ are $f_2, x_4, \tfrac1{12} f_2^2 x_4$, and those of the images of $h_3, f_6, h_7$ vanish; so the only products of degree $\geq 11$ of the images of generators of degree $\leq 8$ that could be nonzero, those of $f_4 f_8$ and $f_8 f_8$, are proportional to $x_4^2 = 0$. Hence $\theta(\mathrm{d}g) = 0 = \mathrm{d}\theta(g)$, and $\theta$ is a homomorphism of dgc-algebras.

$\theta$ is a quasi-isomorphism: By \cref{UpperBoundOnCohomology,CohomologyOfSmlCyc} it suffices to show that $H^n(\theta)$ is surjective for all $n$. Write $\mathrm{pr}_C : \mathcal{P} \to C$ for the projection; by construction, $\mathrm{pr}_C\circ\theta$ sends $f_2 \mapsto f_2$, $f_4 \mapsto x_4$, $f_8 \mapsto \tfrac{1}{12}f_2^2 x_4$, and all other generators to zero.
\begin{itemize}
\item For $n = 2k \leq 8$, the class $[\theta(f_2^k)]$ is nonzero, since $\mathrm{pr}_C$ sends it to $f_2^k \neq 0$; hence it spans $H^{2k}(\mathcal{P})$.
\item For $n = 3$, the class $[\theta(h_3)] = \bigl[(0,0,\tfrac12 e_2\,\mathrm{d}s)\bigr]$ is the image of $\tfrac12 e_2 \notin \operatorname{im}(\rho^2)$ under \cref{MayerVietorisSequence}; hence it spans $H^3(\mathcal{P})$.
\item For $n = 2k \geq 10$, consider
$
  z := f_4 f_6 + 2 h_3 h_7 - 3 f_2 f_8
$.
This is a cocycle,
$
  \mathrm{d}z
  =
  \bracket({ h_3 f_2 f_6 + h_3 f_4^2 })
  -
  \bracket({ h_3 f_4^2 - 2 h_3 f_2 f_6 })
  -
  3 h_3 f_2 f_6
  =
  0
$
(in $\mathrm{CE}(\bIIA)$ it is exact, $z = 2\,\mathrm{d}h_9$ by \cref{BorcherdsTowerOfBianchiIdentities}, but $h_9$ is absent from $\mathrm{CE}(\mathfrak{g})$), and $\theta(z) = \bigl(0,-\tfrac14 f_2^3 x_4, 0\bigr)$. Therefore $\mathrm{pr}_C$ sends the classes $[\theta(f_2^k)]$ and $[\theta(f_2^{k-5}z)]$ to $f_2^k$ and $-\tfrac14 f_2^{k-2}x_4$, which are linearly independent; hence these classes span the 2-dimensional space $H^{2k}(\mathcal{P})$.
\item For all other $n$, $H^n(\mathcal{P}) = 0$.
\end{itemize}

Since $\mathfrak{g}$ is positively graded and of finite type, $\mathrm{CE}(\mathfrak{g})$ is a minimal Sullivan algebra, its differential being quadratic in generators of lower degree. Lifting the quasi-isomorphism $\theta$ along the zigzag of quasi-isomorphisms relating $\mathcal{P}$ to $\Omega^\bullet_{\mathrm{pdR}}\bracket({\SmlCyc S^4})$ \parencite[\S 12]{FHT2000} yields a minimal Sullivan model
\[
  \begin{tikzcd}
    \rho
    :
    \mathrm{CE}(\mathfrak{g})
    \ar[r, "{ \sim }"]
    &
    \Omega^\bullet_{\mathrm{pdR}}\bracket({\SmlCyc S^4})
  \end{tikzcd}
\]
of the simply connected space $\SmlCyc S^4$ of finite rational type, in the sense of \cref{MinimalSullivanModel} \cref{MinimalSullivanModelOfConnected}. By uniqueness of minimal models (\cref{WhiteheadAlgebraInConnectedCase}) we obtain $\mathrm{CE}(\mathfrak{g}) \cong \mathrm{CE}\bracket({\mathfrak{l}\SmlCyc S^4})$, dually an isomorphism of $L_\infty$-algebras $\beta : \mathfrak{l}\SmlCyc S^4 \xrightarrow{\sim} \mathfrak{g}$. Moreover, $\mathfrak{g}$ is \emph{the} Whitehead-bracket dg-Lie algebra of $\SmlCyc S^4$ in the sense of \cref{WHBracketLinfinityAlgebras}: if $\mathfrak{a}$ is any dg-Lie algebra with $\mathrm{CE}(\mathfrak{a}) \cong \mathrm{CE}(\mathfrak{g})$, then $\mathfrak{a}$ has vanishing differential by minimality, so that the linear component of the dual $L_\infty$-isomorphism respects the binary brackets and is a strict isomorphism $\mathfrak{a} \cong \mathfrak{g}$. (As a consistency check, $\dim \mathfrak{g}_{k-1} = 1,1,1,0,1,1,1,0$ for $k = 2,\dots,9$ reproduces the free ranks of $\pi_k\bracket({\SmlCyc S^4})$ from \cite[Prop.~3.12]{GSS26-KfromM}.)

\item 
\label{Step6BorelProjection}
\textbf{The Borel projection.}
Under \cref{eq:homotopy_pushout}, the constant-loop section $s_{w_0} : BS^1 \to \SmlCyc S^4$ at $w_0 \in S^4$ is the inclusion of $\{w_0\}\times BS^1 \subset S^4 \times BS^1$ followed by the structure map of the pushout. It is therefore modeled by $\rho$ followed by restriction, equivalently by $\theta$ followed by $\mathrm{pr}_C$ and by $C \to \R[f_2]$, $x_4 \mapsto 0$. By the formula for $\mathrm{pr}_C\circ\theta$ in \cref{Step5TheComparisonMap}, this composite sends $f_2 \mapsto f_2$ and all other generators to zero. This is the dgc-algebra map dual to the strict map $\R\varphi_1 \to \mathfrak{g}$, $\varphi_1 \mapsto p(\varphi_1)$ \cref{RationalModelsOfBS1AndS3}. 

Moreover, since the Borel projection $\mathrm{pr} : \SmlCyc S^4 \to BS^1$ satisfies $\mathrm{pr}\circ s_{w_0} = \mathrm{id}$, and since $s_{w_0}^\ast$ maps the 1-dimensional space $H^2\bracket({\SmlCyc S^4;\R})$, spanned by $[\rho(f_2)]$, isomorphically onto $H^2(BS^1;\R)$, we have $\mathrm{pr}^\ast f_2 = [\rho(f_2)]$; that is, $\R[f_2] \hookrightarrow \mathrm{CE}(\mathfrak{g}) \xrightarrow{\rho} \Omega^\bullet_{\mathrm{pdR}}\bracket({\SmlCyc S^4})$ represents the Borel projection.

\item 
\label{Step7Normalization}
\textbf{Normalization.}
Write $\SmlCyc S^4 = ES^1 \times_{S^1} L_1 S^4$, with $L_1 S^4$ as in \cite[Def.~3.3]{GSS26-KfromM} and $S^1$ acting by $(t\cdot\ell)(\sigma) = \ell(\sigma+t)$. The loops $\Gamma(\sigma_0, s)$ are round circles of radius $(1+\vert s\vert^2)^{-1/2} \leq 1$ traversed once at constant speed, hence lie in $L_1 S^4$, and so, for a fixed $e_0 \in ES^1$, the $S^3$-family of the lemma is the map
\[
  j
  :
  S^3 \cong \R^3_{\cpt}
  \longrightarrow
  \SmlCyc S^4
  \,,
  \quad
  x \longmapsto [e_0,\Gamma(0,x)]
  \mathrlap{\,.}
\]
Every dgc-algebra map $\mathrm{CE}(\mathfrak{g}) \to \Lambda(h_3)$ (with $\mathrm{d}h_3 = 0$) sends all generators other than $h_3$ to zero, for degree reasons, and sends $h_3 \mapsto \lambda h_3$ for some $\lambda \in \R$, by which it is determined. For the map representing $j$ we have $\lambda = \bigl\langle j^\ast [\rho(h_3)], [S^3]\bigr\rangle$, and we now show that $\lambda \neq 0$.

Let $\pi : ES^1 \times L_1 S^4 \to \SmlCyc S^4$ be the principal $S^1$-bundle of the Borel construction, let $\mathrm{ev} : ES^1 \times L_1 S^4 \to S^4$, $(e,\ell)\mapsto \ell(0)$, and let $\omega := \pi_!\bracket({\mathrm{ev}^\ast x_4}) \in H^3\bracket({\SmlCyc S^4;\R})$ be the fiber integration. The map $\tilde{\jmath} : S^1 \times S^3 \to ES^1\times L_1 S^4$, $(t,x) \mapsto t\cdot\bracket({e_0, \Gamma(0,x)})$, is $S^1$-equivariant over $j$ and exhibits $S^1\times S^3$ as the pullback bundle $j^\ast\bracket({ES^1 \times L_1 S^4})$; by the rotation-equivariance of $\Gamma$ we have $\mathrm{ev}\circ\tilde{\jmath}(t,x) = \Gamma(0,x)(t) = \Gamma(t,x)(0)$. By naturality of fiber integration, it follows that
\[
  \bigl\langle \omega, j_\ast[S^3] \bigr\rangle
  \;=\;
  \pm
  \deg\bracket({
    \mathrm{ev}_0 \circ \Gamma \circ \kappa
  })
  \mathrlap{\,,}
\]
where $\kappa : S^1 \times \R^3_{\cpt} \to (S^1\times\R^3)_{\cpt}$ is the quotient map. This degree is $\pm1$: the last coordinate of $\Gamma(\sigma_0,s)(0)$ equals $-1$ iff $\cos(2\pi\sigma_0) = -1-2\vert s\vert^2$, so the south pole has the single preimage $(\sigma_0,s) = (\tfrac12,0)$ (the points at infinity map to the north pole), at which the differential, in the coordinates $(x^1,\dots,x^4)$ around the south pole, is $\mathrm{diag}(-2\pi,2,2,2)$, hence invertible. (This is the degree condition in \cref{CharacterizationOfMinimalCyclification}; here we only need that it is nonzero.) Since $H^3\bracket({\SmlCyc S^4;\R})$ is spanned by $[\rho(h_3)]$, the class $\omega$ is a multiple of $[\rho(h_3)]$, and therefore $\lambda \neq 0$.

To normalize $\lambda$, note that for $c \in \R^\times$ the presentation \cref{ThePositiveIIABorcherdsAlgebra} provides an automorphism $\tau_c$ of $\bIIA$ with $\varphi_1 \mapsto \varphi_1$ and $\eta_2 \mapsto c\,\eta_2$. It sends $\eta_8 \mapsto c^3 \eta_8$, hence descends to $\mathfrak{g}$, where it fixes $\varphi_1$ and preserves $\mathfrak{k}$. Therefore its dual $\tau_c^\ast$ fixes $f_2$, preserves the span of the remaining generators, and sends $h_3 \mapsto c\,h_3$. Replacing $\rho$ by $\rho\circ\tau^\ast_{1/\lambda}$, and correspondingly $\beta$ by $\tau_{1/\lambda}\circ\beta$, achieves $\lambda = 1$ and leaves \cref{Step6BorelProjection} intact. Now $j$ is represented by $h_3 \mapsto h_3$ with all other generators sent to zero, which is the dgc-algebra map dual to the strict map $\R\eta_2 \to \mathfrak{g}$, $\eta_2 \mapsto p(\eta_2)$ \cref{RationalModelsOfBS1AndS3}.

\item
\label{Step8TheSquare}
\textbf{The square \cref{eq:two-models}.}
Under the equivalence $\MinCyc S^4 \simeq BS^1 \vee S^3$ \cref{MinCycEquivalentToWedgeSum}, the wedge summands are included via the constant loop at $\infty$, that is at $w_0 = (0,0,0,0,1)$, and via the slice $\sigma_0 = 0$, so that their composites with $\TheInclusion$ are $s_{w_0}$ and $j$. By \cref{Step6BorelProjection,Step7Normalization}, the restrictions of $\beta\circ\mathfrak{l}(\TheInclusion)\circ\alpha^{-1}$ along $\R\varphi_1 \hookrightarrow \bIIA$ and $\R\eta_2 \hookrightarrow \bIIA$ are, up to homotopy, the strict maps $\varphi_1 \mapsto p(\varphi_1)$ and $\eta_2 \mapsto p(\eta_2)$, which are also the restrictions of $p$. Since $BS^1 \vee S^3$ is a homotopy coproduct, maps out of it are determined up to homotopy by these two restrictions. Hence $\beta\circ\mathfrak{l}(\TheInclusion) \simeq p \circ \alpha$, which is the commutativity of \cref{eq:two-models}, in the sense in which $\mathfrak{l}(\TheInclusion)$ is defined.
\qedhere

\end{enumerate}

\end{proof}

\subsection
{
  \texorpdfstring
  {For \cref{MassiveBorcherds}}
  {For the Massive IIA Hierachy}
}
\label{ProofOfSecondMainTheorem}

\subsubsection
{Proof of \cref{prop:RationalModelOfMassiveMinCyc}}
\label{proof:RationalModelOfMassiveMinCyc}

\begin{proof}
By \cref{WHBracketLinfinityAlgebras}, the claim \cref{MassiveMinCycWhiteheadAlgebra} consists of the following two statements, both with respect to the component labels $\ell : \MinCyc_m S^4 \mapsto m$ \cref{TheComponentLabels}:
\begin{enumerate}[label=\textbf{\textup{(\Alph*)}}]
  \item
  \label{MassiveMinCycExistence}
  \emph{Existence:}
  $\mathrm{CE}\bigl(\bIIAMassive\bigr)$ admits a comparison map $\rho$ exhibiting it as a minimal Sullivan model over $\mathbb{Z}$ of $\MassMinCyc S^4$ \cref{MassMinCyc}, in the sense of \cref{MinimalSullivanModel} \cref{MinimalSullivanModelOfDisconnected}.

  \item
  \label{MassiveMinCycUniqueness}
  \emph{Uniqueness:}
  For every such minimal Sullivan model $\rho' : \mathrm{CE}(\mathfrak{a}) \to \Omega^\bullet_{\mathrm{pdR}}\bigl(\MassMinCyc S^4\bigr)$ with $\mathfrak{a}$ a dg-Lie algebra \cref{SubcategoryOfDGLieAlgebras}, there is an isomorphism of graded Lie algebras $\mathfrak{a} \cong \bIIAMassive$ fixing $\Romans$, hence dually an isomorphism $\mathrm{CE}\bigl(\bIIAMassive\bigr) \cong \mathrm{CE}(\mathfrak{a})$ fixing $f_0$.
\end{enumerate}

\medskip
\noindent
\emph{Proof of \cref{MassiveMinCycExistence}.}
\begin{enumerate}
\item
\label{StructureOfMassiveIIABorcherds}
\emph{Structure of $\bIIAMassive$.}
Since $\bIIA$ \cref{ThePositiveIIABorcherdsAlgebra} is the free graded Lie algebra on $\varphi_1$ and $\eta_2$ modulo the single relation $[\varphi_1, \varphi_1] = 0$, a derivation of $\bIIA$ is fixed by its values on $\varphi_1$ and $\eta_2$, subject only to preserving this relation. Hence the derivation $D := [\Romans, \text{-}]$ \cref{TheMassiveDerivation} is well-defined, since $D[\varphi_1, \varphi_1] = 2[D\varphi_1, \varphi_1] = 0$; and it squares to zero, since $D^2 = \tfrac{1}{2}[D,D]$ is again a derivation, which vanishes on $\varphi_1$ and $\eta_2$. Therefore the relations in \cref{TheMassiveIIABorcherdsAlgebra} say that $\bIIAMassive$ is the semidirect product
\begin{equation}
  \label{MassiveBorcherdsAsSemidirectProduct}
  \bIIAMassive
  \;\cong\;
  \mathbb{R}\langle \Romans \rangle \ltimes_D \bIIA
  \,,
  \qquad
  [\Romans, \Romans] = 0
  \,,
  \quad
  [\Romans, x] = D x
  \,,
\end{equation}
with vanishing differential. In particular, $\bigl(\bIIAMassive\bigr)_{-1} = \mathbb{R}\langle\Romans\rangle$ and $\bigl(\bIIAMassive\bigr)_{0} = 0$, and $\bIIAMassive$ is $(-1)$-connective and of finite type \cref{ConnectiveFiniteType}, being generated by finitely many elements of degree $\geq -1$.

Dually, write $f_0$ for the CE-generator dual to $\Romans$, and name the CE-generators dual to a homogeneous basis of $\bIIA$ after the corresponding flux densities \cref{UnifiedFluxDensity}. Then \cref{MassiveBorcherdsAsSemidirectProduct} says that
\begin{equation}
  \label{MassiveCEDifferential}
  \mathrm{CE}\bigl(\bIIAMassive\bigr)
  \;\cong\;
  \Bigl(
    \mathbb{R}[f_0] \otimes \mathrm{CE}\bigl(\bIIA\bigr)
    ,\;
    \mathrm{d} = \mathrm{d}_{\mathrm{IIA}} + f_0\, \delta
  \Bigr)
  \,,
  \qquad
  \mathrm{d}\, f_0 = 0
  \,,
\end{equation}
where $\mathrm{d}_{\mathrm{IIA}}$ is the CE-differential of $\bIIA$ and $\delta$ is the derivation dual to $D$, which is linear on generators (with the sign from \cref{CEDifferentialInComponents}). For example, $\mathrm{d}\, f_2 = f_0\, h_3$ and $\mathrm{d}\, h_7 = \tfrac{1}{2} f_4\, f_4 - f_2\, f_6 + f_0\, f_8$, cf. \cref{OverviewDiagramMassive}. Here $\mathrm{d}\, f_0 = 0$ because no bracket in $\bIIAMassive$ has a component along $\Romans$, since that would require a bracket of an element of degree $-1$ with one of degree $0$.

\item
\label{WellOrderingOfMassiveIIAGenerators}
\emph{Minimality.}
Since $\bIIAMassive$ has vanishing differential, the CE-differential \cref{MassiveCEDifferential} has no linear part, and $\mathrm{d}\, f_0 = 0$ was just noted. It remains to exhibit a well-ordered basis as in \cref{ConditionsForMinimalityOverPolynomialDegree0}. To this end, assign to each bracket word its \emph{RR-weight}, namely the number of its letters from $\{\Romans, \varphi_1\}$, so that $\eta_2$ has weight $0$ and all $\varphi_{2k+1}$ have weight $1$. The relations in \cref{TheMassiveIIABorcherdsAlgebra} are homogeneous in this weight, so that the brackets are additive in it. Choose the basis of $\bIIA$ homogeneous in degree and weight, give each dual CE-generator $\alpha$ the weight $\mathrm{wt}(\alpha)$ of its dual basis element, and set
\begin{equation}
  N(\alpha)
  \;:=\;
  \mathrm{deg}(\alpha) + 2\, \mathrm{wt}(\alpha)
  \;\in\;
  \mathbb{N}
  \,.
\end{equation}
Then $N$ strictly decreases from $\alpha$ to every generator appearing in $\mathrm{d}\,\alpha$:
\begin{itemize}
  \item a summand $f_0\, \beta$ in $\mathrm{d}\, \alpha$ has $\mathrm{wt}(\beta) = \mathrm{wt}(\alpha) - 1$ and $\mathrm{deg}(\beta) = \mathrm{deg}(\alpha) + 1$, hence $N(\beta) = N(\alpha) - 1$;
  \item a summand $\beta\, \gamma$ in $\mathrm{d}\, \alpha$ with $\beta, \gamma \neq f_0$ has $\mathrm{wt}(\beta) + \mathrm{wt}(\gamma) = \mathrm{wt}(\alpha)$ and $\mathrm{deg}(\beta) + \mathrm{deg}(\gamma) = \mathrm{deg}(\alpha) + 1$ with $\mathrm{deg}(\gamma) \geq 2$, hence $N(\beta) \leq N(\alpha) - 1$, and symmetrically for $\gamma$.
\end{itemize}
Since $N(\alpha) \geq \mathrm{deg}(\alpha)$, each level of $N$ is finite-dimensional, so that ordering the generators by $N$ (and arbitrarily within each level) gives the required well-ordered basis of $\bigl(\bIIAMassive\bigr)^\vee_{\geq 0}$. Hence $\mathrm{CE}\bigl(\bIIAMassive\bigr)$ is minimal over $\mathbb{R}[f_0]$ at $f_0 = 0$ in the sense of \cref{MinimalCEAlgebra} \cref{MinimalCEAlgebraOverPolynomialDeg0}.
(Ordering by degree alone does not work, due to the summand $f_0\, f_8$ in $\mathrm{d}\, h_7$, while $N(f_8) = 10 < 11 = N(h_7)$.)

\item
\label{FibersOfMassiveIIABorcherds}
\emph{The fibers.}
For $m \in \mathbb{Z}$, setting $f_0 = m$ in \cref{MassiveCEDifferential} gives
\begin{equation}
  \label{FibersOfMassiveCE}
  \begin{aligned}
  &
  \mathrm{CE}\bigl(\bIIAMassive\bigr)\big/(f_0 - m)
  \;\cong\;
  \bracket({
    \mathrm{CE}\bigl(\bIIA\bigr)
    ,\;
    \mathrm{d}_{\mathrm{IIA}} + m\, \delta
  })
  \;=\;
  \mathrm{CE}\bracket({
    \bracket({\bIIA})^{(m)}
  })
  \mathrlap{\,,}
  \\
  &
  \text{where }
  \bracket({\bIIA})^{(m)}
  :=
  \bigl(\bIIA,\; \partial := m D\bigr)
  \,,
  \end{aligned}
\end{equation}
the CE-algebra of the dg-Lie algebra with the brackets of $\bIIA$ and with $\partial\, \eta_2 = -m\, \varphi_1$, $\partial\, \varphi_1 = 0$. (This is the twist of $\bIIAMassive$ by the Maurer--Cartan element $m \Romans$, restricted to non-negative degrees.) The function $N$ from \cref{WellOrderingOfMassiveIIAGenerators} still strictly decreases along $\mathrm{d}_{\mathrm{IIA}} + m\,\delta$, so that \cref{FibersOfMassiveCE} is a (generally non-minimal) Sullivan algebra.

\item
\emph{The fibers model the components.}
By \cref{RationalModelsOfBS1AndS3}, the abelian dg-Lie algebra $\langle \varphi_1 \rangle$ (with $[\varphi_1,\varphi_1] = 0$) is a dg-Lie model of $B S^1$, with $\varphi_1$ representing the generator of $\pi_2(BS^1) \otimes \mathbb{R}$ that pairs to unity with the universal first Chern class. Since $B S^1$ is coformal, a free Quillen model $\mathbb{L}(a_1, a_3, \cdots)$ of $BS^1$ (where $\mathbb{L} := \mathrm{Free}_{\mathrm{dgL}}$) admits a quasi-isomorphism to $\langle \varphi_1 \rangle$ sending $a_1 \mapsto \varphi_1$ and $a_{>1} \mapsto 0$. The attaching map in \cref{MinCycNCellAttachment} has Chern number $m$, hence is represented by the cycle $m\, a_1$, so that, by the Lie model of cell attachments \parencites[\S 24]{FHT2000}[\S 2]{Bubenik2008}, the space $\MinCyc_m S^4$ \cref{MinCycN} is modeled by $\mathbb{L}(a_1, a_3, \cdots) \sqcup \mathbb{L}(b_2)$ with $\partial\, b_2 = m\, a_1$. Extending by $b_2 \mapsto -\eta_2$ gives a dg-Lie homomorphism
\begin{equation}
  \label{QuillenModelOfMassiveComponents}
  \begin{tikzcd}
    \mathbb{L}(a_1, a_3, \cdots) \sqcup \mathbb{L}(b_2)
    \ar[r]
    &
    \langle \varphi_1 \rangle \sqcup \mathbb{L}(\eta_2)
    \;=\;
    \bracket({\bIIA})^{(m)}
    \mathrlap{\,,}
  \end{tikzcd}
\end{equation}
where on the right $\partial\, \eta_2 = -m\, \varphi_1$; it respects the differentials since $\partial\, b_2 = m\, a_1 \mapsto m\, \varphi_1 = \partial(-\eta_2)$. It is a quasi-isomorphism: filter both sides by the number of occurrences of $b_2$, respectively of $\eta_2$, which is bounded in each degree. On the associated graded the differential no longer involves $m$, and the map becomes the free product of the quasi-isomorphism $\inlinetikzcd{\mathbb{L}(a_1, a_3, \cdots) \ar[r] \& \langle \varphi_1 \rangle}$ with the isomorphism $\mathbb{L}(b_2) \to \mathbb{L}(\eta_2)$, which is a quasi-isomorphism since free products preserve quasi-isomorphisms (cf. \cite[proof of Thm.~A.4]{LazarevMarkl2015}).

All dg-Lie algebras in \cref{QuillenModelOfMassiveComponents} are concentrated in positive degrees and of finite type, so that passing to CE-algebras preserves quasi-isomorphisms \cite[\S\S 22--23]{FHT2000}, and the CE-algebra of a Quillen model of a space is quasi-isomorphic to its piecewise de Rham algebra (cf. \cite[\S\S 22--24]{FHT2000}). Therefore the Sullivan algebra $\mathrm{CE}\bigl(\bracket({\bIIA})^{(m)}\bigr)$ \cref{FibersOfMassiveCE} is connected by a zigzag of quasi-isomorphisms to $\Omega^\bullet_{\mathrm{pdR}}\bigl(\MinCyc_m S^4\bigr)$ \cref{EndFormulaForPiecewiseDeRhamAlgebra}. Since Sullivan algebras are cofibrant \cite[\S 12]{FHT2000}, this zigzag may be replaced by a direct quasi-isomorphism
\begin{equation}
  \label{ComponentComparisonMaps}
  \begin{tikzcd}
    \psi_m
    \;:\;
    \mathrm{CE}\bigl(\bracket({\bIIA})^{(m)}\bigr)
    \ar[r, "{ \sim }"]
    &
    \Omega^\bullet_{\mathrm{pdR}}\bigl(\MinCyc_m S^4\bigr)
    \mathrlap{\,.}
  \end{tikzcd}
\end{equation}

\item
\emph{The comparison map.}
Since every singular simplex in a disjoint union lands in a single summand, we have
$
  \Omega^\bullet_{\mathrm{pdR}}\bigl(\MassMinCyc S^4\bigr)
  \cong
  \prod_{m} \Omega^\bullet_{\mathrm{pdR}}\bigl(\MinCyc_m S^4\bigr)
$.
With $q_m$ the quotient maps \cref{FibersOfMassiveCE}, set
\begin{equation}
  \begin{tikzcd}
    \rho
    :=
    \bigl(\psi_m \circ q_m\bigr)_{m \in \mathbb{Z}}
    \;:\;
    \mathrm{CE}\bigl(\bIIAMassive\bigr)
    \ar[r]
    &
    \prod_{m \in \mathbb{Z}}
    \Omega^\bullet_{\mathrm{pdR}}\bigl(\MinCyc_m S^4\bigr)
    \mathrlap{\,.}
  \end{tikzcd}
\end{equation}
This is a dgc-algebra homomorphism with $\rho(f_0) = (m)_{m \in \mathbb{Z}} = \ell$ \cref{TheComponentLabels}, and the map it induces on the $m$th fiber is the quasi-isomorphism $\psi_m$ \cref{ComponentComparisonMaps}, as required in \cref{FiberwiseQuasiIsos}.
\end{enumerate}
Together with \cref{WellOrderingOfMassiveIIAGenerators}, this proves \cref{MassiveMinCycExistence}. On the $0$-component, \cref{FibersOfMassiveCE} recovers $\mathrm{CE}\bigl(\bIIA\bigr)$, the minimal model of $\MinCyc S^4$ from \cref{prop:RationalModelOfMinCyc}.

\medskip
\noindent
\emph{Proof of \cref{MassiveMinCycUniqueness}.}
Let $\rho' : \mathrm{CE}(\mathfrak{a}) \to \Omega^\bullet_{\mathrm{pdR}}\bigl(\MassMinCyc S^4\bigr)$ be a minimal Sullivan model over $\mathbb{Z}$ with respect to $\ell$, with $\mathfrak{a}$ a dg-Lie algebra, and write $\Romans \in \mathfrak{a}_{-1}$ for the element dual to $f_0$.
\begin{enumerate}
\item
\emph{$\mathfrak{a}$ is a semidirect product.}
The linear part of $\mathrm{d}$ on $\wedge^1 \mathfrak{a}^\vee$ is dual to the unary bracket, which hence vanishes by \cref{ConditionsForMinimalityOverPolynomialDegree0}. So $\mathfrak{a}$ is a graded Lie algebra, with $\mathfrak{a}_{-1} = \mathbb{R}\langle \Romans \rangle$ and $\mathfrak{a}_{< -1} = 0$. The condition $\mathrm{d}\, f_0 = 0$ in \cref{ConditionsForMinimalityOverPolynomialDegree0} says that no bracket has a component along $\Romans$, so that $\mathfrak{a}_{\geq 0} \subset \mathfrak{a}$ is an ideal, while $[\Romans, \Romans] \in \mathfrak{a}_{-2} = 0$. Hence
\begin{equation}
  \label{UniquenessSemidirectProduct}
  \mathfrak{a}
  \;\cong\;
  \mathbb{R}\langle \Romans \rangle \ltimes_{D_{\mathfrak{a}}} \mathfrak{a}_{\geq 0}
  \,,
  \qquad
  D_{\mathfrak{a}} := [\Romans, \text{-}]
  \,,
\end{equation}
with $D_{\mathfrak{a}}$ a derivation of $\mathfrak{a}_{\geq 0}$ of degree $-1$.

\item
\emph{The massless part is $\bIIA$.}
Setting $f_0 = 0$ discards exactly the brackets with $\Romans$, so that $\mathrm{CE}(\mathfrak{a})/(f_0) \cong \mathrm{CE}(\mathfrak{a}_{\geq 0})$. By \cref{MinimalityInZeroComponent}, this is a minimal Sullivan model of $\MinCyc_0 S^4 \simeq \MinCyc S^4$ \cref{MinCycAsHomotopyCofiber}, and so, by \cref{prop:RationalModelOfMinCyc} \cref{RationalMinCycIsPositiveBorcherds} and the uniqueness in \cref{WhiteheadAlgebraInConnectedCase}, there is an isomorphism of dgc-algebras $\phi : \mathrm{CE}\bigl(\bIIA\bigr) \xrightarrow{\sim} \mathrm{CE}(\mathfrak{a}_{\geq 0})$. Since both CE-differentials are purely quadratic, the word-length-$2$ component of $\phi \circ \mathrm{d} = \mathrm{d} \circ \phi$ on generators says that the (invertible) linear component of $\phi$ also respects the differentials, hence is dual to an isomorphism of graded Lie algebras
\begin{equation}
  \sigma_0
  \;:\;
  \mathfrak{a}_{\geq 0}
  \xrightarrow{\;\sim\;}
  \bIIA
  \,.
\end{equation}

\item
\emph{The mass derivation is a multiple of $D$.}
The transported derivation $\sigma_0 \circ D_{\mathfrak{a}} \circ \sigma_0^{-1}$ of $\bIIA$ has degree $-1$, while $\bIIA$ is generated by $\varphi_1$ and $\eta_2$, with $\bigl(\bIIA\bigr)_0 = 0$ and $\bigl(\bIIA\bigr)_1 = \mathbb{R}\langle \varphi_1 \rangle$. Hence it sends $\varphi_1 \mapsto 0$ and $\eta_2 \mapsto -\lambda\, \varphi_1$ for a unique $\lambda \in \mathbb{R}$, which is to say that, with $D$ from \cref{StructureOfMassiveIIABorcherds},
\begin{equation}
  \sigma_0 \circ D_{\mathfrak{a}} \circ \sigma_0^{-1}
  \;=\;
  \lambda\, D
  \,.
\end{equation}

\item
\emph{The mass derivation does not vanish.}
If $\lambda = 0$, then $D_{\mathfrak{a}} = 0$, so that the fiber $\mathrm{CE}(\mathfrak{a})/(f_0 - 1) \cong \mathrm{CE}(\mathfrak{a}_{\geq 0}) \cong \mathrm{CE}\bigl(\bIIA\bigr)$ has $H^2 = \mathbb{R}\bracket\langle{ [f_2] }\rangle \neq 0$. But by \cref{FiberwiseQuasiIsos} this fiber is quasi-isomorphic to $\Omega^\bullet_{\mathrm{pdR}}\bigl(\MinCyc_1 S^4\bigr)$, whose cohomology vanishes in degree 2: in the cell structure \cref{MinCycNCellAttachment} of $\MinCyc_1 S^4$, with $B S^1 \simeq \mathbb{C}P^\infty$ having one cell in each even dimension, the 3-cell is attached by a map of degree 1 onto the 2-skeleton $\mathbb{C}P^1 \simeq S^2$, so that $H_2\bigl(\MinCyc_1 S^4; \mathbb{Z}\bigr) = 0$. Hence $\lambda \neq 0$.

\item
\emph{Rescaling.}
Let $\sigma_\lambda$ be the automorphism of $\bIIA$ given by $\varphi_1 \mapsto \varphi_1$, $\eta_2 \mapsto \lambda\, \eta_2$ (well-defined since the relation $[\varphi_1, \varphi_1] = 0$ is homogeneous). Then $\sigma_\lambda \circ (\lambda D) \circ \sigma_\lambda^{-1} = D$, since both sides are derivations that vanish on $\varphi_1$ and send $\eta_2 \mapsto -\varphi_1$. Therefore $\sigma := \sigma_\lambda \circ \sigma_0$ satisfies $\sigma \circ D_{\mathfrak{a}} = D \circ \sigma$, so that
\begin{equation}
  \begin{tikzcd}[
    sep=0pt
  ]
  \mathfrak{a}
  &\overset{\mathclap{\text{%
    \cref{UniquenessSemidirectProduct}
  }}}{\cong}
  &
  \mathbb{R}\langle \Romans \rangle \ltimes_{D_{\mathfrak{a}}} \mathfrak{a}_{\geq 0}
  \ar[
    rr,
    "{ \sim }"
  ]
  &&
  \mathbb{R}\langle \Romans \rangle \ltimes_{D} \bIIA
  &
    \overset{\mathclap{\text{%
      \cref{MassiveBorcherdsAsSemidirectProduct}
    }}}{\cong}
  &
  \bIIAMassive
  \\
  &&
  \Romans 
    &\longmapsto& 
  \Romans
  \\
  &&
  x 
    &\longmapsto& 
  \sigma(x)
  \end{tikzcd}
\end{equation}
is an isomorphism of graded Lie algebras fixing $\Romans$. Its CE-dual is an isomorphism $\mathrm{CE}\bigl(\bIIAMassive\bigr) \xrightarrow{\sim} \mathrm{CE}(\mathfrak{a})$ fixing $f_0$.
\end{enumerate}
This proves \cref{MassiveMinCycUniqueness}. (Only the fibers over $m = 0$ and $m = 1$ entered here, while the condition that $\mathfrak{a}$ is a dg-Lie algebra was essential.)
\end{proof}

\subsubsection
{Proof of \cref{TheMassiveComparisonMap}}
\label{proof:TheMassiveComparisonMap}

\begin{proof}
Write $X_m := \MinCyc_m S^4 \simeq \AttachingSpace{m}$ \cref{MinCycNCellAttachment}, let $\inlinetikzcd{c_m : X_m \ar[r] \& X_m/BS^1 \cong S^3}$ be the map collapsing $BS^1$, and let $\sigma_3 \in H^3(S^3;\mathbb{Z})$ be a generator. For $m = 0$ we use $X_0 = \MinCyc S^4 \simeq BS^1 \vee S^3$ \cref{MinCycEquivalentToWedgeSum}, whose summands are included into $\SmlCyc S^4$ via $\TheInclusion$ as the constant-loop section $s_{w_0}$ at $w_0 = (0,0,0,0,1)$ and as the $S^3$-family $j$ of \cref{lem:minimal_cyclification_and_borcherds}; here $c_0$ is the projection onto $S^3$. We use the notation for universal brane charges introduced in the proof outline after \cref{TheMassiveComparisonMap}, and $L$ denotes the universal complex line bundle over $BS^1$, with $[L] = f_2^C$ \cref{ClassOfL}.

\begin{enumerate}

\item
\label{Step1TheTwist}
\textbf{The twist.}
By \cite[(61)]{GSS26-KfromM}, the pullback of $h_3^C$ along $L_1 S^4 \to \SmlCyc S^4$ is the $B^3\mathbb{Z}$-component of $L_1S^4 \to L S^4 \to L B^4\mathbb{Z} \simeq B^4\mathbb{Z}\times B^3\mathbb{Z}$, where the $B^4\mathbb{Z}$-component is evaluation at $0$. Every such equivalence over the evaluation map has as its $B^3\mathbb{Z}$-component $\pm$ the generator of $H^3(LB^4\mathbb{Z};\mathbb{Z}) \cong \mathbb{Z}$ given by the slant product $\mathrm{ev}^\ast\iota_4/[S^1]$ of the evaluation class with the circle. Since $j$ factors through $L_1 S^4$ via $j' : x \mapsto \Gamma(0,x)$, and since $\mathrm{ev}\bracket({t, j'(x)}) = \Gamma(0,x)(t) = \Gamma(t,x)(0)$, we get
\[
  \bigl\langle j^\ast h_3^C, [S^3] \bigr\rangle
  \;=\;
  \pm
  \deg\bracket({ \mathrm{ev}_0\circ\Gamma\circ\kappa })
  \;=\;
  \pm 1
  \mathrlap{\,,}
\]
with $\kappa : S^1\times\R^3_{\cpt} \to (S^1\times\R^3)_{\cpt}$ the quotient map, by the degree computation in \cref{Step7Normalization} of the proof of \cref{lem:minimal_cyclification_and_borcherds}. We orient $S^3$ such that $j^\ast h_3^C = \sigma_3$; with this orientation, normalization (\cref{InducingQuotient} in \cref{lem:minimal_cyclification_and_borcherds}) says that the generator $h_3$ represents the real image of $h_3^C$.

The twist $h_3^C$ in \cref{MassiveComparisonMap} is, on $X_0$, the composite $h_3^C \circ \TheInclusion$ and, on $X_m$ for $m \neq 0$, the composite of $c_m$ with a map $S^3 \to B^3\mathbb{Z}$ representing $\sigma_3$. In both cases $[h_3^C] = c_m^\ast \sigma_3 \in H^3(X_m;\mathbb{Z})$. We trivialize the twist on the two pieces of the pushout \cref{MinCycNCellAttachment}: on $D^3$ uniquely, on $BS^1$ for $m \neq 0$ by the constancy of $c_m$ there, and on $BS^1$ for $m = 0$ by the trivialization of $h_3^C$ on the constant-loop leg $S^4 \times BS^1 \ni (w_0, -)$ used in \cite[Rem.~3.18]{GSS26-KfromM}. 

The two trivializations differ on $S^2$ by a transition line bundle $G$, whose Chern class $t \in H^2(S^2;\mathbb{Z}) \cong \mathbb{Z}$ is sent to $[h_3^C]$ by the Mayer--Vietoris connecting map $\delta : H^2(S^2;\mathbb{Z}) \to H^3(X_m;\mathbb{Z})$. For $m = 0$, $\delta$ is an isomorphism, the attaching map being null; for $m \neq 0$, $G$ is pulled back along $c_m$ from the corresponding transition bundle of $\sigma_3$ on $S^3 = D^3 \cup_{S^2} \ast$, where $\delta$ is an isomorphism. In either case $t = \epsilon$ for some sign $\epsilon \in \{\pm1\}$, that is
$
  G = H^{\epsilon}
$,
with $H$ the tautological line bundle on $S^2 = \mathbb{C}P^1$.

\item
\label{Step2TwistedKTheory}
\textbf{Twisted K-theory of $X_m$.}
Maps over $B^3\mathbb{Z}$ into $(B\mathrm{U}\times\mathbb{Z})\sslash B\mathrm{U}(1)$ take the homotopy pushout \cref{MinCycNCellAttachment} to a homotopy pullback of mapping spaces. Under the above trivializations, these are $\mathrm{Map}(BS^1, B\mathrm{U}\times\mathbb{Z})$ and $\mathrm{Map}(D^3, B\mathrm{U}\times\mathbb{Z})$ over $\mathrm{Map}(S^2, B\mathrm{U}\times\mathbb{Z})$, the second map being restriction followed by tensoring with $G$. Since $\pi_1\,\mathrm{Map}(S^2, B\mathrm{U}\times\mathbb{Z}) \cong \mathrm{KU}^{-1}(S^2) = 0$ at every base point, $\pi_0$ of this homotopy pullback is the pullback of the $\pi_0$'s:
\[
  \mathrm{KU}^0_{h_3^C}(X_m)
  \;\cong\;
  \bigl\{
    (a, n) \in \mathrm{KU}^0(BS^1)\times\mathbb{Z}
    \;\big\vert\;
    m^\ast a = n\cdot G \in \mathrm{KU}^0(S^2)
  \bigr\}
  \mathrlap{\,,}
\]
with $n$ the rank. Here $\mathrm{KU}^0(BS^1) = \mathbb{Z}\llbracket L-1 \rrbracket$, by the Milnor sequence, since $\mathrm{KU}^1(\mathbb{C}P^k) = 0$; and $\mathrm{KU}^0(S^2) = \mathbb{Z}\oplus\mathbb{Z}(H-1)$ with $(H-1)^2 = 0$, so that $m^\ast L = H^m = 1 + m(H-1)$ and $G = 1 + \epsilon(H-1)$. For $a = \sum_k a_k (L-1)^k$, the condition therefore says $a_0 = n$ and $m\, a_1 = \epsilon\, n$; in particular, the rank of every class on $X_m$ is divisible by $m$. For $m = 0$, this says that restriction to $BS^1$ is the isomorphism \cref{PsiAsLMinusOne}, and we \emph{define} $\Psi_{\vert f_0^C = 0}$ as the preimage of $L - 1$ under it. For $m \neq 0$, we define $\Psi_{\vert f_0^C = m}$ as the class with $(a,n) = \bracket({m + \epsilon(L-1),\, m})$. Taken together, these define $\Psi$ in \cref{MassiveComparisonMap}, and the rank condition \cref{RankOfPhi} holds by construction.

\item
\label{Step3MasslessSector}
\textbf{The massless sector, (\cref{PhiMasslessSector}).}
Here $\Phi$ is the comparison map of \cref{ComparisonMapOnSmlCyc}, constructed in \cite[Rem.~3.18]{GSS26-KfromM}. By \cite[(91)]{GSS26-KfromM}, $\Phi$ restricts on the constant-loop locus $S^4 \times BS^1$ to $2(L-1) + \beta\bracket({[L]+1})$, with respect to the above trivialization. Its further restriction along $s_{w_0}$ is $2(L-1)$, since the Bott generator \cref{BottGenerator} vanishes at a point. Since $\TheInclusion \circ s_{w_0}$ is the inclusion of the summand $BS^1$, the isomorphism \cref{PsiAsLMinusOne} gives $\TheInclusion^\ast[\Phi] = 2[\Psi_{\vert f_0^C = 0}]$, which is \cref{DivisionBy2}. The halving is unique, $\widetilde{\mathrm{KU}}^0(BS^1)$ being torsion-free.

The D6-brane charge of a rank-0 class is its first Chern class, which does not depend on the trivialization of the twist. Since restriction $H^2(X_0;\mathbb{Z}) \to H^2(BS^1;\mathbb{Z})$ is an isomorphism, we get $f_2^K\bracket({\Psi_{\vert f_0^C = 0}}) = c_1(L - 1) = c_1(L) = \TheInclusion^\ast f_2^C$, by \cref{ClassOfL}.

For the D4-brane charge, note that the homotopy fiber of $f_2^C$ on $X_0 = (S^1\times\R^3)_{\cpt}\sslash S^1$ \cref{TheMinimalCyclificationOfS4} is $(S^1\times\R^3)_{\cpt} \simeq S^3 \vee S^4$, mapping by $\Gamma$ to the homotopy fiber $L_1 S^4$ of $f_2^C$ on $\SmlCyc S^4$. Since $H^1(S^3\vee S^4;\mathbb{Z}) = 0 = H^2(S^3\vee S^4;\mathbb{Z})$, the restrictions of $\Psi_{\vert f_0^C = 0}$ and of $\TheInclusion^\ast\Phi$ to this fiber factor uniquely through the homotopy fiber of $f_2^K$, where the D4-brane charge $f_4^K$ is additive \cite[Lem.~3.16]{GSS26-KfromM}. Therefore, by \cref{DivisionBy2} and \cite[(76c)]{GSS26-KfromM},
\[
  2\, f_4^K\bracket({\Psi_{\vert f_0^C = 0}})
  \;=\;
  f_4^K\bracket({\TheInclusion^\ast \Phi})
  \;=\;
  \Gamma^\ast\bracket({2 f_4^C})
  \;\in\;
  H^4(S^3\vee S^4;\mathbb{Z})
  \;\cong\;
  \mathbb{Z}
  \mathrlap{\,,}
\]
whence $f_4^K\bracket({\Psi_{\vert f_0^C = 0}}) = \Gamma^\ast f_4^C$. This is a generator, because $f_4^C$ is pulled back along the evaluation map \cite[(61)]{GSS26-KfromM} and $\mathrm{ev}_0 \circ \Gamma$ has degree $\pm 1$ (\cref{Step1TheTwist}).

\item
\label{Step4ChernCharacter}
\textbf{The Chern character, \cref{ChernCharacterOnMinCycS4}.}
We normalize the generators $f^K_{2k}$ of the minimal model $M_K := \mathrm{CE}\bracket({\mathfrak{l}(B\mathrm{U}\sslash B\mathrm{U}(1))})$ \cref{SullivanModelOfTwistedKTheory} such that they restrict on the fiber $B\mathrm{U}$ to the components $\mathrm{ch}_{2k}$ of the Chern character; this is compatible with $\mathrm{d}f^K_{2k+2} = h_3 f^K_{2k}$, as $\mathrm{ch}(V\otimes L) = \mathrm{ch}(V)\,e^{c_1(L)}$. Then the twisted Chern character of a twisted class is the induced map of Sullivan models relative to $\Lambda(h_3)$, the model of $B^3\mathbb{Z}$ \parencites[(106)--(107)]{GSS26-KfromM}{FSS23-Char}.

On the side of $X_0$, the minimal model is $\mathrm{CE}(\bIIA)$, by \cref{prop:RationalModelOfMinCyc}, with $h_3$ representing the twist (\cref{Step1TheTwist}) and $f_2$ representing $f_2^C = c_1(L)$. The models of the two summand inclusions combine to the quasi-isomorphism
\[
  \begin{tikzcd}[
    sep=0pt
  ]
  \mathrm{CE}(\bIIA)
  \ar[
    rr,
    "{ q }",
    "{ \sim }"{swap}
  ]
  &&
  \R[f_2] \times_{\R} \Lambda(h_3)
  \\
  f_2 &\longmapsto& (f_2, 0)
  \\
  h_3 &\longmapsto& (0, h_3)
  \mathrlap{\,,}
  \end{tikzcd}
\]
with all other generators sent to zero; it is a homomorphism, since $[\varphi_1,\varphi_1] = 0$, and a quasi-isomorphism, since it sends the classes $[f_2^k]$ and $[h_3]$ spanning $H^\bullet\bracket({\mathrm{CE}(\bIIA)}) \cong H^\bullet(BS^1\vee S^3;\R)$ to nonzero classes in a target of the same dimension in each degree.
Since $\bIIA$ is graded by the number of occurrences of $\varphi_1$, and its part of degree 1 in $\varphi_1$ is spanned by the elements $\varphi_{2k+1} = \mathrm{ad}^k_{\eta_2}\varphi_1$ \cref{LowDegreeBorcherdsElements}, the only brackets with a $\varphi_{2k+1}$-component, with respect to a basis homogeneous in this grading as in \cref{LowDegreeBorcherdsElements}, are $[\eta_2,\varphi_{2k-1}] = \varphi_{2k+1}$. Hence $\mathrm{d} f_{2k+2} = h_3 f_{2k}$ for all $k \geq 1$, that is
\[
  F
  \;:=\;
  \textstyle{\sum_{k\geq 1}} f_{2k}
  \qquad
  \text{satisfies}
  \qquad
  \mathrm{d}F = h_3 F
  \mathrlap{\,.}
\]

The class $\Psi_{\vert f_0^C = 0}$, being a map over $B^3\mathbb{Z}$, is modeled by a homomorphism $\psi : M_K \to \mathrm{CE}(\bIIA)$ under $\Lambda(h_3)$, well-defined up to homotopy under $\Lambda(h_3)$ \parencite[\S\S 14--15]{FHT2000}. The two components of $q \circ \psi$ are as follows. Into $\Lambda(h_3)$ it necessarily sends all $f^K_{2k}$ to zero, for degree reasons. Into $\R[f_2]$ (where $h_3 \mapsto 0$) it sends $f^K_{2k} \mapsto \pi_k f_2^k$ for some $\pi_k \in \R$, with $\pi_1 = 1$ because $f^K_2$ represents $c_1\bracket({\Psi_{\vert f_0^C = 0}}) = f_2^C$ (\cref{Step3MasslessSector}). The homomorphism
\[
  \begin{tikzcd}[
    sep=0pt
  ]
  M_K 
  \ar[
    rr,
    "{
      \psi'
    }"
  ]
  &&
  \mathrm{CE}(\bIIA)
  \\
  h_3 &\longmapsto& h_3
  \\
  f^K_{2k} &\longmapsto& c_{2k}
  \mathrlap{\,,}
  \end{tikzcd}
  \;\;\;\;
  \text{where}
  \;\;
  c
  \;:=\;
  \textstyle{\sum_{k\geq1}} \pi_k\, f_2^{k-1} F
  \mathrlap{\,,}
\]
is well-defined, since $\mathrm{d}c = h_3 c$. Moreover, $q\circ\psi' = q \circ \psi$, since $f_2^{k-1}F \mapsto (f_2^k, 0)$. By the relative lifting lemma \parencite[\S 14]{FHT2000}, $\psi' \simeq \psi$ under $\Lambda(h_3)$.

Now let $Y$ be the homotopy fiber of $(f_2^C, h_3^C) : X_0 \to B^2\mathbb{Z}\times B^3\mathbb{Z}$. Since $\pi_1(X_0) = 0$, $\pi_2(X_0) \cong \mathbb{Z}$ (from $BS^1$) and $\pi_3(X_0) \cong \mathbb{Z}$ (from $S^3$), and these classes are isomorphisms on $\pi_2$ and on $\pi_3$ (by \cref{Step1TheTwist}), $Y$ is 3-connected. Hence the twist is trivialized on $Y$ uniquely up to homotopy, and $\Psi_{\vert f_0^C = 0}$ restricts to an untwisted class on $Y$. A model of $Y$ is $\mathrm{CE}(\bIIA)\otimes\Lambda(b_1, b_2)$ with $\mathrm{d}b_1 = f_2$ and $\mathrm{d}b_2 = h_3$ \parencite[\S 15]{FHT2000}, where $b_2$ models the null-homotopy of the twist. Transport along this null-homotopy is modeled by the homotopy
\[
  \begin{tikzcd}[
    sep=0pt
  ]
  h_3 
    &\longmapsto&
  (1-t)\,h_3 - \mathrm{d}t\, b_2
  \\
  f^K_{2k} 
    &\longmapsto& 
  \bracket({ e^{-t b_2} c })_{2k}
  \mathrlap{\,,}
  \end{tikzcd}
\]
which respects the differentials, since $(1-t)h_3 - \mathrm{d}t\,b_2$ is closed and $\mathrm{d}\bracket({e^{-tb_2}c}) = \bracket({(1-t)h_3 - \mathrm{d}t\,b_2})\, e^{-tb_2} c$. At $t = 1$, it gives the model of the untwisted class:
\[
  \mathrm{ch}\bracket({
    \Psi_{
      \vert
      f_0^C, f_2^C, h_3^C = 0
    }
  })
  \;=\;
  \bigl[ e^{-b_2} c \bigr]
  \mathrlap{\,.}
\]
Finally, on $Y$ we have $(\mathrm{d} - h_3)(b_1 F) = f_2 F$, hence $f_2^{k-1}F = (\mathrm{d}-h_3)\bracket({f_2^{k-2}b_1F})$ for $k \geq 2$, and $e^{-b_2}(\mathrm{d}-h_3)(\text{-}) = \mathrm{d}\bracket({e^{-b_2}(\text{-})})$. With $\pi_1 = 1$ this gives $\bigl[e^{-b_2}c\bigr] = \bigl[e^{-b_2}F\bigr]$, whose components are the classes of the untwisted flux densities \cref{UntwistedRRFluxDensities}, with $B_2 = b_2$; and $\widetilde F_2 = f_2 = \mathrm{d}b_1$ is exact. Therefore
\[
  \mathrm{ch}\bracket({
    \Psi_{
      \vert
      f_0^C, f_2^C, h_3^C = 0
    }
  })
  \;=\;
  \widetilde F_4 + \widetilde F_6 + \widetilde F_8 + \widetilde F_{10} + \widetilde F_{12} + \cdots
  \mathrlap{\,,}
\]
which is \cref{ChernCharacterOfPsi}.
\qedhere

\end{enumerate}
\end{proof}

\newpage
\printbibliography

\end{document}